\documentclass[11pt]{amsart}

\usepackage{a4wide}
\usepackage{amssymb}
\usepackage{amsmath} 
\usepackage{amscd} 
\usepackage{graphics}

\theoremstyle{plain}
\newtheorem{theorem}{Theorem} 
\newtheorem{lemma}{Lemma}
\newtheorem{prop}{Proposition} 
\newtheorem{coro}{Corollary}

\theoremstyle{definition}

\newtheorem{remark}{Remark}
\newtheorem{example}{Example}

\newcommand{\ts}{\hspace{0.5pt}}
\newcommand{\nts}{\hspace{-0.5pt}}
\newcommand{\conv}{\!\hspace{0.8pt} *\!}

\newcommand{\RR}{\mathbb{R}\ts}
\newcommand{\QQ}{\mathbb{Q}\ts}
\newcommand{\ZZ}{\mathbb{Z}}
\newcommand{\NN}{\mathbb{N}}
\newcommand{\PP}{\mathbb P}
\newcommand{\EE}{\mathbb{E}\ts}
\newcommand{\XX}{\mathbb{X}}

\newcommand{\gG}{\varGamma}
\newcommand{\gL}{\varLambda}
\newcommand{\gT}{\varTheta}

\newcommand{\cA}{\mathcal{A}}
\newcommand{\cM}{\mathcal{M}}
\newcommand{\cN}{\mathcal{N}}
\newcommand{\cB}{\mathcal{B}}
\newcommand{\cX}{\mathcal{X}}
\newcommand{\cL}{\mathcal{L}}

\newcommand{\dd}{\,\mathrm{d}}

\newcommand{\pd}{\rho}

\newcommand{\supp}{\mathrm{supp}}
\newcommand{\dens}{\mathrm{dens}}
\newcommand{\vol}{\mathrm{vol}}
\newcommand{\card}{\mathrm{card}}

\newcommand{\exend}{\hfill $\Diamond$}

\newcommand{\widec}{\smash{\raisebox{18pt}{\rotatebox{180}{$\widehat{}$}}}}

\newcommand{\widecheck}[1]{\hskip 5pt\widec \hskip -5pt {#1}}

\begin{document}


\title[Diffraction of stochastic point sets]
{Diffraction of stochastic point sets: \\[1.5mm]
Explicitly computable examples}

\author{Michael Baake}
\address{Fakult\"at f\"ur Mathematik,
Universit\"at Bielefeld, \newline
\hspace*{12pt}Postfach 100131, 33501 Bielefeld, Germany}
\email{mbaake@math.uni-bielefeld.de}

\author{Matthias Birkner}
\address{Institut f\"ur Mathematik, Universit\"at Mainz, \newline
\hspace*{12pt}Staudingerweg 9, 55099 Mainz, Germany}
\email{birkner@mathematik.uni-mainz.de}

\author{Robert V.\ Moody}
\address{Department of Mathematics and Statistics, 
University of Victoria, \newline
\hspace*{12pt}Victoria, British Columbia V8W 3P4, Canada}
\email{rmoody@uvic.ca}

\begin{abstract} 
  Stochastic point processes relevant to the theory of long-range 
  aperiodic order are considered that display diffraction
  spectra of mixed type, with special emphasis on explicitly
  computable cases together with a unified approach of reasonable
  generality. The latter is based on the classical theory of point
  processes and the Palm distribution.
  Several pairs of autocorrelation and diffraction
  measures are discussed which show a duality structure
  analogous to that of the Poisson summation formula 
  for lattice Dirac combs.
\end{abstract}

\maketitle

\section{Introduction}

The discoveries of quasicrystals \cite{Shechtman}, aperiodic tilings
\cite{Penrose,KN}, and complex metallic alloys \cite{UF} have greatly
increased our awareness that there is a substantial difference between
the notions of periodicity and long-range order.  Although pinning an
exact definition to the concept of long-range order is not yet
possible (nor perhaps desirable at this intermediate stage, compare
the discussion in \cite{crystal}), there is still some general
agreement that the appearance of a substantial point-like component in
the diffraction of a structure is a strong, though not a necessary,
indicator of the phenomenon.

Mathematically, the diffraction -- say of a point set $\gL$ in $\RR^3$
-- is the measure $\widehat{\gamma}$ on $\RR^3$ which is the Fourier
transform of the volume averaged autocorrelation $\gamma$ of $\gL$
(or, more precisely, of its Dirac comb $\delta_{\gL} = \sum_{x\in\gL}
\delta_x$).  Over the past 20 years or so, considerable effort has
been put into understanding the mathematics of diffraction, especially
conditions under which $\gL$ is pure point diffractive, in the sense
that $\widehat{\gamma}$ is a pure point measure, compare
\cite{Hof,C,Martin2,B,BM2,Gou2,L,LS}. At this point in time, we have a
good collection of models for producing pure point diffraction,
particularly the cut and project sets (or model sets). Under certain
types of discreteness conditions, one can even go as far as to say
that these types of sets essentially characterise the pure point
diffractive point sets \cite{BLM}.

But real life structures are not perfectly pure point diffractive, and
in order to gain further insight into the possible structures of
materials, and more generally into the whole concept of long-range
order, it is necessary to widen the scope of this study to include
mixed diffraction spectra. In particular, this means that one has to
consider structures whose diffraction measures contain at least some
continuous part. Note that distinct structures may have the same
diffraction, which makes the corresponding inverse problem
difficult (and generally unsolvable without further information). 
In fact, this gets worse outside the realm of pure point
spectra, and any improvement requires a better understanding of the
diffraction of structures with some form of randomness.

However, when it comes to mixed spectra, relatively little is known,
although there are many particular examples
\cite{EM,BH,Hof-random,K1,Gui,Wel,DM}.  Even deterministic sets can
have mixed diffraction spectra, and once any randomness is introduced,
this is the norm. Determining the exact nature of the diffraction is
usually difficult and often simply not known. No doubt, the
possibilities, both in Nature and in mathematics, for structures with
long-range order are well beyond what we have presently imagined.
This is also made apparent by systems such as the pinwheel tiling,
compare \cite{Radin} and references therein, which looks like an
amorphous structure in diffraction, in spite of being completely
regular. In particular, except for the trivial Bragg peak at $0$,
there is no pure point part in the diffraction.

\smallskip
The primary goal of this paper is to show how the techniques from the
theory of stochastic point processes may be used when systems with
long-range order are subjected to modifications in the form of
stochastic perturbations.  In as much as our primary goal is the study
of long-range order, the types of point processes of most interest to
us are quite different from those usually studied in stochastic
geometry. For instance, as mentioned above, a phenomenon of
fundamental importance in long-range order is the appearance of Bragg
peaks in diffraction, which refers to a non-trivial pure point part of
the diffraction measure. For ergodic point processes, this implies the
existence of non-trivial (dynamical) eigenvalues and thus excludes
weak mixing \cite{W}. Hence, we are particularly interested in systems
that lie between ergodic and weak mixing. Furthermore, a number of
basic and influential mathematical models of long-range order are
deterministic. Even so, the theory of point processes is relevant
\cite{Gou1} and yields considerable insight.

To elaborate on this a little, consider the classic Penrose tilings
\cite{Penrose}.  Fix a set of two generating Penrose prototiles (with
matching rule markers) and a set of overall orientations for them (so
each prototile comes in $10$ distinct orientations).  The resulting
set of admissible tilings of the plane, even if one vertex of the
first tile laid down is fixed, is uncountable. Replacing each of these
tilings by its corresponding vertex point set and allowing all (global)
translations, collectively we obtain the set $\XX$ of all Penrose point 
sets in the given orientation. The set $\XX$, which is also called the
\emph{hull}, has a natural topology in which it is compact, and it
carries a unique (and hence ergodic) stationary Borel probability
measure $\mu$. The pair $(\XX,\mu)$ can now be viewed both as a
dynamical system and as a stationary ergodic point process, which
permits the powerful tools from both subjects to be applied.

In particular, the diffraction and the dynamical spectrum are linked
by the dynamics \cite{BLM}, while the diffraction is also the
Fourier transform of the first
moment measure of the Palm measure of the point process
\cite{Gou1}. This type of scenario, which applies to many models of
long-range order, deterministic or otherwise, has not yet been
investigated in much detail.  It would seem desirable then, as a first
step, to establish methods, capable of being explicitly computable,
that would cover typical and much-studied situations and also suggest
ways in which to generalise what is known, and even move into yet
unexplored territory.

\smallskip
As already implicitly mentioned, mathematical diffraction theory is
based on the approach set out by Hof in \cite{Hof,Hof-random}, namely
via autocorrelations and their measures.  Our paper is primarily
guided by examples, set in as great a generality as we can manage
without becoming too technical. The examples are selected under the
consistent theme that they are computable, while they unify and extend
existing results in a systematic way.  Briefly, the types of
situations that we consider are these:
\begin{itemize}
\item[\rm (i)] renewal processes on the real line (Sec.~3), as a
  versatile, elementary approach to one-dimensional phenomena; the
  main result here is Theorem~\ref{non-lattice-gives-ac};
\item[\rm (ii)] randomisation of a given point set $\gL$ (with a
  certain discreteness restriction) whose diffraction is known, by
  complex, identically distributed, finite random measures that
  are independently centred at each point of $\gL$ (Sec.~4); see
  Theorem~\ref{pprm};
\item[\rm (iii)] randomisation of a random point process $\varPhi$
  (with known law)  by identically distributed, finite, random 
  measures (positive or signed) which are independently centred 
  at each point of any realisation of $\varPhi$ (Sec.~5.1--5.4);
  the main results are formulated in Theorems~\ref{thm:palm} and 
  \ref{compound-thm};
\item[\rm (iv)] equilibria of critical branching Brownian motions
  (Sec.~5.5, with Theorem~\ref{cbbm}).
\end{itemize}

\medskip

\noindent
While (i) and (ii) have a bit of a review character, (iii) and (iv)
are new in this context. The results for (i) and (ii) are included
with details and several examples because they have immediate
applications to practical diffraction analysis, while the results are
scattered over the literature or only covered implicitly.  Since the
transfer of methods and results from stochastic geometry to
mathematical diffraction theory requires a somewhat unusual view on
measures and their Fourier transforms, we begin with a brief
recapitulation of concepts needed here.  For the sake of completeness
and readability (as well as lack of reference), some details on the
ergodic theorem that we need for (iii) are added as an appendix.

\section{Some recollections from Fourier analysis and diffraction
  theory}
\label{sec:prelims}

Throughout the paper, we need various standard notions and results
from Fourier analysis and measure theory, which we summarise here,
introducing our notation at the same time.

First, let $\mu$ be a finite, regular (and possibly complex) Borel
measure on $\RR^d$; compare \cite{Rud} for background.  Its Fourier
(or Fourier-Stieltjes) transform is a uniformly continuous function 
on $\RR^d$, defined by
\[
    \widehat{\mu} (k) \; = \; \int_{\RR^d} e^{-2\pi ikx}
    \dd \mu (x) \ts ,
\]
see \cite{R} for details. This definition includes the Fourier
transform of arbitrary Schwartz functions or integrable functions (the
corresponding spaces being denoted by $\mathcal{S} (\RR^d)$ and $L^{1}
(\RR^d)$) by viewing them as Radon-Nikodym densities for Lebesgue
measure $\lambda$, hence as finite measures.  In this version of the
Fourier transform, with the factor $2\pi$ included in the exponent,
there is no need for prefactors (though the factor $2\pi$ reappears up
front under differentiation). In particular, one has the usual
convolution theorem in the form $\widehat{\mu * \nu} =
\widehat{\mu}\,\widehat{\nu}$, where
\[
   (\mu * \nu) (g) = \int_{\RR^d \times\ts \RR^d}
   g(x+y) \dd \mu (x) \dd \nu (y)
\]
for continuous functions $g \in C_{0} (\RR^{d})$ (note that we
identify finite, complex, regular Borel measures on $\RR^d$ with
continuous linear functionals on $C_{0} (\RR^{d})$, in line with the
Riesz-Markov representation theorem \cite[Thm.~6.19]{Rud}). 

Below, we need to go beyond the situation of finite measures.  For the
introduction of unbounded measures, the linear functional point of
view is advantageous. Here, an \emph{unbounded measure} is thus
understood as a linear functional on the space $C_{\mathsf{c}}
(\RR^{d})$ of continuous functions with compact support such that, for
any compact set $K\subset \RR^{d}$, there is a constant $a=a^{}_{K}$
with $\lvert \mu (g)\rvert \le a^{}_{K} \| g \|_{\infty}$ for all
continuous $g$ with support in $K$. The corresponding space $\cM
(\RR^{d})$ is thus equipped with the vague topology, see
\cite[Ch.~XIII]{Dieu} for details. As before, we can identify these
measures with the locally finite, complex, regular Borel measures on
$\RR^{d}$, by an appropriate version of the Riesz-Markov
representation theorem; see \cite[Thm.~69.1]{Ber} for a formulation
for positive measures and use the polar representation
\cite[Thm.~13.16.3]{Dieu} for an extension to complex measures.  The
absolute value $\lvert \mu \rvert$ of $\mu$, also known as the
\emph{total variation} measure of $\mu$, is the smallest positive
measure such that $\lvert \mu (g)\rvert \le \lvert \mu \rvert (\lvert
g \rvert )$ holds for all $g\in C_{\mathsf{c}} (\RR^{d})$.

When an unbounded measure $\mu$ also defines a tempered distribution,
via $\mu(\varphi) = \int_{\RR^d} \varphi\dd\mu$ for $\varphi\in
\mathcal{S} (\RR^d)$, it is called a \emph{tempered measure}. Its
Fourier transform (as a distribution) is then defined via
$\widehat{\mu} (\varphi) = \mu (\widehat{\varphi})$ as usual
\cite{RS}, so that $\widehat{\mu}$ is a tempered distribution. Below,
we only consider situations where $\widehat{\mu}$ is also a measure,
hence a linear functional on $C_{\mathsf c} (\RR^d)$. Recall that a
(complex) measure $\mu$ is called \emph{translation bounded} when
\[
         \sup_{t\in\RR^{d}} \lvert \mu \rvert (t+K) < \infty
\]
holds for arbitrary compact sets $K\subset\RR^d$.  Translation
boundedness is a sufficient (though not a necessary) criterion for a
measure to be tempered, see \cite{RS} for details.  In this setting,
we call a measure \emph{transformable} when it is tempered and when
its Fourier transform (as a distribution) is again a
measure. Transformability of a measure is a difficult question in
general; see \cite{GdeL} and references therein for details.

A measure $\mu\in\cM (\RR^{d})$ is called \emph{positive definite},
when $\mu (g * \widetilde{g}\ts ) \ge 0$ holds for every function
$g\in C_{\mathsf c} (\RR^d)$; here, $\widetilde{g}$ is the function
defined by $\widetilde{g} (x) = \overline{g(-x)}$. Positive definite
measures have various nice properties, some of which can be
summarised as follows; see \cite[Sec.~4]{BF} for details and proofs.

\begin{lemma} \label{pos-def-props}
  For $\mu\in\cM (\RR^{d})$, the following properties hold.
\begin{itemize}
\item[\rm (i)] If $\mu$ is positive and positive definite, it is
        translation bounded;
\item[\rm (ii)] If $\mu$ is positive definite, it is Fourier
    transformable, and $\widehat{\mu}$ is a positive,
    translation bounded measure;
\item[\rm (iii)] A transformable measure $\mu$ is positive
    definite if and only if\/ $\widehat{\mu}$ is a positive measure;
\end{itemize}
  Moreover, the mapping $\mu \mapsto \widehat{\mu}$ defines a
  bijection between the positive definite and the transformable 
  positive measures on $\RR^{d}$.  \qed
\end{lemma}

Let us consider some examples that will reappear later.  If
$\gG\subset\RR^d$ is a \emph{lattice} (meaning a discrete subgroup of
$\RR^d$ with compact factor group $\RR^d/\gG$), we write $\delta_{\gG}
:= \sum_{x\in\gG} \delta_x$ for the corresponding Dirac comb, with
$\delta_x$ the normalised point measure at $x$.  It is well-known that
$\delta_{\gG}$ is a tempered measure, whose Fourier transform is again
a tempered measure. The latter is explicitly given by the Poisson
summation formula (PSF) in its version for lattice Dirac combs
\cite[Ex.~6.22]{BF},
\begin{equation} \label{PSF-1}
   \widehat{\delta}_{\gG} \; = \; \dens(\gG)\,
   \delta_{\gG^*} ,
\end{equation}
where $\gG^{*} := \{ x\in\RR^d \mid x\cdot y\in\ZZ \mbox{ for all }
y\in\gG \}$ is the \emph{dual lattice} of $\gG$; see \cite{C} for
details.  The density of $\gG$ is well-defined and given by
$\dens(\gG)= 1/\lvert \det(\gG) \rvert$, where $\det(\gG)$ is the
oriented volume of a (measurable) fundamental domain of $\gG$. It can
most easily be calculated as the determinant of a lattice
basis. Observing $\lvert \det (\gG^*)\rvert = 1/ \lvert \det(\gG)
\rvert$, a more symmetric version of the PSF reads
\begin{equation} \label{PSF-2}
  \bigl( \sqrt{\lvert \det(\gG)\rvert} \; 
  \delta_{\gG} \bigr)^{\!\widehat{\hphantom{w}}} 
  \, = \;\, \sqrt{\lvert \det(\gG^*)\rvert} \; \delta_{\gG^*} .
\end{equation}
In particular, one has $\widehat{\delta}_{\ZZ^d} = \delta_{\ZZ^d}$,
so that the lattice Dirac comb of $\ZZ^d$ is self-dual in this sense.

\begin{remark} \textsc{Radially symmetric PSF}.
As an aside of independent interest, let us recall the following
related formula for a radially symmetric situation in $\RR^d$, which
emerges from a simplified model of powder diffraction \cite{BFG}.
Let $\gG$ and $\gG^*$ be as before, and let $\eta^{}_{\gG} (r)$
and $\eta^{}_{\gG^*} (r)$ denote the numbers of points of $\gG$ and 
$\gG^*$ on centred spheres $\partial B_r (0)$ of radius $r$. The
(non-zero) numbers $\eta^{}_{\gG} (r)$ are also called the
\emph{shelling numbers} of the lattice $\gG$. If $\mu_r$ denotes
the uniform probability measure on $\partial B_r (0)$, with
$\mu^{}_{0} = \delta^{}_{0} $, one has
the following radial analogue of the PSF in \eqref{PSF-1},
\begin{equation} \label{radial-PSF}
  \Bigl(\sum_{r\in\mathcal{D}^{}_{\!\gG}} \eta^{}_{\gG} (r)
  \, \mu_{r} \Bigr)^{\!\widehat{\hphantom{m}}}
  = \; \dens (\gG)\sum_{r\in\mathcal{D}^{}_{\!\gG^*}} 
  \eta^{}_{\gG^*} (r)\, \mu_{r} \/ ,
\end{equation}
where $\mathcal{D}^{}_{\!\gG} = \{r\ge 0\mid \eta^{}_{\gG} (r)>0\}$
and analogously for $\mathcal{D}^{}_{\!\gG^*}$, see \cite{BFG} for
a proof and further details. The formula can also be
brought to a more symmetric form, as in Eq.~\eqref{PSF-2}.
\exend
\end{remark}

Another simple, but important, pair of mutual Fourier transforms
follows from the relations $\widehat{\delta}_0 = \lambda$ and
$\widehat{\lambda}=\delta_0$, with $\lambda$ being Lebesgue measure,
so that we have
\begin{equation} \label{poisson-lebesgue}
   \bigl( \delta_0 + \lambda \bigr)^{\!\widehat{\hphantom{w}}}
   \, = \;\, \delta_0 + \lambda .
\end{equation}
We shall meet this self-dual pair of measures below in 
Examples~\ref{poisson-on-the-line} and \ref{ex-poisson},
in connection with the Poisson process.

\smallskip
A little less obvious is the following result.
\begin{lemma} \label{lemma-1}
  Let $\lambda$ denote Lebesgue measure on $\RR^d$ and\/ $0 < \alpha <
  d$. The function $x\mapsto 1/\lvert x\rvert^{d-\alpha}$ is locally
  integrable and, when seen as a Radon-Nikodym density for $\lambda$,
  defines an absolutely continuous and translation bounded
  measure on $\RR^d$.  This measure satisfies the identity
\[
   \left( \frac{\Gamma\bigl(\tfrac{d-\alpha}{2} \bigr)}
   {\pi^{\tfrac{d-\alpha}{2}}}\; 
   \frac{\lambda}{\lvert x\rvert^{d-\alpha}}
   \right)^{\!\!\widehat{\hphantom{w}}} \! (k) \; = \; \;
   \frac{\Gamma\bigl(\tfrac{\alpha}{2} \bigr)}
   {\pi^{\tfrac{\alpha}{2}}}\; 
   \frac{\lambda}{\lvert k\rvert^{\alpha}} ,
\]
   where the transformed measure is again translation bounded
   and absolutely continuous. Moreover, both measures
   are positive and positive definite.
\end{lemma}

\begin{proof}
  Local integrability of both measures on $\RR^d$ rests upon that of
  their densities around $0$, which follows from rewriting the volume
  element in polar coordinates, $\dd \lambda(x) = r^{d-1} \dd r \dd
  \varOmega$, with $\dd \varOmega$ the standard surface element of the
  unit sphere in $\RR^d$. Absolute continuity and translation
  boundedness are then clear, while the Fourier identity follows from
  a calculation with the heat kernel, see \cite[Sec.\ 2.2.3]{Pinsky}.
  As both measures are clearly positive, they are also positive
  definite by the Bochner-Schwartz theorem \cite[Thm.~IX.10]{RS},
  compare Lemma~\ref{pos-def-props}.
\end{proof}

Incidentally, dividing the identity in Lemma~\ref{lemma-1} by
$\Gamma(\alpha/2)/\pi^{\alpha/2}$ shows that
\begin{equation}\label{lemma2-limit-formula}
   \frac{\Gamma\bigl(\tfrac{d-\alpha}{2}\bigr)\,\pi^{\tfrac{\alpha}{2}}}
   {\Gamma\bigl(\tfrac{\alpha}{2}\bigr)\,\pi^{\tfrac{d-\alpha}{2}}}\; 
   \frac{\lambda}{\lvert x\rvert^{d-\alpha}}
   \; \; \xrightarrow{\,\alpha\ts\to\ts 0}\;\; \delta_{0}
\end{equation}
in the vague topology, which follows from the corresponding Fourier 
transforms of the left hand side converging vaguely to $\lambda$.

\smallskip Let us now briefly review the concept of the diffraction
measure of a complex measure $\omega$ as the Fourier transform of the
autocorrelation $\gamma$ of $\omega$. It motivation comes from the
physics of diffraction \cite{Cowley}, while its precise mathematical
formulation was pioneered by Hof \cite{Hof}.

In general, a complex measure $\omega$ need not be transformable, and
may thus not be a good object for harmonic analysis. In view of
Lemma~\ref{pos-def-props}, it seems appealing to first attach a
positive definite measure to $\omega$, which is possible as follows.
If $\omega_r$ denotes the restriction of $\omega$ to the open ball
$B_r$ of radius $r$ around $0$, the natural \emph{autocorrelation
  measure} $\gamma = \gamma^{}_{\omega}$ is defined as
\begin{equation} \label{def-autocorr}
     \gamma := \lim_{r\to\infty}
     \frac{\omega_r *\ts \widetilde{\omega_r}}{\vol (B_r)} \ts ,
\end{equation}
provided the limit exists. Here, $\widetilde{\mu}$ denotes the measure
given by $\widetilde{\mu} (g) = \overline{ \mu (\widetilde{g}\ts )}$ for
$g\in C_\mathsf{c} (\RR^{d})$, with $\widetilde{g}$ as before. If
$\omega$ is translation bounded, the one-parameter family of finite
measures $\bigl\{\frac{\omega_r *\ts \widetilde{\omega_r}}{\vol (B_r)}
\mid r>0\bigr\}$ is uniformly translation bounded and hence precompact
in the vague topology by \cite[Prop.~2.2]{Hof}. One can thus always
select converging subsequences to define \emph{an} autocorrelation
(which then depends on the sequence of averaging sets). As long as
balls are used, one speaks of \emph{natural} autocorrelations. More
generally, one may work with any \emph{averaging sequence}
\[
   \cA = \{ A_n \mid n\in\NN \}
\] 
of relatively compact, open sets $A_n \subset \RR^{d}$ that satisfy
$\,\overline{\! A_{n}\!}\, \subset A_{n+1}$ for all $n\in\NN$ together
with $\bigcup_{n\in\NN} A_n = \RR^{d}$. Again, for translation bounded
measures $\omega$, the corresponding limit in \eqref{def-autocorr}
exists, at least along suitable subsequences.

An important further ingredient is the concept of a van Hove sequence,
which is an averaging sequence with a restricted `surface to volume'
ratio. To formalise this, let $K,C\subset \RR^{d}$ be compact and
define
\begin{equation}\label{vH1}
   \partial^{K} C := \bigl( ( C + K) \setminus C^{\circ} \bigr)
   \cup \bigl( (\, \overline{\RR^{d}\setminus C} - K)
   \cap C \bigr),
\end{equation}
which may be viewed as a $K$-thickened boundary of $C$.
Then, $\cA$ is called \emph{van Hove} when, for every compact
$K\subset\RR^{d}$,
\begin{equation}\label{vH2}
      \lim_{n\to\infty}
     \frac{\vol (\partial^{K} \,\overline{\! A_n \!}\, )}
      {\vol (A_n)} = 0 \ts .
\end{equation}
Now, the comparison of limits taken along different averaging
sequences makes sense, and becomes independent of $\cA$ for ergodic
systems; compare \cite[Lemma~1.1]{Martin2}.  Also, as follows from
\cite[Lemma~1.2]{Martin2}, translation bounded measures satisfy the
relation
\begin{equation} \label{convolution-freedom}
   \lim_{n\to\infty}
  \frac{\omega_n *\ts \widetilde{\omega_n}}{\vol (A_n)}
   \, = \lim_{n\to\infty}
  \frac{\omega_n *\ts \widetilde{\omega}}{\vol (A_n)} \ts ,
\end{equation}
provided that $\cA$ is van Hove and one of the limits exists. Here,
$\omega_n = \omega |^{}_{A_n}$, and $\omega_n * \widetilde{\omega}$ is
well-defined by \cite[Prop.~1.13]{BF}. This freedom will be used
several times below.

The general situation for a translation bounded measure $\omega$
is as follows. The van Hove property of $\cA$ implies that $\lvert
\omega\rvert (A_n) \le c \, \vol(A_n)$ with a constant $c>0$. An
obvious modification of \cite[Prop.~2.2]{Hof} in conjunction with
\cite[Lemma~1.2]{Martin2} then gives the following result.
\begin{lemma} \label{lem:utb} 
  Let $\omega$ be a translation bounded measure, and $\cA$ a van Hove
  averaging sequence. With $\gamma^{}_{n}:= \frac{\omega_{n} *\,
  \widetilde{\omega_{n}}}{\vol (A_n)}$ and
  $\gamma^{}_{n;\mathrm{mod}}:= \frac{\omega^{}_{n} *\,
  \widetilde{\omega}}{\vol (A_n)}$, the families
  $\{\gamma^{}_{n} \mid n\in\NN \}$ and
  $\{\gamma^{}_{n;\mathrm{mod}} \mid n\in\NN \}$ are
  uniformly translation bounded and hence precompact in
  the vague topology. Any accumulation point of either family,
  of which there is at least one, is also an accumulation
  point of the other family, and a translation bounded,
  positive definite measure.    \qed
\end{lemma}

Lemma~\ref{pos-def-props} applies to any autocorrelation measure, and
the corresponding measure $\widehat{\gamma}$ is then a positive,
translation bounded measure. It is called the \emph{diffraction
  measure} of $\omega$, relative to the averaging sequence $\cA$.  In
ergodic situations, we have no dependence on $\cA$ and thus suppress
it. Then, the diffraction measure is also related to the Bartlett
spectrum known from stochastic geometry, though there are important
differences to be discussed later; see Remark~\ref{rem:Bartlett}
below.

In general, an interesting initial question concerns the spectral type
of $\widehat{\gamma}$, which follows from the spectral decomposition
\begin{equation} \label{diffrac-1}
   \widehat{\gamma} \; = \;
   \bigl(\widehat{\gamma}\bigr)_{\sf pp} +
   \bigl(\widehat{\gamma}\bigr)_{\sf sc} +
   \bigl(\widehat{\gamma}\bigr)_{\sf ac}
\end{equation}
of $\widehat{\gamma}$ into its pure point, singular continuous and
absolutely continuous parts relative to $\lambda$, the latter being
the Haar measure on $\RR^{d}$ with $\lambda \bigl( [0,1]^{d} \bigr) =
1$.  Lattices and regular model sets \cite{Martin2,BM2} are examples
with $\widehat{\gamma} = (\widehat{\gamma})^{}_{\sf pp}$, while the
Thue-Morse and the Rudin-Shapiro sequence show singular continuous and
absolutely continuous components, respectively; compare \cite{HB} and
references given there. Absolutely continuous components appearing as
a result of stochastic influence are the main theme below.

\section{Renewal processes in one dimension}

An illustrative class of examples is provided by the classical renewal
process on the real line, defined by a probability measure $\varrho$
on $\RR_+ = \{ x>0 \}$ of finite mean as follows. Starting from some
initial point, at an arbitrary position, a machine moves to the right
with constant speed and drops a point on the line with a random
waiting time that is distributed according to $\varrho$.  When this
happens, the clock is reset and the process resumes.  In what follows,
we assume that both the velocity of the machine and the expectation
value of $\varrho$ are $1$, so that we end up (in the limit that we
let the initial point move to $-\infty$) with realisations that are
almost surely point sets in $\RR$ of density $1$.

Clearly, the process just described defines a stationary process.  It
can thus be analysed by considering all realisations which contain the
point $0$.  Moreover, there is a clear (distributional) symmetry
around this point, so that we can determine the autocorrelation (in
the sense of \eqref{def-autocorr}) of almost all realisations from
studying what happens to the right of $0$ (we will make this
approach rigorous in Proposition~\ref{renewal-1} below). Indeed, if we 
want to know the frequency per unit length of the occurrence of two points
with distance $x$ (or the corresponding density), we need to sum
the contributions that $x$ is the first point after $0$, the second
point, the third, and so on. In other words, we almost surely obtain
the autocorrelation
\begin{equation} \label{auto-1}
    \gamma \; = \; \delta_0 + \nu + \widetilde{\nu}
\end{equation}
with $\nu = \varrho + \varrho \conv \varrho + \varrho \conv \varrho
\conv \varrho + \ldots = \sum_{n=1}^{\infty} \varrho^{*n}$ 
and $\widetilde{\nu}$ as defined above,
provided that the sum in Eq.~\eqref{auto-1} converges properly. Note
that the point measure at $0$ simply reflects that the almost sure
density of the resulting point set is $1$. In the slightly more
general case of a probability measure $\varrho$ on $\RR_{+} \cup
\{0\}$, one has the following convergence result. It is essentially a
measure theoretic reformulation of the main lemma in
\cite[Sec.~XI.1]{Feller}, but we prefer to give a complete proof that
is adjusted to our setting.

\begin{lemma} \label{convergence}
  Let $\varrho$ be a probability measure on $\RR_{+}\cup \{0\}$,
  with $\varrho (\RR_{+})>0$. Then, $\nu^{}_n := \varrho +
  \varrho \conv \varrho + \ldots + \varrho^{*n}$ with $n\in\NN$
  defines a sequence of positive measures that converges
  towards a translation bounded measure $\nu$ in the vague
  topology.
\end{lemma}

\begin{proof}
  Note that the condition $\varrho (\RR_{+})>0$ implies $0\le
  \varrho(\{0\}) < 1$, hence excludes the case $\varrho = \delta_0$.
  When $\varrho=\delta_{a}$ for some $a>0$, one has $\nu_{n}^{}=
  \sum_{m=1}^{n} \delta_{ma}^{}$ by a simple convolution calculation,
  and the claim is obvious. In all remaining cases,
  it is possible to choose some $a\in\RR_{+}$ with
  $\varrho(\{a\})=0$ and $0<\varrho([0,a))=p < 1$, so that also
  $\varrho([a,\infty))=1-p  <  1$.  Since the sequence $\nu^{}_n$ is
  monotonically increasing, the claimed vague convergence follows from
  showing that $\limsup_{n\to\infty} \nu^{}_n ([0,x))$ is bounded by
  $C_1 + C_2 x$ for some constants $C_i$.  As there are at most
  countably many points $y$ with $\varrho(\{y\})>0$, it is sufficient
  to show these estimates for all $x\in\RR_{+}$ with
  $\varrho(\{x\})=0$. In a second step, we then demonstrate that
  $\sum_{n=1}^{\infty} \varrho^{*n} ([b,b+x))$ is bounded by $1+ C_1 +
  C_2 x$, independently of $b$, which establishes translation
  boundedness.

  If $(X_i)^{}_{i\in\NN}$ denotes a family of i.i.d.\ random variables,
  with common distribution according to $\varrho$ (and thus values in
  $\RR_{+}\cup \{0\}$), one has
\[
     \PP \bigl( X_1 + \ldots + X_m < x \bigr) \; = \;
     \varrho^{*m} ([0,x)).
\]
On the other hand, for the $a$ chosen above, one has the inequality
\[
    \PP ( X_1 + \ldots + X_m < x)  \; \le \;
    \PP \bigl( \mathrm{card} \{1\le i\le m \mid
    X_i \ge a \} \le x/a \bigr) 
    \; = \; \sum_{\ell=0}^{[x/a]} \binom{m}{\ell}\,
    (1-p)^\ell \, p^{m-\ell} ,
\]
where $\binom{m}{\ell}=0$ whenever $\ell > m$.
Observing $\sum_{m=1}^{\infty} p^m = p/(1-p)$ and
\[
   \sum_{m=1}^{\infty} \binom{m}{\ell}\,
   (1-p)^\ell\, p^{m-\ell} \; = \;
   (1-p)^\ell \, \frac{1}{\ell !} \frac{\dd^\ell}{\dd p^\ell}
   \sum_{m=0}^{\infty} p^m \; = \; \frac{1}{1-p}
\]
for all $\ell\ge 1$, the previous inequality
implies, for arbitrary $n\in\NN$,
\[
    \nu^{}_n ([0,x)) \; \le \;
    \sum_{m=1}^{\infty} \sum_{\ell=0}^{[x/a]}
    \binom{m}{\ell}\, (1-p)^\ell \, p^{m-\ell}
    \; = \; \frac{p+[x/a]}{1-p} \; \le \;
    \frac{p}{1-p} + \frac{1}{a (1-p)}\, x ,
\]
which establishes the first claim.

For the second estimate, we choose $b\ge 0$, $x>0$ and observe
\[
\begin{split}
   \sum_{n=1}^\infty & \varrho^{*n} \bigl([b,b+x)\bigr) 
   \, = \sum_{n=1}^\infty \PP \bigl(b 
   \le X_1+\cdots+X_n < b+x \bigr) \\
   & = \sum_{n=1}^\infty \sum_{k=1}^n 
   \PP \bigl( X_1+\cdots+X_{k-1} \le b \le X_1+\cdots+X_{k} 
      \text{ and } b \le  X_1+\cdots+X_n < b+x \bigr) \\
   & \le  \sum_{n=1}^\infty \sum_{k=1}^n 
    \PP\bigl(X_1+\cdots+X_{k-1} \le b \le X_1+\cdots+X_{k}\bigr) 
    \; \PP\bigl(X_{k+1}+\cdots+X_n < x \bigr) \\
   & = \sum_{k=1}^\infty \PP\bigl( X_1+\cdots+X_{k-1} 
     \le b \le X_1+\cdots+X_{k}\bigr) 
   \sum_{n=k}^\infty \PP \bigl(X_{k+1}+\cdots+X_n < x \bigr) \\
   & =  1+ \sum_{m=1}^\infty \PP \bigl(X_1+\cdots+X_m < x \bigr),
\end{split}
\]
with the convention to treat empty sums of random variables
as $0$. The last step
used the i.i.d.\ property of the random variables together with
$\PP (0\le X) = 1$ and 
\[
   \sum_{k=1}^{\infty} \PP \bigl(X_1+\cdots+X_{k-1} \le b 
   \le X_1+\cdots+X_{k}\bigr) = 1 \ts .
\]
In conjunction with our previous estimate, this completes the proof.
\end{proof}

When $\varrho(\{0\}) > 0$, we are outside the realm of (renewal) point
processes, and formula \eqref{auto-1} for the autocorrelation no
longer applies. This case might nevertheless be analysed with the
methods of Sections \ref{elementary} and \ref{general}, see
Example~\ref{random-weight} and Corollary~\ref{compound-diffraction}
in particular.  For the remainder of this section, we assume
$\varrho(\{0\}) =0$, so that $\varrho$ is a measure on $\RR_{+}$; see
Remark~\ref{renewal-revisited} below for an alternative approach via
random counting measures, or \cite[Ch.~XI.9]{Feller}.

\begin{prop} \label{renewal-1} 
  Consider a renewal process on the real line, defined by a
  probability measure $\varrho$ on\/ $\RR_{+}$ with mean $1$.  This
  defines a stationary stochastic process, whose realisations are
  point sets that almost surely possess the autocorrelation measure
  $\gamma = \delta_{0} + \nu + \widetilde{\nu}$ of\/ $\eqref{auto-1}$.
  
  Here, $\nu = \sum_{n=1}^{\infty} \varrho^{*n}$ is a translation
  bounded positive measure. It satisfies the renewal equations
\[
   \nu \; = \; \varrho + \varrho * \nu \qquad\text{and}\qquad
   (1-\widehat{\varrho}\ts\ts )\, \widehat{\nu} 
    \; = \; \widehat{\varrho}\, ,
\]
  where $\widehat{\varrho}$ is a uniformly continuous function
  on\/ $\RR$. In this setting, the measure $\gamma$ is both 
  positive and positive definite.
\end{prop}

\begin{proof}
  The renewal process is a classic stochastic process on the real line
  which is known to be stationary and ergodic; compare
  \cite[Ch.~VI.6]{Feller} for details.  Consequently, the measure of
  occurrence of a pair of points at distance $x + \dd x$ (or the
  corresponding density) can be calculated by fixing one point at $0$
  (due to stationarity) and then determining the ensemble average for
  another point at $x + \dd x$ (due to ergodicity). This is the
  justification for the heuristic reasoning given above, prior to
  Eq.~\eqref{auto-1}.

  By Lemma~\ref{convergence}, $\nu$ is a translation bounded measure,
  so that the convolution $\varrho * \nu$ is well defined by
  \cite[Prop.~1.13]{BF}.  The first renewal identity is then clear
  from the structure of $\nu$ as a limit, while the second follows by
  Fourier transform and the convolution theorem. The autocorrelation
  is a positive definite measure by construction, though this is not
  immediate here on the basis of its form as a sum, see \cite{A} for a
  related discussion. 
\end{proof}

Let us now consider the spectral type of the resulting diffraction
measure for the class of point sets generated by a renewal process.
This requires a distinction on the basis of the support of
$\varrho$. To this end, the second identity of
Proposition~\ref{renewal-1} is helpful, because one has
\begin{equation} \label{series-1}
     \widehat{\nu} (k) \; = \; \frac{\widehat{\varrho} (k)}
     {1-\widehat{\varrho} (k) }
\end{equation}
at all positions $k$ with $\widehat{\varrho} (k) \neq 1$.  This is in
line with summing $\widehat{\nu}$ as a geometric series, which gives
the same formula for $\widehat{\nu} (k)$ for all $k$ with
$\lvert\widehat{\varrho} (k)\rvert < 1$ and has \eqref{series-1} as
the unique continuous extension to all $k$ with
$\lvert\widehat{\varrho} (k)\rvert = 1\neq\widehat{\varrho} (k)$. In
fact, one sees that $\widehat{\nu} (k)$ is a continuous function on
the complement of the set $\{ k\in\RR\mid \widehat{\varrho} (k) =
1\}$.  For most $\varrho$, the latter set happens to be the singleton
set $\{0\}$.

In general, a probability measure $\mu$ on $\RR$ is called
\emph{lattice-like} when its support is a subset of a translate of a
lattice, see \cite{Gne} for details. We need a slightly stronger
property here, and call $\mu$ \emph{strictly lattice-like} (called
\emph{arithmetic} in \cite{Feller}) when its support is a subset of a
lattice. So, the difference is that we do not allow any translates
here; see \cite{B} for related results.
   
\begin{lemma} \label{lattice-like}
  If $\mu$ is a probability measure on $\RR$, its
  Fourier transform, $\widehat{\mu} (k)$, is a uniformly
  continuous and positive definite function on $\RR$, with
  $\lvert\ts \widehat{\mu} (k)\rvert \le \widehat{\mu} (0)
  = 1$. 

  Moreover, the following three properties are equivalent.
\begin{itemize}
\item[\rm (i)] $\mathrm{card} \{k\in\RR \mid
  \widehat{\mu} (k) = 1 \} > 1 ;$\vspace{1mm}
\item[\rm (ii)] $\mathrm{card} \{k\in\RR \mid
  \widehat{\mu} (k) = 1 \} = \infty ;$\vspace{1mm}
\item[\rm (iii)] $\mathrm{supp} (\mu)$ is strictly lattice like.
\end{itemize}
\end{lemma}

\begin{proof}
One has $\widehat{\mu} (k) = \int_{\RR} e^{-2\pi i k x} \dd\mu (x)$,
whence the first claims are standard consequences of Fourier
analysis; compare \cite[Prop.~5.2.1]{Pinsky} and \cite[Sec.~1.3.3]{R}.

If $\mu = \sum_{x\in\gG} p(x) \delta_x$ for a lattice $\gG\subset\RR$,
with $p(x)\ge 0$ and $\sum_{x\in\gG} p(x) = 1$, one has
\[
   \widehat{\mu} (k) \; = \;
   \sum_{x\in\gG} p(x)\, e^{-2\pi i k x} ,
\]
so that $\widehat{\mu} (k) = 1$ for any $k\in\gG^*$. In particular,
$\gG^*\subset \{k\in\RR \mid \widehat{\mu} (k) = 1 \}$, so that we
have the implications (iii) $\Rightarrow$ (ii) $\Rightarrow$
(i).

Conversely, if $\widehat{\mu} (k) =1$ for some $k\neq 0$, one has 
$\int_{\RR} e^{-2\pi i k x} \dd\mu (x) = 1$ and hence
\begin{equation} \label{zero-integral}
   \int_{\RR}\bigl( 1- \cos (2\pi kx)\bigr) \dd\mu (x) \; = \; 
   \int_{\mathrm{supp}(\mu)}\bigl( 1- \cos (2\pi kx)\bigr) 
   \dd\mu (x) \; = \; 0 ,
\end{equation}
where $\mathrm{supp} (\mu)$, the support of the probability measure
$\mu$, is a closed subset of $\RR$ and measurable. The integrand is a
continuous non-negative function that, due to $k\neq 0$, vanishes
precisely on the set $\frac{1}{k}\ZZ$, which is a lattice.

Write $\mathrm{supp}(\mu)=A \dot{\cup} B$ as a disjoint union of
measurable sets, with $A=\mathrm{supp}(\mu) \cap \frac{1}{k}\ZZ$ and
$B=\mathrm{supp} (\mu)\cap (\RR\setminus\frac{1}{k}\ZZ)$. We can now
split the second integral in \eqref{zero-integral} into an integral
over $A$, which vanishes because the integrand does, and one over the
set $B$, which would give a positive contribution by standard
arguments, unless $B=\varnothing$.  But this means $\mathrm{supp}(\mu)
= A \subset \tfrac{1}{k} \ZZ$, so that (i) $\Rightarrow$ (iii), which
establishes the result.
\end{proof}

At this point, we can state the main result of this section, the
diffraction properties of renewal processes; compare 
\cite[Ex.~8.2\ts (b)]{DVJ1} for a special case.

\begin{theorem} \label{non-lattice-gives-ac} 
  Let $\varrho$ be a probability measure on $\RR_+$ with mean $1$, and
  assume that $\varrho$ is not strictly lattice-like. Assume further
  that a moment of $\varrho$ of order $1+\varepsilon$ exists for some
  $\varepsilon > 0$. Then, the point sets obtained from the
  stationary renewal process based on $\varrho$ almost surely have a
  diffraction measure of the form $\widehat{\gamma}  = \delta_0 + 
    \bigl( \widehat{\gamma} \bigr)_{\sf ac}$ with
\[
    \bigl( \widehat{\gamma} \bigr)_{\sf ac} \; = \;
    \frac{1-\lvert\widehat{\varrho}(k)\rvert^2}
    {\lvert 1- \widehat{\varrho}(k)\rvert^2}\,\lambda
    \; = \; (1-h)\,\lambda ,
\]
where $h$ is a continuous function on\/
$\RR\setminus \{0\}$ that is locally integrable. It is given by
\[
    h(k) \; = \; \frac{2\,\bigl(\lvert\widehat{\varrho} 
    (k)\rvert^2 - \mathrm{Re} (\widehat{\varrho}(k))\bigr)}
    {\lvert 1 - \widehat{\varrho} (k)\rvert^2}
\]
and measures the difference from a constant background as
described by $\lambda$.
\end{theorem}

When $\varrho$ is strictly lattice-like, the pure point part
becomes a lattice Dirac comb, and the behaviour of $h$ at $0$
repeats at each point of the underlying lattice, see
Remark~\ref{renewal-ext} for details.

\begin{proof}
  The process has a well-defined autocorrelation $\gamma$, by an
  application of Proposition~\ref{renewal-1}, in the sense that almost
  every realisation of the process is a point set $\gL$ with this
  autocorrelation. Since $\gamma$ is a positive definite measure, it
  is Fourier transformable by Lemma ~\ref{pos-def-props}(ii), with
  $\widehat{\gamma}$ being a positive measure on $\RR$.

  The point measure at $0$ with intensity $1$ reflects the fact that
  the resulting point set $\gL$ almost surely has density $1$. To see
  this, define $g_n = \frac{1}{n}
  \mathbf{1}_{[-\frac{n}{2},\frac{n}{2}]}$ and $h_n = g_n *
  \widetilde{g_n}$. Here, $h_n$ is a positive definite, tent-shaped
  function with support $[-n,n]$ and maximal value $\frac{1}{n}$ at
  $0$. It has (inverse) Fourier transform $ \widecheck{h_n} (k) =
  \bigl(\frac{\sin(\pi k n)}{\pi k n}\bigr)^2$, which is a
  non-negative function (with maximum value $1$ at $k=0$) that
  concentrates around $0$ as $n\to\infty$. Let $\omega_r :=
  \delta^{}_{\gL\ts\cap\ts [-r,r]}$.  Using \eqref{def-autocorr}
  together with $\bigl(\omega_r * \widetilde{\omega_r}\bigr) (g_n *
  \widetilde{g_n}) \ge 0$, it is not difficult to see that $\gamma
  (h_n) \xrightarrow{\, n\to\infty\,} \bigl( \dens (\gL)\bigr)^2$,
  which is almost surely $1$ (for this, assume first that
  $r\gg n\gg 1$, then take the limit $r\to\infty$ followed 
  by the limit $n\to\infty$).  On the other hand, one has
\[
     \gamma \ts (h_n)  \, = \widecheck{\,\widehat{\gamma}} \ts (h_n)
     \, = \, \widehat{\gamma} \ts (\widecheck{h_n})
     \; \xrightarrow{\, n\to\infty\,} \; \widehat{\gamma} 
     \bigl( \{0\} \bigr),
\]  
  due to the concentration property of $\widecheck{h_n}$ (in
  particular, for all $\varepsilon > 0$, one verifies the relation
  $\widehat{\gamma} \bigl(B_{\varepsilon} (0)\bigr) \ge \bigl( \dens
  (\gL)\bigr)^2 > 0$, which proves the existence of a point measure at
  $0$).
  
  Due to the assumption
  that $\mathrm{supp} (\varrho)$ is not contained in a lattice, we may
  invoke Lemma~\ref{lattice-like} to see that $\widehat{\varrho}
  (k)\neq 1$ whenever $k\neq 0$, so that we have pointwise convergence
\[
    \widehat{\nu}^{}_n (k) \; \xrightarrow{n\to\infty} \;
    \widehat{\nu} (k) \; = \; 
    \frac{\widehat{\varrho} (k)}{1 - \widehat{\varrho} (k)}
\]
on $\RR\setminus\{0\}$, and similarly for $\widehat{\widetilde{\nu}}$.
Since $\widehat{\varrho}\ts$ is uniformly continuous on $\RR$ and
$\widehat{\varrho} (k) \neq 1$ on $\RR\setminus\{0\}$, both
$\widehat{\nu}$ and $\widehat{\widetilde{\nu}}$ are represented, on
$\RR\setminus\{0\}$, by continuous Radon-Nikodym densities. Writing
$(\delta_{0} + \nu + \widetilde{\nu} \, )^{\widehat{\hphantom{w}}}\!
= (1-h)\ts\lambda$, hence $(\nu + \widetilde{\nu} \,
)^{\widehat{\hphantom{w}}}\! = - h\ts \lambda $, the formula for $h$
follows from $\widehat{\widetilde{\nu}}=\overline{\widehat{\nu}}$.

It remains to show that $1-h$ is locally integrable near $0$. Let $X$
be a random variable with distribution $\varrho$. Since the latter
has mean $1$ and our assumption guarantees that
$\langle X^{1+\varepsilon}\rangle = \int_{0}^{\infty} x^{1+\varepsilon}
\dd \varrho (x) < \infty$, we have the Taylor series expansion
\[
    \widehat{\varrho} (k) \, = \, 1 -2\pi i k + \mathcal{O} \bigl(
    \vert k \rvert^{1+\varepsilon}\bigr)\ts , \quad
    \text{as $\lvert k \rvert \to 0$}\ts ,
\]
by an application of \cite[Thm.~1.5.4]{Ush}.  Inserting this into
the expression for $h$ results in
\[
     h(k) \, = \, 2 + \mathcal{O} (k^{-1+\varepsilon})\ts ,
     \quad \text{as $\lvert k \rvert \to 0$}\ts ,
\]
which establishes integrability around $0$, and thus absolute
continuity of the measure $(1-h)\lambda$.

As the contribution to the central peak is already completely
accounted for by the term $\delta_{0}$, the claim follows.
\end{proof}

\begin{remark}\textsc{Asymptotic behaviour of $h$}.
  When, in the setting of
  Theorem~\ref{non-lattice-gives-ac}, the second moment of $\varrho$
  exists, one obtains from \cite[Thm.~1.5.3]{Ush} the slightly
  stronger expansion
\[
    \widehat{\varrho} (k) \, = \, 1 -2\pi i k 
    -2\pi^2 \langle X^2 \rangle\, k^2 +\,
    {\scriptstyle \mathcal{O}} 
    \bigl(\vert k \rvert^{2}\bigr)\ts , \quad
    \text{as $\lvert k \rvert \to 0$}\ts .
\]
This leads to the asymptotic behaviour
\[
    h (k) \, = \, 2 - \langle X^2 \rangle +
    {\scriptstyle \mathcal{O}} (1) \ts ,
    \quad \text{as $\lvert k \rvert \to 0$}\ts ,
\]
which implies that $h$ is bounded and can continuously be extended to
$h(0) = 2 - \langle X^2 \rangle = 1- \sigma^2$, where $\sigma^2$ is
the variance of $\varrho$.  Clearly, the existence of higher moments
implies stronger smoothness properties.  \exend \end{remark}

\begin{remark} \textsc{Complement of Theorem~\ref{non-lattice-gives-ac}}.
\label{renewal-ext}
When $\varrho$ happens to be strictly lattice like, the $\ZZ$-span of
the finite or uniformly discrete\footnote{Recall that a set $S \subset
  \RR^{d}$ is called \emph{uniformly discrete} when there is a
  number $s>0$ such that the distance between any two distinct 
  points of $S$ is at least $s$.} set
$\supp (\varrho)$ is a lattice of the form $\gG = b\ts \ZZ$, where
$b>0$ is unique (in other words, $\gG$ is the coarsest lattice in
$\RR$ that contains $\supp (\varrho)$). Then, one finds the
diffraction
\[
    \widehat{\gamma} = \delta^{}_{\ZZ/b} + (1-h) \lambda \ts ,
\]
with the function $h$ from Theorem~\ref{non-lattice-gives-ac}.  Note
that $h$ is well-defined (and continuous) on $\RR\setminus\gG^*$, with
$\gG^* = \ZZ/b$ being the dual lattice of $\gG$. Moreover, it is
locally integrable around all points of $\gG^*$, so that
$(1-h)\ts\lambda$ is again an absolutely continuous measure.  Note
that, since the underlying point set is always a subset of $\gG$, the
diffraction measure is periodic, with $\gG^*$ as its lattice of
periods; compare \cite{B} for general results in this direction.

When $\supp (\varrho)$ is a finite set, one is in the situation of a
random tiling with finitely many prototiles. A more detailed
discussion, together with an explicit calculation of $h$ for this
case, is given in \cite[Thm.~2]{BH}; see also
Example~\ref{random-tiling} and Remark~\ref{rem:Meyer}.
\exend \end{remark}

Let us turn to some examples, for which we employ the
Heavyside function,
\begin{equation} \label{heavyside}
   \gT (x) \; := \;
   \begin{cases} 1, & \text{if $x>0$}, \\
              \tfrac{1}{2}, & \text{if $x=0$}, \\
                 0, & \text{if $x<0$}.
   \end{cases}
\end{equation}
This formulation of $\gT$ has some advantage for formal calculations
around generalised functions and their Fourier transforms.

\begin{example} \textsc{Poisson process on the real line}.
\label{poisson-on-the-line}
The probably best-known stochastic process is the classical
(homogeneous) Poisson process on the line, with intensity $1$, where
$\varrho = f\lambda$ is given by the density
\[
    f(x) \; = \; e^{-x}\,\gT(x).
\]
It is easy to check that the convolution of $n+1$ copies of
this function yields $e^{-x} x^n \gT(x) /n!$, which gives
$\nu = \gT\lambda$. As the intensity is $1$, this results 
in the autocorrelation
\[
    \gamma \; = \; \delta_0 + \nu + \widetilde{\nu} 
    \; = \; \delta_0 + \lambda
\]
and thus in the diffraction $\widehat{\gamma} = \gamma$,
compare Eq.~\eqref{poisson-lebesgue}.
\exend \end{example}

\begin{remark} \textsc{Characterisation of Poisson processes}.
\label{rem-one} 
  Let $N$ denote a homogeneous Poisson process on the real line, so
  that, for any measurable $A \subset \mathbb{R}$, $N(A)$ is the
  number of renewal points that fall into $A$.  It is well-known that
  $N(A)$ is then Poisson-($\lambda(A)$)-distributed, which means that
\[
   \PP ( N(A)= k ) \; = \; 
    \frac{e^{-\lambda(A)}\, (\lambda(A))^k}{k \ts !}
\]
with $k\in\NN_0$, and that, for any collection of pairwise disjoint
sets $A_1, A_2, \dots, A_m$, the random numbers $N(A_1), \dots,
N(A_m)$ are independent.  In fact, this property characterises the
Poisson process (compare \cite[Ch.~2.1]{DVJ1}), and it can serve as a
definition in higher dimensions or in more general measure spaces, to
which the renewal process cannot be extended.
\exend \end{remark}

\begin{figure}
\centerline{\includegraphics{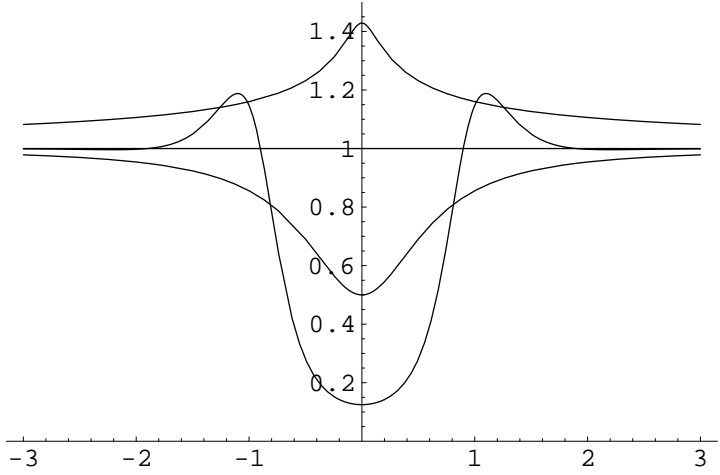}}
\caption{Absolutely continuous part of the diffraction measure
from Example~\ref{general-gamma}, for $\alpha=0.7$ (top
curve), $\alpha=1$ (horizontal line,
which also represents Example~\ref{poisson-on-the-line}), $\alpha=2$ 
(see also Example~\ref{example-two}) and $\alpha=8$ (overshooting
curve).}
\label{fig-renewal}
\end{figure}

\begin{example} \textsc{Renewal process with repulsion}. 
\label{example-two}
A perhaps more interesting example in this spirit is given
by the density 
\[
    f(x) \; = \;   4 x\, e^{-2x}\, \gT(x) .
\]
It is normalised and has mean $1$, as in
Example~\ref{poisson-on-the-line}, but models a repulsion of points
for small distances. Note that this distribution can be realised out
of Example~\ref{poisson-on-the-line} by taking only every second point,
followed by a rescaling of time. 

By induction (or by using well-known properties of the gamma
distributions, compare \cite[Sec.~II.2]{Feller}), one checks that
\[
    f^{*n} (x) \; = \;
    \tfrac{4^n}{(2n-1)!}\, x^{2n-1}\, e^{-2x} \gT(x) ,
\]
which finally results in the autocorrelation
\[
    \gamma \; = \; \delta_0 + (1-e^{-4\lvert x\rvert})\, \lambda
    \; = \; \delta_0 + \lambda - e^{-4\lvert x\rvert}\, \lambda
\]
and in the diffraction measure
\[
    \widehat{\gamma} \; = \;
    \delta_0 + \frac{2 + (\pi k)^2}{4 + (\pi k)^2}\, \lambda
    \; = \; \delta_0 + \lambda - \frac{2\ts \lambda}{4 + (\pi k)^2} .
\]
This is illustrated in Figure~\ref{fig-renewal}. The `dip' in the
absolutely continuous part around $0$, and thus the deviation from the
previous example, reflects the effectively repulsive nature of the
stochastic process when viewed from the perspective of neighbouring
points.
\exend \end{example}

\begin{example} \textsc{Renewal process with gamma law of mean $1$}.
\label{general-gamma}
The previous two examples are special cases of the gamma family of
measures. For fixed mean $1$, they are parametrised by a real number
$\alpha > 0$ via $\varrho_{\alpha} = f_{\alpha}\ts\lambda$ and the
density
\begin{equation} \label{gamma-1}
   f_{\alpha} (x) \; := \; \frac{\alpha^{\alpha}}{\Gamma (\alpha)}
   \, x^{\alpha - 1}\, e^{-\alpha x}\, \gT (x) .
\end{equation}
While $\alpha=1$ is the `interaction-free' Poisson process of
Example~\ref{poisson-on-the-line}, the density implies an effectively
attractive (repulsive) nature of the process for $0<\alpha <1$ (for
$\alpha>1$).  When $\alpha = k \in \NN$, the process can also be
interpreted as a modified Poisson process where one keeps only every
$k$th point (followed by an appropriate rescaling).

Observing $f_{\alpha}^{*n} (x) =
\tfrac{\alpha^{n\alpha}}{\Gamma(n\alpha)}\, x^{n\alpha -1} \,
e^{-\alpha x}\, \gT (x)$ for $n\in\NN$, this leads to the measure
\begin{equation} \label{gamma-2}
   \nu^{}_{\alpha} \; = \; g^{}_{\alpha}\,\gT \, \lambda
   \qquad\mbox{with}\qquad
   g^{}_{\alpha} (x) \; = \; \alpha \, e^{-\alpha x}
   \sum_{n=1}^{\infty}
   \frac{(\alpha x)^{n\alpha - 1}}{\Gamma(n\alpha)}
    \,  .
\end{equation}
Note that, for fixed $\alpha$, one has $\lim_{x\to\infty}\, g_{\alpha}
(x) = 1$.  The calculations result in the autocorrelation
\[
    \gamma_{\alpha} \; = \; 
    \delta_0 + g_{\alpha}(\lvert x\rvert)\,\lambda
\]
and in the diffraction $\widehat{\gamma}_{\alpha} = \delta_0 +
(1-h_{\alpha})\,\lambda $, where $h_\alpha$ is the symmetric function
defined by
\[
    h_{\alpha} (k) \; = \; \frac{2\,\bigl(1 - \mathrm{Re}
    ((1+2\pi ik/\alpha)^\alpha) \bigr)}
    {\big\lvert 1 - (1+2\pi ik/\alpha)^\alpha \big\rvert^2}\, .
\]
The latter follows from the general form of $h$ in
Theorem~\ref{non-lattice-gives-ac}, together with the observation that
$\widehat{f}_{\alpha} (k) = (1 + 2\pi i k/\alpha)^{-\alpha}$.

It is easy to see that $\lim_{k\to\pm\infty}\, h_{\alpha}(k) = 0$, for
any fixed $\alpha > 0$, which makes the role of $h_{\alpha}$ as the
deviation from the Poisson process diffraction more transparent, where
$\alpha=1$ and $h_1 \equiv 0$. Note also that $\lim_{\alpha\to\infty}
\widehat{\gamma}_{\alpha} = \delta_{\ZZ}$ in the vague topology, in
line with the limits mentioned before.  This can nicely be studied in
a series of plots of the diffraction with growing value of the
parameter $\alpha$. Figure~\ref{fig-renewal} shows some initial cases.
\exend \end{example}

\begin{remark} \textsc{Construction of Delone sets}.
  Of particular interest in the applications are \emph{Delone sets}
  (which are point sets that are both uniformly discrete and
  relatively dense), because points (representing atoms, say) should
  neither be too close nor too far apart. Such sets can also arise
  from a renewal process.  In fact, if one considers a probability
  measure $\varrho$ on $\RR_+$, the resulting point sets are always
  Delone sets when $\mathrm{supp} (\varrho)\subset [a,b]$ with $0<a\le
  b<\infty$, and conversely.  This equivalence does not depend on the
  nature of $\varrho$ on $[a,b]$, while the local complexity of the
  resulting point sets does. In particular, if $\varrho$ is absolutely
  continuous, the point sets will not have finite local complexity
  (see below for a definition).  
\exend \end{remark}

It is clear that no absolutely continuous $\varrho$ is lattice-like,
hence certainly not strictly lattice-like, so that all these examples
match Theorem~\ref{non-lattice-gives-ac}.  Probability
measures $\varrho$ with $\mathrm{supp} (\varrho)$ contained in a
lattice are covered by Remark~\ref{renewal-ext}.
They are of interest because they form a link to point sets and 
tilings of finite local complexity, which have only finitely many
patches of a given size (up to translations). Let us consider some
examples.

\begin{example} \textsc{Deterministic lattice case}. 
\label{latt-ex}
The simplest case is $\varrho = \delta_1$. From
$\delta_1 * \delta_1 = \delta_2$, one sees that
$\nu = \delta^{}_{\NN}$ and hence
\[
   \gamma \; = \; \delta^{}_0 + \delta^{}_{\NN} + \delta^{}_{-\NN}
          \; = \; \delta^{}_{\ZZ} \, ,
\]
which is a lattice Dirac comb, with Fourier transform
\[
   \widehat{\gamma} \; = \; \delta^{}_{\ZZ}
\]
according to the Poisson summation formula \eqref{PSF-1}.
This is the deterministic case of the integer lattice $\ZZ$,
covered in this setting.
\exend \end{example}

\begin{remark} \textsc{Deterministic limit of Example~\ref{general-gamma}}.
  The last example can also be seen as a limiting case of the
  measure $\varrho_\alpha$ defined by Eq.~\eqref{gamma-1}.  In
  particular, one has $\lim_{\alpha\to\infty}\,\varrho_\alpha
  =\delta_1$ and $\lim_{\alpha\to\infty}\,\nu_{\alpha} =
  \delta_{\NN}$, with $\nu_{\alpha}$ as in \eqref{gamma-2} and both
  limits to be understood in the vague topology. This can also be
  seen by means of the strong law of large numbers. For each
  $n\in\NN$, by well-known divisibility properties of the family of 
  Gamma distributions, $\varrho_n$ is the distribution of
  \[
       \frac{1}{n} \sum_{i=1}^{n} X_i \ts ,
  \]
  where the $X_i$ are independent and exponentially
  distributed random variables with mean $1$. This sum
  then concentrates around $1$, with a standard deviation 
  of order $1/\sqrt{n}$.
\exend \end{remark}

\begin{example} \textsc{Random tilings with finitely many prototiles}.
\label{random-tiling}
Consider the measure
\[
    \varrho \; = \; \alpha \delta_{a} + (1\! -\!\alpha) \delta_{b} ,
\]
with $\alpha\in (0,1)$ and $a,b > 0$, subject to the restriction
$\alpha a + (1\! - \!\alpha)b=1$ to ensure density $1$.  Each
realisation of the corresponding renewal process results in a point
set that can also be viewed as a random tiling on the line with two
prototiles, of lengths $a$ and $b$. As before, place a normalised
point measure at each point of the realisation.  Then, the diffraction
(almost surely) has a pure point and an absolutely continuous part,
but no singular continuous one. The pure point part can be just
$\delta_0$ (when $b/a$ is irrational) or a lattice comb (see
Remark~\ref{renewal-ext}); details are given in \cite{BH}, including
an explicit formula for the AC part.

This has a straight-forward generalisation to any finite number of
prototiles, with a similar result. Also in this case, there is an
explicit formula for the diffraction measure, which was derived in
\cite{BH} by a direct method, without using the renewal process.
\exend \end{example}

\begin{remark} \textsc{Continuous diffraction with `needles'}.
\label{rem:Meyer}
  Looking back at Lemma~\ref{lattice-like}, one realises that
  Example~\ref{random-tiling} revolves around the lattice condition in
  an interesting way. Namely, even if $\varrho$ is \emph{not} strictly
  lattice-like, $\mathrm{supp} (\varrho)$ for a random tiling
  example with finitely many prototiles is a finite set, and thus a
  subset of a \emph{Meyer set} (which is a relatively dense set $\gL$
  whose difference set $\gL-\gL$ is uniformly discrete). We then know
  from the harmonic analysis of Meyer sets, compare \cite{Moody} and
  references therein, that $\widehat{\varrho}\ts (k)$ will come
  $\varepsilon$-close to $1$ with bounded gaps in $k$. This means that
  the diffraction measure, though it is absolutely continuous apart
  from the central peak at $k=0$, will develop sharp `needles' that
  are close to point measures in the vague topology --- a phenomenon
  that was also observed in \cite{BH} on the basis of the explicit
  solution.  
\exend \end{remark}

\section{Arbitrary dimensions: Elementary approach}
\label{elementary}

Let us now develop some intuition for the influence of randomness on
the diffraction of point sets and certain structures derived from them
in Euclidean spaces of arbitrary dimension. In this section, our point
of view is from a single point set $\gL\subset\RR^d$ that is being
modified randomly, by replacing each point by a complex, finite,
random cluster. This is still relatively easy as long as $\gL$ is
sufficiently `nice'. In Section~\ref{general}, we revisit this
situation from the point of view of a stationary ergodic point
process, which treats almost all of its realisations at once and
permits a larger generality for the sets $\gL$, though the clusters
will then be restricted to positive or signed measures.

\smallskip Let $\gL\subset\RR^d$ be a fixed point set, which we assume
to be of \emph{finite local complexity} (FLC). By definition, this
means that there are only finitely many distinct patches of any given
size (up to translations) in $\gL$.  This property is equivalent to
the difference set $\gL-\gL$ being locally finite \cite{Martin2}, the
latter saying that $K\cap (\gL - \gL)$ is a finite set for all compact
$K\subset \RR^{d}$.  In particular, since $0$ is then isolated in
$\gL-\gL$, the set $\gL$ itself is uniformly discrete; see
Remark~\ref{rem:ext-thm2} for a possible extension.  Attached to $\gL$
is its \emph{Dirac comb} $\delta_{\gL} = \sum_{x\in\gL} \delta_x$,
which is a translation bounded measure, as a consequence of the FLC
property.  We associate to $\delta_{\gL}$ the autocorrelation and the
diffraction measure as explained in Section~\ref{sec:prelims}, for a
suitably chosen averaging sequence $\cA = \{A_n\mid n\in\NN\}$ of van
Hove type. A natural choice is $A_n = B_{r_n} (0)$, with $B_r (0)$
denoting the open ball of radius $r$ around $0$, for a non-decreasing
series of radii with $r_n \xrightarrow{n\to\infty}\infty$
(alternatively, nested cubes are also quite common).

Set $\gL_n = \gL\cap A_n$ (so that $\gL_n \! \stackrel{\scriptstyle
  \nearrow}{} \!\!  \gL$ in the obvious local topology \cite{Martin2})
and consider
\[
    \gamma^{}_{\gL,n}  \, := \; 
    \frac{\delta_{\gL_n} \! * \widetilde{\delta_{\gL_n}}}{\vol (A_n)}
    \; = \,\frac{1}{\vol (A_n)} \sum_{x,y\in\gL_n} \delta_{x-y} \ts .
\]
We now make the assumption that the limit
\begin{equation} \label{limit-exists}
  \lim_{n\to\infty} \gamma^{}_{\gL,n} \; =: \; \gamma^{}_{\gL}
\end{equation}
exists in the vague topology, which is then the autocorrelation
measure of the set $\gL$ relative to the averaging sequence $\cA$.

\begin{remark} \textsc{Accumulation points versus limits}.
\label{accu-versus-limits}
Due to translation boundedness of $\delta_{\gL}$, the sequence of
measures $\gamma^{}_{\gL,n}$ always has points of accumulation; see
\cite[Prop.~2.2]{Hof} and Lemma~\ref{lem:utb}. Consequently, one can
always select a subsequence of $\mathcal{A}$ for which the assumption
\eqref{limit-exists} is satisfied. This remains true even if we relax
the nesting condition for $\cA$. In this sense, when the
autocorrelation is not unique (as in the example of the visible
lattice points without nesting \cite{BMP}), we simply select
\emph{one} of the possible autocorrelations by a suitable choice of
$\mathcal{A}$. Our results below apply to any autocorrelation of this
kind separately.  In this sense, the assumption made in
\eqref{limit-exists} is not restrictive.  \exend
\end{remark}  

As briefly explained in Section~\ref{sec:prelims}, see
Lemma~\ref{lem:utb}, the van Hove property of $\mathcal{A}$ in the
context of \eqref{limit-exists} implies that one also has
\begin{equation} \label{pos-def-limit}
   \lim_{n\to\infty} \gamma^{}_{\gL,n;\mathrm{mod}}\, := 
   \lim_{n\to\infty} \frac{\delta_{\gL_n} \! * 
         \widetilde{\delta_{\gL}}}{\vol (A_n)}
   \, = \, \gamma^{}_{\gL} \/ ,
\end{equation}
the difference between the two approximating measures in
\eqref{limit-exists} and \eqref{pos-def-limit} being a `surface term'
that vanishes in the infinite volume limit $n\to\infty$. 
Eq.~\eqref{limit-exists} explicitly shows that the measure
$\gamma^{}_{\gL}$ is positive definite (hence transformable by
Lemma~\ref{pos-def-props}), while \eqref{pos-def-limit} is easier to
work with for (pointwise) calculations in the presence of random
modifications as introduced below.

Since $\gL-\gL$ is locally finite by assumption, Eq.~\eqref{pos-def-limit} 
is equivalent to the existence of all the pointwise limits
\begin{equation} \label{eta-limits}
   \lim_{n\to\infty}  \eta^{}_n (z) \, =: \, \eta(z) \ts ,
\end{equation}
with the approximating coefficients
\[
     \eta^{}_{n} (z) =    
      \frac{\card \{x\in\gL_n \mid x-z \in\gL\}}{\vol (A_n)} \ts ,
\]
where $\eta(z) = 0$ for any $z\not\in\gL-\gL$.  Clearly, the measure
$\gamma^{}_{\gL}$ as well as the coefficients $\eta(z)$ may (and
generally will) depend on the averaging sequence; compare
Remark~\ref{accu-versus-limits}. 

The next step consists in modifying $\gL$ by a random process in a
local way. To come to a reasonably general formulation that includes
several notions of randomness known from lattice theory, compare
\cite{Gui,Wel}, we employ a formulation with finite, random, complex
measures. Let $\varOmega$ denote a measure-valued random variable, and
$Q$ the corresponding law, which is itself a probability measure on
$\cM_{\sf bd} = \cM_{\sf bd} (\RR^d)$, the space of finite complex
measures on $\RR^d$.  To keep the notation compact, we use the symbol
$\EE_Q$ for the various expectation values that arise in connection
with $(\varOmega, Q)$.  In particular,
we write $\EE_Q (\varOmega) = \int_{\cM_{\sf bd}} \omega \dd Q
(\omega)$, where $\omega$ refers to the realisations of $\varOmega$ as
usual. Note that we also refer via the index $Q$ to the underlying law
for one random variable for more complicated expectation values,
rather than using the underlying (though hidden) probability
space. This will be explained in more detail in Section~\ref{general}
below.

To proceed, we need a version of the strong law of large numbers (SLLN)
for measures. 
\begin{lemma} \label{measure-SLLN}
  Let $(\varOmega_i)_{i\in\NN}$ be a sequence of integrable, finite, 
  i.i.d.\ random measures, with common law $Q$. Then, with
  probability $1$, one has
\[
   \frac{1}{n} \sum_{i=1}^{n} \varOmega_i 
   \;\xrightarrow{n\to\infty} \;\EE_Q (\varOmega_1)
\]
  in the vague topology.
\end{lemma}
\begin{proof}
  By definition, integrability means that $\EE_Q (\lvert \varOmega_i
  \rvert)$, which is independent of $i\in\NN$, is a finite measure. As
  the space of continuous functions $C_{\mathsf{c}} (\RR^{d})$ is
  separable, the almost sure convergence of the measures follows from
  the almost sure convergence of $\frac{1}{n} \sum_{i=1}^{n}
  \varOmega_i (\varphi)$ for an arbitrary (but fixed) bounded,
  continuous function $\varphi$. This, in turn, follows from the
  conventional SLLN \cite{Ete}, possibly after splitting the sums into
  their real and imaginary parts and applying the SLLN twice.
\end{proof}

Recall that $\widetilde{\omega}$ is the measure defined by
$\widetilde{\omega}(\varphi) = \overline{\omega(\widetilde{\varphi}\ts)}$.
Let $\varOmega$ and $\varOmega{\ts}'$ be two independent random measures,
with the same law $Q$, and such that $\EE_Q (\lvert \varOmega \rvert)$
is a finite measure, and also assume the second moment condition
$\EE_Q \bigl( (\ts\lvert\varOmega\rvert (\RR^d))^2 \bigr)<\infty$. Then,
the convolution $\varOmega\conv\varOmega{\ts}'$ is
well defined, and one obtains from elementary calculations
the important relations
\begin{equation} \label{conv-expect}
   \EE_Q (\widetilde{\varOmega}) \, = \, \widetilde{\EE_Q (\varOmega)}
   \quad\mbox{and}\quad
   \EE_Q (\varOmega\conv\widetilde{\varOmega}{\ts}') \, = \,
   \EE_Q (\varOmega)*\widetilde{\EE_Q (\varOmega{\ts}')}\/ ,
\end{equation}
the second due to the assumed independence.

\smallskip 
Let us now fix an FLC set $\gL$, which is assumed to possess the
autocorrelation measure $\gamma^{}_{\gL}$ relative to the van Hove
averaging sequence $\cA$ chosen, and consider the family
$(\varOmega_x)^{}_{x\in\gL}$ of integrable, complex, i.i.d.\ random
measures, with common law $Q$ and subject to the moment conditions
mentioned above. When $\varOmega$ is any representative of these
random measures, $\EE_Q (\lvert\varOmega\rvert)$ is a finite measure
by assumption, and the measure-valued expectations $\EE_Q (\varOmega)$
and $\EE_Q (\varOmega\conv\widetilde{\varOmega})$ exist (note that
also $\EE_Q (\lvert\varOmega\conv \widetilde{\varOmega}\rvert)$ is a
finite measure, due to the condition on the second moment). We are now
interested in the random object
\begin{equation} \label{random-comb}
    \delta^{(\varOmega)}_{\gL} =
         \sum_{x\in\gL} \varOmega_x \! * \delta_x \ts ,
\end{equation}
which is almost surely a locally finite measure (though not
necessarily translation bounded). 

To see this, we observe that, for any
bounded Borel set $B\subset \RR^{d}$, the sum $\sum_{x\in\gL} \lvert
(\varOmega_x * \delta_x) (B) \rvert$ converges almost surely, since
\[
    \EE^{}_{Q} \bigl( \lvert (\varOmega_x * \delta_x) (B) \rvert
    \bigr) = \EE^{}_{Q} \bigl( \lvert \varOmega (B-x)\rvert \bigr)
    \le \bigl( \EE^{}_{Q} (\lvert\varOmega \rvert )\bigr) (B-x)
\]
and the convolution $\delta^{}_{\!\gL} * \EE^{}_{Q}
\bigl(\lvert\varOmega\rvert \bigr)$ is a well-defined locally finite
measure due to the translation boundedness of $\delta^{}_{\!\gL}$
(note that the summands in $\sum_{x\in\gL} \lvert \varOmega_x *
\delta_x \rvert(B)$ are non-negative, hence convergence of the means
implies almost sure convergence).  As the Borel $\sigma$-algebra on
$\RR^{d}$ is countably generated, we can find a set of $Q$-measure $1$
on which the sum \eqref{random-comb} converges (absolutely) for each
Borel set $B$, and the limit is a measure.

Let $\cA$ be fixed and assume for simplicity that each $A_n$ is
invariant under $x\mapsto -x$. The ($n$-th) approximating
autocorrelation of $\delta_\gL^{(\varOmega)}$ reads
\begin{equation} \label{eq:random-auto-corr}
   \gamma^{(\varOmega)}_{\gL,n} \, = \, \frac{1}{\vol (A_n)} \;
   \delta_\gL^{(\varOmega)}{}_{\textstyle |^{}_{A_n}} \!\! * \,
   \widetilde{\delta_\gL^{(\varOmega)}{}_{\textstyle |^{}_{A_n}}}
   = \, \frac{1}{\vol (A_n)}\,
   \sum_{x \in\gL} \bigl( \varOmega_x * \delta_x 
   \bigr)_{\textstyle |^{}_{A_n}} \!\! * \, 
   \sum_{y \in \gL} \bigl( \widetilde{\ts \varOmega_y} * 
   \delta_{-y} \bigr)_{\textstyle |^{}_{A_n}}  \ts . 
\end{equation}
For certain pointwise calculations and arguments, it will be more
convenient below to consider the modified approximating
autocorrelation
\begin{equation}
\label{eq:modautocorrOmega}
    \gamma^{(\varOmega)}_{\gL,n;\ts\mathrm{mod}} \, := \, 
    \frac{1}{\vol (A_n)}\,
    \Bigl(\sum_{x\in\gL_n}\varOmega_x \conv\ts \delta_x\Bigr) *
    \Bigl(\sum_{y\in\gL} \widetilde{\varOmega_y} \conv\ts
    \delta_{-y} \Bigr). 
\end{equation}
To this end, we need a probabilistic analogue of
Eq.~\eqref{convolution-freedom}.
\begin{prop} \label{prop:modautocorr}
   Almost surely, $\gamma^{(\varOmega)}_{\gL,n}$ of\/
   \eqref{eq:random-auto-corr} and $\gamma^{(\varOmega)}_{\gL,n;\ts
   \mathrm{mod}}$ of\/ \eqref{eq:modautocorrOmega} define sequences of
   locally finite random measures. Moreover, we can choose a strictly
   increasing subsequence $(n_k)^{}_{k\in\NN}$ such that,
   in the vague topology, we almost surely have  
\[
    \gamma^{(\varOmega)}_{\gL,n_k} - 
    \gamma^{(\varOmega)}_{\gL,n_k;\ts \mathrm{mod}}
    \;\xrightarrow{k\to\infty}\;  0 \ts .
\]
   In particular, if\/ $\gamma^{(\varOmega)}_{\gL,n}$ or\/
   $\gamma^{(\varOmega)}_{\gL,n;\ts \mathrm{mod}}$ almost surely
   converges to $\gamma$ along $\cA$ or along a subsequence of it, we can
   choose a subsequence $\cA'$ of $\cA$ so that both sequences almost
   surely converge to $\gamma$ along $\cA'$.
\end{prop}
\begin{proof}
  We abbreviate $\mu(\cdot) := \EE\big( |\varOmega|(\cdot)\big)$,
  $\nu(\cdot) := \EE\big( |\varOmega|(\RR^d) \,
  |\varOmega|(\cdot)\big)$.  Due to the assumption $\EE\bigl((\ts
  |\varOmega|(\RR^d))^2 \big) < \infty$, both $\mu$ and $\nu$ are
  finite, positive measures.  Consequently, since $\gL$ is uniformly
  discrete, $\mu * \delta_\gL$ and $\nu * \delta_\gL$ are translation
  bounded by \cite[Prop.~1.13]{BF} (and thus certainly locally finite).

  Let us first verify that the expression in
  \eqref{eq:modautocorrOmega} almost surely defines a locally finite
  measure (the estimate for \eqref{eq:random-auto-corr} is completely
  analogous, with the same upper bound). If $B\subset\RR^{d}$ is a
  bounded Borel set, we have
\begin{align*} 
   \vol(A_n) \, \EE_{Q} \bigl( \lvert
   \gamma^{(\varOmega)}_{\gL,n;\ts \mathrm{mod}}\rvert \ts (B) \bigr)\,
   & \le \sum_{\substack{x\in \gL_n \\ y \in \gL }} 
   \EE_{Q} \left[ \int_{\RR^{d}\nts \times \RR^d}
   \mathbf{1}^{}_B (u+v) \dd \bigl( 
   \lvert\Omega_x \rvert * \delta_x \bigr)(u) 
   \dd \bigl( \lvert \widetilde{\Omega_y}\rvert * 
   \delta_{-y}\bigr)(v) \right] \\
   & \le \sum_{\substack{x\in \gL_n \\ y \in \gL \setminus\{x\}}} 
     \int_{\RR^{d}\nts \times \RR^{d}} 
     \mathbf{1}^{}_B (u+v) \dd ( \mu * \delta_x)(u) 
     \dd (\widetilde{\mu} * \delta_{-y})(v) \\
   & \hspace{2em} + \sum_{x \in \gL_n} 
     \EE_{Q} \left[ \int_{\RR^{d}\nts \times \RR^{d}}  
     \mathbf{1}^{}_{B}(u+v) \dd |\Omega_x|(u) 
     \dd |\widetilde{\Omega_x}|(v) \right] \\
   & \le \Bigl(\bigl( \mu * \delta^{}_{\!\gL_n} \bigr) * 
     \widetilde{\bigl( \mu * \delta^{}_{\! \gL} \bigr)}\Bigr) (B) 
     \, + \, \nu(\RR^d)\, \card (\gL_n) 
      \; < \; \infty \ts . 
\end{align*}
Thus, by arguments analogous to those used before, the sum on the
right hand side of \eqref{eq:modautocorrOmega} almost surely converges
absolutely when applied to any bounded Borel set $B$.  Again, since
the Borel $\sigma$-algebra on $\RR^d$ is countably generated, this
suffices to ensure that $\gamma^{(\varOmega)}_{\gL,n;\ts
  \mathrm{mod}}$ is a locally finite, random measure. 

Furthermore, again for a bounded Borel set $B$, one has
\begin{equation} \label{eq:ref}
  \begin{split} 
   \lefteqn{ \vol(A_n)  \, \EE_{Q} \bigl( \lvert 
     \gamma^{(\varOmega)}_{\gL,n} -
     \gamma^{(\varOmega)}_{\gL,n;\ts \mathrm{mod}}\rvert\ts (B) \bigr)} \\[2mm] 
& \le \sum_{x,y \in \gL} \EE_{Q} \left[\int_{\RR^{d}\nts\times\RR^{d}}\!\!
      \mathbf{1}^{}_B(u \! + \! v)\, 
      \bigl| \mathbf{1}^{}_{\nts A_n}(u) \mathbf{1}^{}_{\nts A_n}(v)
       - \mathbf{1}^{}_{\nts A_n}(x) \bigr| 
      \dd \bigl( \lvert\Omega_x \rvert * \delta_x \bigr) (u) 
      \dd \bigl( \lvert\widetilde{\Omega_y} \rvert * 
          \delta_{-y}\bigr) (v) \right] \\[2mm]
& = \sum_{\substack{x,y \in \gL \\ x\neq y}} \int_{\RR^{d}\nts\times\RR^{d}} 
    \!\! \mathbf{1}^{}_B(u+v)\, 
    \bigl| \mathbf{1}^{}_{\nts A_n}(u) \mathbf{1}^{}_{\nts A_n}(v) 
    - \mathbf{1}^{}_{\nts A_n}(x) \bigr| \dd ( \mu * \delta_x)(u) 
    \dd (\widetilde{\mu} * \delta_{-y})(v) \\
& \hspace{1em} + \sum_{x \in \gL} 
     \EE_{Q} \left[ \int_{\RR^{d}\nts\times\RR^{d}} \!\!
     \mathbf{1}^{}_B (u+v) \,
     \bigl| \mathbf{1}^{}_{\nts A_n}(u)
     \mathbf{1}^{}_{\nts A_n}(v) - 
     \mathbf{1}^{}_{\nts A_n}(x) \bigr|  
     \dd \bigl( \lvert\Omega_x \rvert * \delta_x \bigr) (u) 
     \dd \bigl( \lvert\widetilde{\Omega_x} \rvert * 
     \delta_{-x}\bigr) (v)\right] . 
  \end{split}
\end{equation}

We estimate the term in the third line of \eqref{eq:ref} as follows:
\[
  \begin{split}
\sum_{\substack{x,y \in \gL \\ x\neq y}}  \int_{\RR^{d}\nts\times\RR^{d}} 
   & \! \mathbf{1}^{}_B(u+v)\, 
    \bigl| \mathbf{1}^{}_{\nts A_n}(u) \mathbf{1}^{}_{\nts A_n}(v) 
    - \mathbf{1}^{}_{\nts A_n}(x) \bigr| \dd ( \mu * \delta_x)(u) 
    \dd (\widetilde{\mu} * \delta_{-y})(v) \\
= & \sum_{x \in \gL_n} \sum_{\substack{y \in \gL\\y \neq x}}
  \int_{\RR^{d}\nts\times\RR^{d}} 
    \!\! \mathbf{1}^{}_B(u+v)\, 
    \bigl| \mathbf{1}^{}_{\nts A_n}(u) \mathbf{1}^{}_{\nts A_n}(v) 
    - 1 \bigr| \dd ( \mu * \delta_x)(u) 
    \dd (\widetilde{\mu} * \delta_{-y})(v) \\
& {} + \sum_{x \in \gL \setminus \gL_n} \sum_{\substack{y \in \gL\\y \neq x}}
  \int_{\RR^{d}\nts\times\RR^{d}} 
    \!\! \mathbf{1}^{}_B(u+v)\, 
    \mathbf{1}^{}_{\nts A_n}(u) \mathbf{1}^{}_{\nts A_n}(v) 
    \dd ( \mu * \delta_x)(u) 
    \dd (\widetilde{\mu} * \delta_{-y})(v) \\
   \leq & \int_{\RR^{d}\nts\times\RR^{d}} 
     \mathbf{1}^{}_B (u+v)\, 
     \bigl| \mathbf{1}^{}_{\nts A_n}(u)
            \mathbf{1}^{}_{\nts A_n}(v) - 1 \bigr| 
   \dd \bigl( \mu * \delta^{}_{\!\gL_n} \bigr)(u) 
   \dd \bigl( \widetilde{\mu * \delta^{}_{\!\gL}} \bigr)(v) \\[2mm]
   & {} + \int_{\RR^{d}\nts\times\RR^{d}} 
     \mathbf{1}^{}_B (u+v) \, 
     \mathbf{1}^{}_{\nts A_n}(u) \mathbf{1}^{}_{\nts A_n}(v) 
     \dd \bigl( \mu * \delta^{}_{\!\gL \setminus \gL_n} \bigr)(u) 
     \dd \bigl( \widetilde{\mu * \delta^{}_{\!\gL}} \bigr)(v) \\[3mm]
    \leq & 
     \; \Bigl(\bigl( \mu * \delta^{}_{\!\gL_n}\bigr)
    {}_{\textstyle |^{}_{\RR^d \setminus A_n}} \!\! * \,
    \bigl( \widetilde{\mu * \delta^{}_{\!\gL}} \bigr)\Bigr) (B) 
    + \Bigl( \bigl( \mu * \delta^{}_{\!\gL_n}\bigr) 
    * \bigl( \widetilde{\mu * \delta^{}_{\!\gL}} \bigr)
    {}_{\textstyle |^{}_{\RR^d \setminus A_n}}\Bigr) (B) \\[2mm]
    & {\,} + \Bigl(
    \bigl( \mu * \delta^{}_{\!\gL \setminus \gL_n}\bigr)
    {}_{\textstyle |^{}_{A_n}} * 
    \bigl( \widetilde{\mu * \delta^{}_{\!\gL}} \bigr)
    {}_{\textstyle |^{}_{A_n}} \Bigr) (B),
  \end{split}
\] 
where, in the first inequality, we have used the fact that removing
the restriction $x\neq y$ in the summation only adds positive terms
(note that $\mu$ is a positive measure by definition), and employed
the estimate $\bigl| \mathbf{1}^{}_{\nts A_n}(u) \mathbf{1}^{}_{\nts
  A_n}(v) - 1 \bigr| \leq \mathbf{1}^{}_{\RR^d \setminus \nts
  A_n}(u) + \mathbf{1}^{}_{\RR^d \setminus \nts A_n}(v)$ for the
second inequality.  There is a constant $c^{}_B$ that depends on $B$
(as well as on $\mu$ and the averaging sequence) and a sequence
$d_n\to 0$ that is independent of $B$ such that the sum in the last
two lines above is bounded by $d_n \ts c^{}_{B}\, \vol(A_n)$.
This comes from the fact that there are only contributions from 
`surface terms'; compare the arguments in \cite{Martin2}.

By way of example, we verify that, for any $R>0$,
\begin{align*} 
\Bigl(\bigl(  \mu \, * & \, \delta^{}_{\!\gL_n}\bigr)
    {}_{\textstyle |^{}_{\RR^d \setminus A_n}} \!\! * \,
    \bigl( \widetilde{\mu * \delta^{}_{\!\gL}} \bigr)\Bigr) (B) \\
   \leq &  \Bigl( \mu(\RR^d)\, \bigl| \{ x \in \gL_n \mid 
   d(x,\RR^d\setminus A_n) \leq R \} \bigr|  + 
   \mu(\RR^d \setminus B_R)\ts \lvert\gL_n \rvert \Bigr) 
   \sup_{x\in\RR^d} \bigl( \widetilde{\mu * \delta^{}_{\!\gL}} 
   \bigr)(B+x).
\end{align*} 
Note that this (together with analogous statements for the two other 
summands above) yields the claim by the van Hove property of the 
averaging sequence $\cA$. 

Observe next that, for a (possibly) complex measure $\xi$ with 
$|\xi|(\RR^d) < \infty$, a translation bounded measure $\nu$ 
and a bounded Borel set $B$, we have (by an application of
\cite[Prop.~2.2]{Hof} and its proof) the estimate
$\lvert (\xi * \nu)(B)\rvert \leq \lvert\xi\rvert (\RR^d) 
\sup_{x \in \RR^d} \lvert \nu \rvert(B+x) < \infty$.
Finally, note that $\widetilde{\mu * \delta^{}_{\!\gL}}$ is
translation bounded by \cite[Prop.~1.13]{BF}, and that
\begin{align*} 
   \bigl(  \mu \, * \, \delta^{}_{\!\gL_n}\bigr)\;
    {}_{\textstyle |^{}_{\RR^d \setminus A_n}} \!\! (\RR^d) 
   = & \sum_{x\in\gL_n} \bigl( \mu_{\textstyle |^{}_{B_R}}\!\! * 
    \, \delta_x \bigr) (\RR^d \setminus A_n) + 
  \sum_{x\in\gL_n} \bigl( \mu_{\textstyle |^{}_{\RR^d \setminus B_R}}\!\! 
  * \, \delta_x \bigr)   (\RR^d \setminus A_n) \\[1mm]
   \leq &\, \mu(\RR^d)\ts \bigl| \{ x \in \gL_n 
   \mid d(x,\RR^d\setminus A_n) 
   \leq R \} \bigr|  + \mu(\RR^d \setminus B_R)
   \ts \lvert \gL_n \rvert \ts .
\end{align*}

Similarly, the term in the fourth line of \eqref{eq:ref} is bounded 
from above by
\begin{align} 
\nonumber
   \sum_{x \in \gL_n} & \EE_{Q} 
     \left[ \int_{\RR^{d}\nts\times\RR^{d}} 
     \mathbf{1}^{}_B(u+v)\, \bigl(
      \mathbf{1}^{}_{\RR^d \setminus A_n}(u) + 
      \mathbf{1}^{}_{\RR^d \setminus A_n}(v) \bigr)
   \dd \bigl( \lvert\Omega_x \rvert * \delta_x \bigr) (u) 
   \dd \bigl( \lvert\widetilde{\Omega_x} \rvert * 
       \delta_{-x}\bigr) (v) \right] \\
\nonumber
   & {} + \sum_{x \in \gL \setminus \gL_n} 
      \EE_{Q} \left[\int_{\RR^{d}\nts\times\RR^{d}} 
      \mathbf{1}^{}_B(u+v) 
      \mathbf{1}^{}_{\nts A_n}(u)
      \mathbf{1}^{}_{\nts A_n}(v)
    \dd \bigl( \lvert\Omega_x \rvert * \delta_x \bigr) (u) 
    \dd \bigr( \lvert\widetilde{\Omega_x} \rvert * 
        \delta_{-x}\bigr) (v) \right] \\
\nonumber
   & \le \sum_{x \in \gL_n} 
   \EE_{Q} \Bigl( \bigl| \widetilde{\varOmega_x} \bigr| (\RR^d) 
   \int_{\RR^{d}} \mathbf{1}^{}_{\RR^d \setminus A_n}(u) 
   \dd \bigl( \lvert \Omega_x \rvert * \delta_x \bigr)(u) 
   \Bigr) \\
\nonumber
   & {} \hspace{1.5em} + \sum_{x \in \gL_n} 
   \EE_{Q} \Bigl( \bigl| {\varOmega_x} \bigr| (\RR^d) 
   \int_{\RR^{d}} \mathbf{1}^{}_{\RR^d \setminus A_n}(v) 
   \dd \bigl( \lvert \widetilde{\Omega_x} \rvert * 
   \delta_{-x}\bigr) (v) \Bigr) \\
\nonumber
   & {} \hspace{1.5em}  + \sum_{x \in \gL \setminus \gL_n} 
   \EE_{Q} \Bigl( \bigl|\widetilde{\varOmega_x} \bigr| (\RR^d) 
   \int_{\RR^{d}} \mathbf{1}^{}_{\nts A_n}(u) 
   \dd \bigl( \lvert \Omega_x \rvert * \delta_x \bigr)(u) 
   \Bigr) \\
\label{eq:lemma.appr.term4}
   & = \, \bigl( \nu * \delta^{}_{\!\gL_n} \bigr) (\RR^d \setminus A_n) 
    + \bigl( \widetilde{\nu * \delta^{}_{\!\gL_n}} \bigr) 
          (\RR^d \setminus A_n)
    + \bigl( \nu * \delta^{}_{\!\gL \setminus \gL_n} \bigr) (A_n) 
    \; \le \;  d_{n}^{\ts\prime} \vol(A_n)\ts ,
\end{align}
with a sequence $d_n'\longrightarrow 0$. Combining the above
estimates, we obtain
\begin{equation}
   \EE_{Q} \Bigl( \frac{ \bigl| \gamma^{(\varOmega)}_{\gL,n}-
    \gamma^{(\varOmega)}_{\gL,n;\ts \mathrm{mod}} \bigr|(B)} 
    {\vol(A_n)}\Bigr) \, \le \, (c^{}_B + 1)\, d''_n
\end{equation}
for a sequence $d_n''\longrightarrow 0$; hence, for $\varepsilon>0$,
\begin{equation} \label{eq:lemma.appr.bd1}
   \PP\Bigl( \frac{ \bigl|\gamma^{(\varOmega)}_{\gL,n} -
   \gamma^{(\varOmega)}_{\gL,n;\ts \mathrm{mod}}\bigr| (B)}
   {\vol(A_n)} > \varepsilon \Bigr) 
   \, \le \, \frac{c^{}_B + 1}{\varepsilon}\, d_n'' 
\end{equation}
by Markov's inequality. If we choose $(n_k)^{}_{k\in\NN}$ such that
$\sum_{k} d_{n_k}'' < \infty$, we obtain from
\eqref{eq:lemma.appr.bd1} and the (first) Borel-Cantelli lemma that
\begin{equation} \label{eq:lemma.appr.lim} 
   \frac{ \bigl| \gamma^{(\varOmega)}_{\gL,n_k} -
   \gamma^{(\varOmega)}_{\gL,n_k;\ts \mathrm{mod}} \bigr| (B) }
   {\vol(A_{n_k})}\; \longrightarrow \; 0 \, , \quad 
   \mbox{almost surely as $k\to\infty\ts$}. 
\end{equation}
By \eqref{eq:lemma.appr.bd1}, we may choose the subsequence 
$(n_k)^{}_{k\in\NN}$ independently of $B$ in such a way that, 
for each bounded Borel set $B$, \eqref{eq:lemma.appr.lim} holds 
almost surely. Finally, since the Borel $\sigma$-algebra on 
$\RR^d$ is countably generated, this implies 
the main claim of the lemma. The last statement is then obvious.
\end{proof}

Let us now resume our study of $\gamma^{(\Omega)}_{\gL}$. Invoking
Proposition~\ref{prop:modautocorr} and replacing the averaging sequence 
$\cA=(A_n)^{}_{n\in\NN}$ by the subsequence chosen there, we may use the  
modified measures $\gamma^{(\varOmega)}_{\gL,n;\ts \mathrm{mod}}$ 
as our approximating  measures. Observing
\[
  \widetilde{\delta^{(\varOmega)}_{\gL}} \; = \;
  \sum_{y\in\gL} \widetilde{\varOmega}_y \nts * \delta_{-y} \/ ,
\]
the modified autocorrelation approximant reads
\begin{equation} \label{random-approx}
  \begin{split}
    \gamma^{(\varOmega)}_{\gL,n; \ts \mathrm{mod}} 
    & \; = \; \frac{1}{\vol (A_n)}\,
    \Bigl(\sum_{x\in\gL_n}\varOmega_x \conv\ts \delta_x\Bigr) *
    \Bigl(\sum_{y\in\gL} \widetilde{\varOmega}_y \conv\ts
    \delta_{-y} \Bigr) \\[1mm]
    & \; =  \sum_{z\in\gL-\gL} \Bigl(\frac{1}{\vol (A_n)}
    \sum_{\substack{x\in\gL_n \\ x-z\in\gL}}
    \varOmega_x \conv \widetilde{\varOmega}_{x-z} \Bigr)
    * \delta_z \; =:  \sum_{z\in\gL-\gL}
    \zeta^{(\varOmega)}_{z,n}  \! * \delta_z \, ,
  \end{split}
\end{equation}
where we now need to analyse the behaviour of the random measures
$\zeta^{(\varOmega)}_{z,n}$.

Let us first look at $z=0$, where we obtain
\begin{equation}\label{zero-limit}
   \zeta^{(\varOmega)}_{0,n} \; = \;
   \frac{\card(\gL_n)}{\vol (A_n)} \,
   \frac{1}{\card(\gL_n)} \,\sum_{x\in\gL_n}
   \varOmega_x * \widetilde{\varOmega}_x 
   \; \xrightarrow{n\to\infty} \;
   \dens(\gL)\cdot \EE_Q (\varOmega * \widetilde{\varOmega})
   \qquad \mbox{(a.s.)}
\end{equation}
by an application of Lemma~\ref{measure-SLLN}. Note that
$\dens(\gL)=\eta(0)$ as introduced in Eq.~\eqref{eta-limits}.
Next, assume $z\in\gL-\gL$ with $z\neq 0$. Then, we split
$\zeta^{(\varOmega)}_{z,n}$ into two sums,
\begin{equation}\label{splitting}
    \zeta^{(\varOmega)}_{z,n} \; = \;
    \frac{1}{\vol (A_n)} \, \Bigl(
    {\sum_{\substack{x\in\gL_n \\ x-z\in\gL}}}^{\!\!\!(0)}
    \,\varOmega_x * \widetilde{\varOmega}_{x-z}\; + 
    {\sum_{\substack{y\ts\in\gL_n \\ y-z\in\gL}}}^{\!\!\!(1)}
    \,\varOmega_y * \widetilde{\varOmega}_{y-z}\Bigr),
\end{equation}
where the upper index stands for the following additional
restriction:\ Given $z$, our point set $\gL$ is the disjoint
union of countably many maximal linear chains of the form
\[
     \{ \ldots, x+2z,x+z,x,x-z,x-2z,\ldots \} 
\]
with \emph{all} points lying in $\gL$, and $x$ being chosen as its
representative. Such a chain may be finite or infinite, but has no
gaps by construction.  For each of these chains, the random measures
$\varOmega_{x+mz} * \widetilde{\varOmega}_{x+(m-1)z}$, are identically
distributed, but not independent (due to the index overlap).  However,
those with $m$ \emph{even} (type (0)) are mutually independent, as are
those with $m$ \emph{odd} (type (1)). This way, each element of $\gL$
inherits the type as a label, and the terms in \eqref{splitting} are
distributed to the two sums according to their type. This approach
guarantees that the terms for $\zeta^{(\varOmega)}_{z,n}$ which
already showed up in $\zeta^{(\varOmega)}_{z,n-1}$ end up in the same
sum as before, no matter what the detailed structure of the (nested)
averaging sequence might be. We also split the number of terms
\[
   \card \{x\in\gL_n\mid x-z\in\gL \} \; = \;
   N^{(0)}_{n} + N^{(1)}_{n}
\]
accordingly.  We can now rewrite our previous expression in the form
\begin{equation} \label{intermediate}
    \zeta^{(\varOmega)}_{z,n} \; = \;
    \frac{\card\{x\in\gL_n\mid x-z\in\gL \}}{\vol (A_n)}\,
    \Bigl( \frac{N^{(0)}_{n}}{N^{(0)}_{n} + N^{(1)}_{n}}\,
    \frac{{\sum}^{(0)}}{N^{(0)}_{n}} \, +
    \frac{N^{(1)}_{n}}{N^{(0)}_{n} + N^{(0)}_{n}}\,
    \frac{{\sum}^{(1)}}{N^{(1)}_{n}} \Bigr),
\end{equation}
where the term in brackets is a convex combination of two
random measures $\sum^{(0)}\!\! /N^{(0)}_{n}$ and 
$\sum^{(1)}\!\! /N^{(1)}_{n}$. By \eqref{eta-limits},
the factor in front of the bracket converges to $\eta(z)$.
When this limit is non-zero, we know that $N^{(i)}_{n}
\xrightarrow{n\to\infty}\infty$ for $i\in\{0,1\}$, so that
Lemma~\ref{measure-SLLN} and Eq.~\eqref{conv-expect}
imply
\begin{equation} \label{sum-one}
   \frac{1}{N^{(i)}_{n}} \, {\sum}^{(i)} \;
   \xrightarrow{n\to\infty} \;
   \EE_Q (\varOmega) * \widetilde{\EE_Q (\varOmega)}
   \qquad \mbox{(a.s.)}.
\end{equation}
Although we do not know whether the rational prefactors in
\eqref{intermediate} converge, we have a convex combination of two
sequences that each almost surely converge to the same limit, which
must then also be the limit of their convex combination. Put together,
this gives
\begin{equation}\label{pw-limit}
    \zeta^{(\varOmega)}_{z,n} \; 
    \xrightarrow{n\to\infty} \;
    \eta(z)\cdot \EE_Q (\varOmega) * 
    \widetilde{\EE_Q (\varOmega)} \qquad \mbox{(a.s.)}
\end{equation}
for all $z\in\gL-\gL$ with $z\neq 0$.

These considerations will be sufficient when the random measures 
almost surely have a (deterministic) compact support. To formulate the
main result of this section in greater generality, we need one further
technical property. For brevity, we write $B^{\mathsf{c}}_{r} =
\RR^d \setminus B^{}_{r} (0)$.
\begin{lemma} \label{lem:uni-bound}
    If $\cM'$ is a set of uniformly translation bounded, positive measures 
    and $\nu$ a finite, positive measure on $\RR^d$, there is a sequence
    $R_k \nearrow \infty$ such that
\[
    \sum_{k=1}^{\infty} \sup_{\omega\in\cM'}
    \bigl( \omega{}_{\textstyle |_{B^{\ts \mathsf{c}}_{R_k}}}
    \!\! \! * \nu \bigr)
    (K) \, < \, \infty
\]    
    holds for any compact set $K\subset\RR^d$.
\end{lemma}
\begin{proof}
Since $\nu$ is a finite, positive measure and $\nu (B^{\mathsf{c}}_{r})$ a 
decreasing function that tends to $0$ as $r\to\infty$, we can choose 
radii $R_k$ with $\nu (B^{\mathsf{c}}_{R_k}) < 1/k^2$, so that
$\sum_{k=1}^{\infty} \nu (B^{\mathsf{c}}_{R_k}) < \pi^2/6 < \infty$.
Moreover, we may do this in such a way that the differences
between consecutive radii do not decrease, meaning that
$R_{2} \ge 2 \ts R_{1}$ and
$R_{k+2} - R_{k+1} \ge R_{k+1} - R_{k}$ for all $k\in\NN$.

Uniform translation boundedness of $\cM'$ means that, for any compact
$K\subset\RR^d$, there is a positive constant $\alpha^{}_{K}$ with
$\omega (x+K)\le \alpha^{}_{K}$, simultaneously for all $x\in\RR^d$
and all $\omega\in\cM'$. If $K\subset\RR^d$ is compact, we have
$K\subset B_{r} (0)$ for some $r>0$. If $R>r$, one has
\[
   \begin{split}
   \bigl( \omega{}_{\textstyle |_{B^{\ts \mathsf{c}}_{R}}} \! * \nu \bigr) (K)
   & = \int_{\RR^{d} \nts\times \RR^{d}} \mathbf{1}^{}_{K} (x+y)\,
       \mathbf{1}^{}_{B^{\ts \mathsf{c}}_{R}} (x) \dd \omega (x) \dd \nu (y) \\
   & = \int_{\RR^d} \omega\bigl( (K-y) \cap B^{\ts \mathsf{c}}_{R}\bigr)
       \dd \nu (y) \, \le \, \alpha^{}_{K}\, \nu(B^{\ts \mathsf{c}}_{R-r}) \ts ,
   \end{split}
\]   
where the last step follows because $\omega\bigl( (K-y) \cap 
B^{\ts \mathsf{c}}_{R}\bigr)$ vanishes whenever $\lvert y \rvert \le R-r$.
For radii $0\le R \le r$, the bound is simply given by
$\alpha^{}_{K} \, \nu(\RR^d)$, which is finite.

Now, for some $m\in\NN$, we have $R_m - r \ge R_{1}$, with the
sequence of radii chosen before. The additional difference property
of the radii makes sure that the radii $R_{m+k}$ with $k\in\NN$ give
a summable contribution, while the remaining terms are finite by
construction. Since this argument is uniform in $\omega\in\cM'$
and holds for all compact $K\subset\RR^d$, our claim follows.
\end{proof}

\begin{theorem} \label{pprm} 
  Let $\gL\subset\RR^d$ be an FLC point set such that its Dirac comb
  $\delta_{\gL}$ possesses the autocorrelation measure
  $\gamma^{}_{\gL}$ of\/ $\eqref{limit-exists}$, relative to the fixed
  averaging sequence $\mathcal{A}$, and thus the diffraction measure\/
  $\widehat{\gamma}^{}_{\gL}$. Let $(\varOmega_x)_{x\in\gL}$ be a
  family of integrable, complex, i.i.d.\ random measures with
  common law $Q$ and finite second moment measure, with 
  $\varOmega$ being any representative of this family, 
  and consider the random measure $\delta^{(\varOmega)}_{\gL}\!$
  of \eqref{random-comb}.

  Then, possibly after replacing $\cA$ by a suitable subsequence
  $\cA'$, the sequence of approximating measures
  $\gamma^{(\varOmega)}_{\gL,n}$ of\/ \eqref{eq:random-auto-corr}
  almost surely converges, as $n\to\infty$, to
  the positive definite, translation bounded autocorrelation measure
\[
   \gamma^{}_{\gL,Q} \; = \;
   \bigl(\EE_Q (\varOmega) * \widetilde{\EE_Q (\varOmega)}\bigr)
   * \gamma^{}_{\gL} + \dens (\gL)\,
   \bigl(\EE_Q (\varOmega\conv\widetilde{\varOmega})
   - \EE_Q (\varOmega) * \widetilde{\EE_Q (\varOmega)}\bigr) \! *
   \delta_0 \/ .
\]
   This measure has the Fourier transform
\[
   \widehat{\gamma}^{}_{\gL,Q} \; = \;
   \bigl\lvert \widehat{\EE_Q (\varOmega)}\bigr\rvert^2
   \cdot \widehat{\gamma}^{}_{\gL} + \dens (\gL)\,
   \bigl(\EE_Q (\varOmega\conv
   \widetilde{\varOmega}) - \EE_Q (\varOmega) * 
   \widetilde{\EE_Q (\varOmega)}\bigr)^{\!\widehat{\hphantom{m}}}
   \! \cdot \lambda \ts , 
\]
   which is the almost sure diffraction measure of the random
   measure\/ $\delta^{(\varOmega)}_{\gL}$ relative to $\cA'$.
\end{theorem}
\begin{proof}
  The previous calculations establish the individual almost sure
  convergence of the (countably many) measures
  $\zeta^{(\varOmega)}_{z,n}$, with the limits as given in
  Eqs.~\eqref{zero-limit} and \eqref{pw-limit}. Our assumptions on
  $\varOmega$ ensure that $\EE_Q (\varOmega) * \widetilde{\EE_Q
    (\varOmega)}$ in \eqref{pw-limit} is a finite, positive definite
  measure, which is concentrated at $0$ in the sense that $\bigl|
  \EE_Q (\varOmega) * \widetilde{\EE_Q (\varOmega)}\bigr| \bigl( \RR^d
  \setminus B_r (0)\bigr) \xrightarrow{\,r\to\infty\,} 0$, while
  $\bigl(\EE_Q (\varOmega) * \widetilde{\EE_Q (\varOmega)} \bigr)
  (g*\widetilde{g}) > 0$ for all $0\ne g\in C_{\mathsf{c}} (\RR^d)$.
  In view of \eqref{zero-limit} and \eqref{pw-limit}, our (claimed) 
  almost sure limit $\gamma^{}_{\gL,Q}$ inherits translation boundedness 
  from $\gamma^{}_{\gL}$. 
  
  The (deterministic) measure $\gamma^{}_{\gL,Q}$ is positive definite, 
  hence transformable by Lemma~\ref{pos-def-props}. Its Fourier transform
  has the form claimed as a result of the convolution theorem
  \cite[Excs.~4.18]{BF}. The latter is applicable here because all
  expectation measures involved are finite measures, so that their
  Fourier transforms are represented by uniformly continuous functions
  on $\RR^d$.
  
  It remains to establish the limit property.  Let us first assume
  that there is a (deterministic) compact set $C$ so that $\supp
  (\varOmega_x) \subset C$ almost surely. This implies $\supp
  (\zeta^{(\varOmega)}_{z,n}) \subset C-C$ for all $n$ and $z$, so
  that only terms from finitely many $z\in \gL - \gL$ contribute to
  $\gamma^{(\varOmega)}_{\gL,n;\mathrm{mod}}$ on any compact $K\subset
  \RR^d$. In this case, we may use an elementary pointwise calculation
  to see that $\gamma^{(\varOmega)}_{\gL,n;\mathrm{mod}}$ tends to the
  claimed limit, and Proposition~\ref{prop:modautocorr} gives the
  assertion.
  
  In the general case, this simple argument is not conclusive, and we
  need to estimate putative contributions from distant points
  $z\in\gL-\gL$ to $\gamma^{(\varOmega)}_{\gL,n;\mathrm{mod}}$.  For
  bounded $K, B \ni 0$, with $\mu(\cdot) := \EE\big(
  |\varOmega|(\cdot)\big)$ as above, we have
 \begin{equation}\label{eq:bt18.1}
\begin{aligned}
    \EE_{Q} \biggl[ \Bigl| & \sum_{\substack{z\in\gL-\gL \\ z \not\in B}} 
    \frac{1}{\vol (A_n)} \sum_{\substack{x\in\gL_n \\ x-z\in\gL}}
    \bigl(\varOmega_x \conv \widetilde{\varOmega}_{x-z} \conv \delta_z
    \bigr)(K) \Bigr| \biggr] \\
    \leq & \sum_{\substack{z\in\gL-\gL \\ z \not\in B}}
    \frac{1}{\vol (A_n)} \sum_{\substack{x\in\gL_n \\ x-z\in\gL}}
   \EE_{Q} \Bigl[ \bigl( |\varOmega_x|  \conv |\widetilde{\varOmega}_{x-z}| 
   \conv \delta_z \bigr)(K)\Bigr] = \!\!
   \sum_{\substack{z\in\gL-\gL \\ z \not\in B}}
   \frac{1}{\vol (A_n)} \sum_{\substack{x\in\gL_n \\ x-z\in\gL}}
   \bigl(\mu \conv \widetilde{\mu} \conv \delta_z \bigr)(K) \\
   = & \, \biggl( \mu \conv \widetilde{\mu}  \conv 
   \Bigl( \frac{1}{\vol (A_n)}  
   \sum_{\substack{z\in\gL-\gL, \, x \in \gL_n \\ x-z \in \gL, \, z \not\in B}} 
   \hspace{-0.7em} \delta_z \Bigr)\biggr)(K) 
   = \Bigl( \mu \conv \widetilde{\mu} \conv 
      \bigl( \gamma_{\gL,n;\ts \mathrm{mod}}{}_{\textstyle |_{\RR^d \setminus B}} 
      \bigr)  \Bigr)(K)  \, =: \, \phi_n (K,B) \ts .
\end{aligned}
\end{equation}
Since $\{ \gamma_{\gL,n;\ts \mathrm{mod}} \mid n \in \NN\}$ are uniformly 
translation bounded by Lemma~\ref{lem:utb}, we can choose a sequence 
of radii $R_k \nearrow \infty$ according to Lemma~\ref{lem:uni-bound} such 
that, for all compact $K \subset \RR^d$,
\begin{equation}  \label{eq:bt18.2}
     \sum_{k=1}^{\infty}  \, \sup_{n \in\NN} \ts
      \phi_n \bigl( K, B_{R_k} (0) \bigr) < \infty. 
\end{equation}
 
We have proved above that, for each $z \in \gL-\gL$,
$\zeta^{(\varOmega)}_{z,n} \xrightarrow{\, n\to\infty \,}
\zeta^{(\varOmega)}_{z,\infty}$, almost surely in the vague topology,
where $\zeta^{(\varOmega)}_{0,\infty} = \dens(\gL)\cdot \EE_Q
(\varOmega * \widetilde{\varOmega})$ and
$\zeta^{(\varOmega)}_{z,\infty} = \eta(z)\cdot \EE_Q (\varOmega) *
\widetilde{\EE_Q (\varOmega)}$ for $z\neq 0$.  Possibly after passing
to another subsequence, we may now assume that the convergence is so
fast that, for any $g\in C_{\mathsf{c}} (\RR^d)$,
\begin{equation} \label{eq:nochduennerefolge}
  \Bigl| \!
  \sum_{\substack{z\in\gL-\gL \\ \lvert z \rvert < R_n }} \!  \Bigl( \bigl(
  \zeta^{(\varOmega)}_{z,n} - \zeta^{(\varOmega)}_{z,\infty}\bigr)
  \conv \, \delta_z \Bigr)(g) \, \Bigr| \; \xrightarrow{\, n\to\infty
    \,}\; 0 \qquad \mbox{(a.s.)} \ts .
\end{equation}
Using Markov's inequality and \eqref{eq:bt18.1}, we conclude
for any bounded $B$ and $\varepsilon>0$ that
\begin{equation}
    \PP\Bigl(\!  \sum_{\substack{z\in\gL-\gL \\ \lvert z \rvert \ge R_n }} 
    \! \bigl| \zeta^{(\varOmega)}_{z,n} \conv \, \delta_z \bigr|(K)
    \geq \varepsilon \Bigr)
    \, \le \, \frac{\phi_n\bigl(K, B_{R_n} (0) \bigr)}{\varepsilon} \ts ,
\end{equation}
which is summable by \eqref{eq:bt18.2}. Hence, by Borel-Cantelli, 
\begin{equation} 
     \sum_{\substack{z\in\gL-\gL \\ \lvert z \rvert \ge R_n }} 
     \zeta^{(\varOmega)}_{z,n} \conv \, \delta_z
     \; \xrightarrow{\, n\to\infty \,} \; 0 \qquad \mbox{(a.s.)} \ts .
\end{equation}
Combining this with \eqref{eq:nochduennerefolge} shows that the limit
is the expected one (from the pointwise calculation) also in this
case, which yields the claim.
\end{proof}

Note that our argument is based on the potential selection of a
subsequence of the original (deterministic) $\cA$. However, it also
shows that the limit derived in Theorem~\ref{pprm} is the only point
of accumulation along any deterministic subsequence of $\cA$.

\begin{remark} \textsc{Randomisation of Meyer sets}.
\label{Meyer-deform}
A particularly relevant class of point sets in the theory of aperiodic
order are \emph{Meyer sets}, which are relatively dense sets $\gL$
such that $\gL - \gL$ is uniformly discrete. Such sets always have a
diffraction measure with a non-trivial pure point part, with a
relatively dense supporting set \cite{Nicu,B}, despite the fact that
Meyer sets can have entropy\footnote{The binary random tilings of
  Example~\ref{random-tiling} produce Meyer sets whenever
  $b/a\in\QQ$.}.  If modified by a family of random measures according
to Theorem~\ref{pprm}, the resulting diffraction still shows the
original diffraction with its non-trivial pure point component,
modulated by the function $\big| \widehat{\EE_{Q}
  (\varOmega)}\big|^2$, in addition to the diffuse background
originating from the added randomness.  \exend \end{remark}

Let us look at consequences of Theorem~\ref{pprm} in terms of
some examples. 

\begin{example} \textsc{Deterministic clusters}.
\label{det-cluster}
Let $S\subset\RR^d$ be a finite point set, and consider
$\varOmega\equiv\delta_S = \sum_{x\in S}\delta_x$. Clearly, this
completely deterministic case gives $\EE_Q (\lvert\varOmega\rvert) =
\EE_Q ( \varOmega ) = \delta_S$ and $\EE_Q
(\varOmega\conv\widetilde{\varOmega}) = \delta_S *
\widetilde{\delta_S}$, so that Theorem~\ref{pprm} gives
$\gamma^{(\varOmega)}_{\gL} = \bigl(\delta_S * \widetilde{\delta_S}
\bigr) * \gamma^{}_{\gL}$ and $\widehat{\gamma}^{(\varOmega)}_{\gL} =
\lvert \widehat{\delta_S}\rvert^2\cdot\widehat{\gamma}^{}_{\gL}$,
which is always true (rather than almost always)
in this case. A particularly simple instance of
this emerges from $S=\{a\}$, which effectively means a global
translation by $a$. This leads to the relations
$\gamma^{(\varOmega)}_{\gL} = \gamma^{}_{\gL}$ and
$\widehat{\gamma}^{(\varOmega)}_{\gL} = \widehat{\gamma}^{}_{\gL}$, as
it must.  \exend \end{example}

\begin{example} \textsc{Random weight model}.
\label{random-weight}
Here, we consider $\varOmega = H \delta_0$, where $H$ is a
complex-valued random variable with a law $\mu$ that 
satisfies $\EE_{\mu} (\lvert H \rvert^2) < \infty$
(hence also  $\EE_{\mu} (\lvert H \rvert) < \infty$).  
Clearly, this gives $\EE_Q (\varOmega) =
\EE_{\mu} (H)\,\delta_0$ and $\EE_Q
(\varOmega\conv\widetilde{\varOmega}) = \EE_{\mu} (\lvert H
\rvert^2)\,\delta_0$, so that Theorem~\ref{pprm} results in the
diffraction formula
\[
      \widehat{\gamma}^{(\varOmega)}_{\gL}  \; = \;
      \lvert \EE_{\mu} (H) \rvert^2 \cdot \widehat{\gamma}^{}_{\gL}
      + \dens (\gL) \, \bigl( \EE_{\mu} (\lvert H \rvert^2)
      - \lvert \EE_{\mu} (H) \rvert^2 \bigr)\cdot\lambda
      \qquad \mbox{(a.s.)} \/ .
\]
The autocorrelation is clear from Theorem~\ref{pprm}.
\exend \end{example}

\begin{remark} \textsc{Interpretation as particle gas}.
\label{Lambda-gas}
A widely used special case of Example~\ref{random-weight} is the
random occupation model, or `$\!\gL$-gas'. Here, $\varOmega$
may take the value $\delta_0$ (with probability $p$, for
`occupied') or $0$ (with probability $1-p$, for `empty').
This gives the diffraction
\[
    \widehat{\gamma}^{(\varOmega)}_{\gL}  \; = \;
    p^2 \cdot \widehat{\gamma}^{}_{\gL} +
    \dens (\gL) \cdot p (1-p) \cdot \lambda 
    \qquad \text{(a.s.)},
\]
which was derived in a similar setting in \cite{BM1}, and later
generalised to Bernoulli and Markov systems \cite{BH}, to systems with
finite range Gibbs measures \cite{BZ}, and beyond \cite{K1,K2}.
\exend \end{remark}

The results of Examples~\ref{det-cluster} and \ref{random-weight} can
be extended in many ways, some of which will be met later on.  One
further possibility consists in replacing a point by a `profile', as
described by an integrable function, or by a finite collection of such
profiles, which could represent different types of atoms.  The
corresponding formulas for the autocorrelation and the diffraction are
then easy analogues of the ones given so far.

\begin{example} \textsc{Random displacement model}.
\label{random-displacement}
Consider the random measure $\varOmega = \delta^{}_{X}$, where $X$ is
an $\RR^d$-valued random variable with law $\nu$. If $A\subset\RR^d$ 
is a Borel set, one has
\[
     \bigl(\EE_Q (\varOmega)\bigr) (A) \; = \; \int_{\RR^d} \delta_x (A)
     \dd\nu (x) \; = \; \int_{\RR^d} \mathbf{1}_{A} (x) \dd\nu (x) 
     \; = \; \nu (A) \/ ,
\]
which shows that $\EE_Q (\varOmega) = \nu$. One also
finds $\EE_Q (\varOmega\conv\widetilde{\varOmega}) =
\nu (\RR^d)\, \delta_0  = \delta_0$. Theorem~\ref{pprm} 
now results in the equations
\[
   \begin{split}
      \gamma^{(\varOmega)}_{\gL} & \; = \;
      (\nu\conv\widetilde{\nu}\ts)*\gamma^{}_{\gL} 
      + \dens (\gL)\, (\delta_0 - \nu\conv\widetilde{\nu}\ts)
      \qquad \mbox{(a.s.)} \/ , \\[1mm]
     \widehat{\gamma}^{(\varOmega)}_{\gL} & \; = \;
     \lvert \widehat{\nu}\rvert^2 \cdot \widehat{\gamma}^{}_{\gL}
      \; + \; \dens (\gL)\, (1 -  \lvert \widehat{\nu}\rvert^2)\cdot
     \lambda \ts \ts \qquad \mbox{(a.s.)} \/ ,
   \end{split}
\]
which recovers Hof's result on the diffraction at high temperature
\cite{Hof-random}.
\exend \end{example}

In comparison, Hof's approach to the random displacement model
\cite{Hof-random} also uses the SLLN, but does not require the FLC 
property. Instead, he needs an ergodicity assumption on
the underlying point set; compare also \cite{L}.

\begin{remark} \textsc{Extension of Theorem~\ref{pprm}}.
\label{rem:ext-thm2}
The argument above is shown for FLC sets in a pointwise 
fashion, to make the result more transparent. 
However, it is clear that one does not need the FLC property itself.
Indeed, it is sufficient to assume that the fixed point set $\gL$,
relative to a chosen van Hove averaging sequence $\cA$, possesses
an autocorrelation that is a pure point measure of the form
$\gamma = \sum_{z\in F} \eta(z) \ts \delta_z$ with $F$ a locally finite
point set. An argument with local test functions will then still connect
to the SLLN and thus avoid the need for ergodic assumptions on
the underlying set $\gL$.
\exend \end{remark}

With hindsight, it is rather clear that the formulas of
Theorem~\ref{pprm} are robust, and should also hold for other point
sets, such as those coming from a homogeneous Poisson process or from
a model set based particle gas, as introduced in \cite{BM1}. So, to
complement our approach of this section, let us now consider ergodic
point processes instead, meaning that also the set $\gL$ becomes part
of the random structure.

\section{Arbitrary dimensions: Point process approach}
\label{general}

Here, we are interested in the diffraction of certain random subsets
of $\mathbb{R}^d$, where we restrict ourselves to the situation that
these subsets are self-averaging in a suitable way. This will be
guaranteed by the ergodicity of the underlying stochastic process.
One further benefit is that we are freed from details of the averaging
sequence and the potential selection of subsequences thereof.  It is
convenient to start by putting ourselves in the context of random
counting measures, which we now summarise in a way that is tailored to
diffraction theory.

\subsection{Random measures and point processes}

Let $\cM^+$ denote the set of all locally finite, positive measures
$\phi$ on $\RR^d$ (where we mean to include the $0$ measure).  That
$\phi$ is locally finite (some authors say $\phi$ is boundedly finite
or that $\phi$ is a Radon measure) means that, for all bounded Borel
sets $A$, $\phi(A) < \infty$. The space $\cM^+$ is closed in the
topology of vague convergence of measures
(in fact, $\cM^+$ is a complete separable metric space,
see \cite[A 2.6]{DVJ1}\footnote{We refer to the second edition
of this work throughout, which comes in two volumes \cite{DVJ1,DVJ2}.
All results we need are also contained in the original one volume
edition \cite{DVJ}, sometimes with a slightly different numbering.}). 
We let $\varSigma_{\cM^+}$
denote the $\sigma$-algebra of Borel sets of $\cM^+$.  The latter can
be described as the $\sigma$-algebra of subsets of $\cM^+$ generated
by the requirement that, for all Borel sets $A\subset\RR^d$, the
mapping $\phi\mapsto\phi(A)$ is measurable; compare \cite[Chs.~1.1 and
1.2]{Karr} for background.

A \emph{random measure} on $\RR^d$ is a random variable $\varPhi$ from
a probability space $(\varTheta, \mathcal{F}, \pi)$ into $(\cM^+,
\varSigma_{\cM^+})$.
Let us write $\mathcal{P} (\cM^+)$ for the convex set of probability
measures on $\cM^+$.  The \emph{distribution} of a random measure
$\varPhi$ is the probability measure $P=P^{}_{\varPhi} \in \mathcal{P}
(\cM^+)$, defined by $P=\pi\circ\varPhi^{-1}$. In other words, $P$ is
the \emph{law} of $\varPhi$, written as $\mathcal{L} (\varPhi) = P$.
Note that, as soon as $P$ is given or determined, one can usually
ignore the underlying probability space.

{}For each $t\in\RR^d$, let $T_t$ denote the translation operator on
$\RR^d$, as defined by the mapping $x\mapsto t+x$. Clearly, one has
$T_t \ts T_s = T_{t+s}$, and the inverse of $T_t$ is given by
$T^{-1}_t = T^{}_{-t}$. For functions $f$ on $\RR^d$, the
corresponding translation action is defined via $T_t f = f\circ
T_{-t}$, so that $T_t f (x) = f(x-t)$. Similarly, for $\phi \in
\cM^+$, let $T_x \phi := \phi \circ T_{-x}$ be the image measure under
the translation, so that $(T_x \phi)(A) = \phi(T_{-x}(A)) = \phi(A-x)$
for any measurable subset $A \subset \mathbb{R}^d$, and $(T_{x} \phi)
(f) = \int_{\RR^d} f(y) \dd (T_x \phi) (y) = \int_{\RR^d} f(x+z) \dd
\phi(z) = \phi (T_{-x} f)$ for functions.  This means that there is a
translation action of $\RR^d$ on $\cM^+$. Finally, we also have a
translation action on $\mathcal{P} (\cM^+)$, via $(T_x Q)(A) =
Q(T_{-x}A)$ for any measurable $A\subset \cM^{+}$.

Our primary interest is in random counting measures. A measure $\phi$
on $\RR^d$ is called a \emph{counting measure} if $\phi (A)\in\NN_{\ts
  0}$ for all bounded Borel sets $A$. These are positive,
integer-valued measures of the form $\phi = \sum_{i\in I}
\delta_{x_i}$, where the index set $I$ is (at most) countable and the
support of $\phi$ is a locally finite subset of $\RR^d$.  The
(positive) counting measures form a subset $\cN^+ \! \subset \cM^+$.
We can repeat the above discussion of $\cM^+$ by restricting
everything to $\cN^+$. The vague topology on $\cN^+$ is the same as
its topology inherited from $\cM^+$, and its $\sigma$-algebra of Borel
sets $\varSigma_{\cN^+}$ consists of the intersections of the elements
of $\varSigma_{\cM^+}$ with $\cN^+$. The concepts of the law of a
random measure and the translation action by $\RR^d$ carry over. In
particular, for $x \in \mathrm{supp}(\phi)$ with $\phi \in \cN^+$,
$T_{-x}\phi$ corresponds to the counting measure obtained from $\phi$
by translating its \emph{support} so that $x$ is shifted to the origin.

A \emph{point process} on $\RR^d$ is a random variable $\varPhi$ from
a probability space $(\varTheta, \mathcal{F}, \pi)$ into $(\cN^+,
\varSigma_{\cN^+})$. Alternatively, a point process is a random
measure for which $\pi$-almost all $\theta\in\varTheta$ are counting
measures. Furthermore, it is called \emph{simple} when, for
$\pi$-almost all $\theta\in\varTheta$, the atoms of
$\phi=\varPhi(\theta)$ have weight (or multiplicity) $1$.

In many instances, the point processes we are dealing with are
\emph{simple}.  Whenever this happens, we feel free to identify point
measures with their supports.  In this case, the measures almost
surely are Dirac combs of the form $\phi=\delta_S$ with
$S\subset\RR^d$ locally finite. Later on, we create compound processes
in which an underlying point process is decorated with a random finite
measure, and this will take us from $\cN^+$ to $\cM^+$, which is also
the reason why we introduced random measures above.

\smallskip

For a random measure (or a point process) $\varPhi$ with
law $P$, the expectation measure $\mathbb{E}_P (\varPhi)$
is defined by
\begin{equation} 
    \bigl(\mathbb{E}_P (\varPhi)\bigr)  (A) \; = \;
    \mathbb{E}_P \bigl( \varPhi(A) \bigr) \; = \;
    \int_{\cN^+} \phi(A)  \dd P(\phi),  \quad 
    \mbox{for } A \subset \mathbb{R}^d \mbox{ Borel.}
\end{equation} 
It is a measure on $\mathbb{R}^d$ which gives the \emph{expected} 
mass (or number of points) that $\varPhi$ has in $A$. More precisely,
in terms of the  underlying probability space $(\varTheta,\mathcal{F},\pi)$, 
one writes
\[
   \mathbb{E}_P \bigl( \varPhi(A) \bigr) \; = \; 
   \int_{\varTheta} \varPhi(\theta) (A)
   \dd \pi(\theta) \; = \;  \int_{\cN^+} \varPhi(A)  \dd P(\varPhi) \ts .
\]
It is common in the probability literature (and we adopt this slight
abuse of notation here, too) to suppresses the explicit dependence on
$(\varTheta,\mathcal{F},\pi)$ by simply writing $\varPhi$ for the
general instance $\varPhi(\theta)$ of the process $\varPhi$. The
latter is called \emph{stationary} when its law $P$ is translation
invariant, which means that $T_t P = P\circ T_{-t} = P$ holds for all
$t\in\RR^d$.

\begin{remark} \textsc{Intensity of a process}.
\label{stationary-case} If $P$ is stationary, we have
  $T_t\ts \mathbb{E}_P ( \varPhi) = \mathbb{E}_P (\varPhi)$ for all
  $t\in\RR^d$, whence $\mathbb{E}_P (\varPhi)$ must be a multiple of
  Lebesgue measure (the latter being Haar measure on $\RR^d$).
  Consequently, 
  \[
      I_P (\varPhi) \, = \, \mathbb{E}_P (\varPhi) 
      \, = \, \pd\ts \lambda\ts ,
  \]     
  where $\pd\! \in [0,\infty]$ is usually called the \emph{intensity}
  of $P$.  Unfortunately, this term is already in use for the
  positive weights of Bragg peaks in diffraction theory.  
  In the setting of simple point processes, $\pd$ also has the meaning
  of a point density, averaged over all realisations of the process.
  In the ergodic case (see below for a definition), it is then almost
  surely the density of a given realisation in the usual sense.  We
  thus often prefer to call $\pd$ the \emph{point density} of the simple
  point process or the \emph{density} of the random measure.  \exend
\end{remark}

{}From now on, we always assume that $\pd$ is \emph{finite}.  Let
$\varPhi\! : (\varTheta, \mathcal{F}, \pi) \longrightarrow
(\cX,\varSigma_{\cX})$ be a stationary random measure (where $\cX =
\cM^+$) or point process ($\cX = \cN^+$), with law $P$. Then,
$(\cX,\varSigma_{\cX}, P)$ is a probability space with translation
invariant probability measure $P$. In fact, we will usually simply
assume that $(\cX,\varSigma_{\cX}, P)$ is itself the probability space
(or basic process) we are dealing with. In general, there will be
several different spaces, and to keep track of the processes, we use
the law of the basic process as an index.

Let us recall that the random measure or point process $\varPhi$ is
called $\emph{ergodic}$ when $(\cX,\varSigma_{\cX}, P)$ is ergodic as
a dynamical system \cite{W} under the translation action of $\RR^{d}$,
see below for more. In particular, we do not refer to strict
ergodicity this way.

\subsection{Palm distribution and autocorrelation} 

Let $P \in \mathcal{P}(\cN^+)$ be stationary with finite density $0 <
\pd < \infty$. The assumption $\rho>0$ is no restriction, since it is
easy to see that a realisation of a stationary point process with
intensity $\rho=0$ almost surely is the zero measure.

Let $\mathbf{1}_\cB$, as usual, denote the characteristic function of
the set $\cB\subset \cN^+$, and choose a Borel set $A \subset \mathbb{R}^d$ 
with $0 < \lambda(A) < \infty$. The \emph{Palm distribution} 
$P^{}_0$ is the probability measure on $\cN^+$ that satisfies
\begin{equation}  \label{Palm-one}
  P^{}_0(\cB) \; = \; \frac{1}{\mathbb{E}_P ( \varPhi(A) )} 
   \int_{\cN^+} \sum_{x \in A\cap\supp(\varPhi)} \varPhi(\{x\}) \,
   \mathbf{1}_\cB\big( T_{-x}\varPhi \big)  \dd P(\varPhi)
\end{equation}
for any $\cB \in \varSigma_{\cN^+}$, compare \cite[Ch.~4.4]{SKM} or
\cite[Ch.~3]{KMM} for background. Due to stationarity,
Remark~\ref{stationary-case} applies to $\mathbb{E}_P ( \varPhi(A) )$,
whence the prefactor simplifies to $(\pd\ts\lambda(A))^{-1}$.  Note
that the sum under the integral runs over at most countably many
points. Moreover, the definition does \emph{not} depend on the actual
choice of $A$.  Intuitively, $P^{}_0$ describes the configuration
$\varPhi$ as seen from a typical point in $\supp (\varPhi)$, with that
point translated to the origin. Alternatively, in the case of simple
point processes, one can think of $P^{}_0$ as the distribution of
$\varPhi$, conditioned on having a point measure at $0$. 
This actually amounts to condition properly on an event of probability 
$0$, which might need some further explanation.

The first point of view can be made precise, at least in the ergodic
case, as a limit, via sampling points in $\varPhi$ over larger and
larger balls, see \cite[Thm.~3.6.6]{KMM} or \cite[Prop.~13.4.I and
Prop.~13.4.IV]{DVJ2} as well as Eq.~\eqref{ergodic} below.  
The second interpretation can be corroborated by
conditioning $\varPhi$ to have a point in a small ball around $0$
and then again taking a limit, see \cite[Thm.~13.3.IV]{DVJ2}. In more 
precise terms, $P^{}_0$ would be called the \emph{Palm distribution 
with respect to\/ $0 \in \mathbb{R}^d$}, compare \cite[Ch.~10]{K} or
\cite[Ch.~13.1]{DVJ2}. Since we will mostly be dealing with the
stationary scenario, we refrain from spelling out the full name.

\smallskip

There is an alternative approach to the Palm distribution, which also
applies to the random measure case, compare \cite[Chs.~13.1 and
13.2]{DVJ2}.  Let $\varPhi \!: (\varTheta, \mathcal{F}, \pi)
\longrightarrow (\cM^+,\varSigma_{\cM^+})$ be a stationary random
measure with law $P$ and finite mean density $\pd < \infty$.  Then,
the \emph{Palm distribution} is the unique probability measure $P_0$ on
$\cM^+$ that satisfies
\begin{equation} \label{PalmGeneral}
    \EE_{P} \left(\int_{\RR^d}  g(x,\varPhi) 
    \dd \varPhi(x)\right) 
    \; = \; \pd
    \int_{\RR^d}  \int_{\cM^+} g(x, T_x\psi) 
    \dd P_0(\psi) \dd x
\end{equation}
for all non-negative functions $g$ on $\RR^d \times \cM^+$ for which
$\int_{\RR^d} \int_{\cM^+} g(x,\phi) \dd\phi(x) \dd P(\phi)$ is
finite. When dealing with point processes, all this reduces to $\cN^+$
by simply replacing $\cM^+$ with $\cN^+$ throughout
Eq.~\eqref{PalmGeneral}, compare \cite[Ch.~13.2 and
Thm.~13.2.III]{DVJ2}.

If  $\varPhi$ is an ergodic, stationary random measure,
an application of the ergodic theorem implies
\begin{equation} \label{ergodic} 
     \frac{1}{\lambda (B_n)} \int_ {B_n} F(T_{-x}\varPhi) \dd \varPhi(x) 
     \; \xrightarrow{n\to\infty} \; 
     \pd \int_{\cM^+} F(\varPsi)  \dd P^{}_0 (\varPsi) 
     \qquad \text{(a.s.)},
\end{equation}
for any non-negative measurable function $F \! : \;
\cM^+ \to \mathbb{R}$, see \cite[Prop.~13.4.I]{DVJ2} or the
proof of \cite[Thm.~3.6.6]{KMM}. Here and below, we write
$B_n$ for $B_n (0)$ and $\lambda (B_n)$ for $\vol (B_n(0))$.

\smallskip

In the literature, the probability measure $P^{}_0$ is usually called
the \emph{Palm distribution} of $P$ (with respect to $0$), while the
term \emph{Palm measure} is also in use for the unnormalised version
$\pd P^{}_{0}$, a convention we adopt here.  The first moment measure
of the latter coincides with the \emph{autocorrelation measure} of the
underlying \emph{process} and is denoted by $\gamma^{}_{P}$.  This is
motivated by the following result on the autocorrelation
$\gamma_{P}^{(\varPhi)}$ of a given \emph{realisation}, which is
somewhat implicit in the literature. Its importance in our present
context was first emphasised by Gouer\'{e} in \cite{Gou1}; see also
\cite{LS} for complementary aspects.

\begin{theorem} \label{thm:palm} Let $\varPhi$ be a stationary,
  ergodic, positive random measure with distribution $P$. Assume that
  $P$ has finite density $\pd$, and that $P$ has locally finite second
  moments in the sense that $\EE_P ( \varPhi(A)^2 ) < \infty$ for any
  bounded $A \subset \mathbb{R}^d$ $($this follows for instance from
  the condition $\EE_P ( \varPhi(B_r(x))^2 ) < \infty$ for all
  $x\in\RR^{d}$ and some fixed radius $r)$.  Let $\varPhi_n :=
  \varPhi|_{B_n(0)}$ denote the restriction of\/ $\varPhi$ to the
  centred ball of radius $n$.  Then, the\/ \emph{natural
    autocorrelation} $\gamma_{P}^{(\varPhi)}$ of\/ $\varPhi$,
  defined via an averaging sequence of centred nested balls, almost
  surely exists and satisfies
\[ 
   \gamma_{P}^{(\varPhi)} \, := \,
   \lim_{n\to\infty} \, \frac{\varPhi_n \! * 
   \widetilde{\varPhi_n}} {\vol(B_n(0))} \; = \;
   \lim_{n\to\infty}\, \frac{\varPhi_n \! * 
   \widetilde{\varPhi}} {\vol(B_n(0))}
   \; = \; \pd\ts I^{}_{\! P^{}_0} \; = \; \gamma^{}_{P} \, ,
\] 
where the limit refers to the vague topology on $\cM^+$.  Here,
$I_{P^{}_0}$ is the first moment measure of the Palm distribution,
\[ 
    I_{P^{}_0}(A) \; = \; \int_{\cM^+}
   \varPsi(A) \dd P^{}_0(\varPsi), \qquad 
   \mbox{for } A \subset \mathbb{R}^d \mbox{ Borel}\ts . 
\] 
\end{theorem} 
\begin{proof}
As test function, fix a non-negative continuous
function $g\! :\; \mathbb{R}^d \to [0,\infty)$ with compact support. 
With $B_n^{\mathsf c} := \RR^d \backslash B_n$, we have
\begin{eqnarray*} 
  \lefteqn{ \frac{1}{\lambda(B_n)} \int_{\RR^d} 
    g(x) \dd \bigl( \varPhi_n *
    \widetilde{\varPhi_n} \bigr)(x) 
    \; = \; \frac{1}{\lambda(B_n)}
    \int_{B_n \times B_n} \!\! g(x-y) 
    \dd\varPhi(x)\dd \varPhi(y) } \\[1mm]
   & = & \frac{1}{\lambda(B_n)} \int_{B_n} 
    \biggl( \, \int_{\RR^d} g(x-y) \dd \varPhi(y)  \, -
    \int_{B_n^{\mathsf c}}   g(x-y) \dd \varPhi(y) \biggr) 
    \dd \varPhi (x)\\[1mm] 
    & = & \frac{1}{\lambda(B_n)}
    \int_{B_n} F_g \bigl( T_{-x} \varPhi \bigr) 
     \dd \varPhi (x) \; - \; R_n (g)
\end{eqnarray*}
(note that both integrals inside the big brackets in the second
line are finite because $g$ has compact support), where $\phi
\mapsto F_g (\phi) = \int_{\RR^d} g(-z) \dd \phi(z)$ defines a 
measurable function, and the remainder is given by
\[ 
  R_n (g) \; = \; \frac{1}{\lambda(B_n)} 
  \int_{B_n}\int_{B_n^{\mathsf c}} g(x-y) 
   \dd \varPhi (y) \dd \varPhi (x)\, .
\]
Note that $R_n$, which is a random measure, is precisely the
difference between the elements of the two approximating sequences of
random measures in the claim.  In view of \eqref{ergodic}, it thus
remains to show that $\lim_{n\to\infty} R_n = 0$ almost surely. Choose
$k$ so that $g(x) = 0$ for $|x| > k$, and fix some $\varepsilon > 0$.
We then have, for $n>k / \varepsilon$,
\[
  R_n (g) \; \le \;  \frac{\| g \|_\infty}{\lambda(B_n)} 
  \int_{B_n}  \varPhi\bigl(B_n^{\mathsf c} \cap  (x+ B_k)\bigr) 
  \dd \varPhi (x) \; \le \; \frac{\| g \|_\infty}{\lambda(B_n)} 
   \int_{B_n \backslash B_{ (1-\varepsilon) n}} 
   \!\!\!\!\! \varPhi (x+B_k) \dd \varPhi (x) \, ,
\]
where $\phi\mapsto G(\phi) := \phi(B_k)$ is again measurable. Hence we
obtain
\[
  \begin{split}
    R_n (g) & \, \le \, \frac{\| g \|_\infty}{\lambda(B_n)} 
    \int_{B_n}  G\bigl(T_{-x}\varPhi \bigr) \dd \varPhi (x)
     \, - \, \frac{\lambda (B_{(1-\varepsilon)n})}{\lambda(B_n)} 
     \frac{\| g \|_\infty}{\lambda (B_{(1-\varepsilon)n})} 
     \int_{B_{(1-\varepsilon)n}} G\bigl(T_{-y}\varPhi \bigr) 
      \dd \varPhi (y) \\[1mm]
     & \!\!\!\!\! \xrightarrow{n\to\infty} \,
     \big( 1 - (1-\varepsilon)^d \big)\, \| g \|_\infty \;
     \pd\! \int_{\cM^+} G(\varPsi)  \dd P^{}_0(\varPsi) 
     \, = \, \bigl( 1 - (1-\varepsilon)^d \bigr) \, \| g \|_\infty \;
     \pd \, I^{}_{\! P^{}_0}(B_k) \ts ,
  \end{split}
\]
almost surely by \eqref{ergodic}. 
Now take $\varepsilon\, {\scriptstyle \searrow}\, 0$ to conclude. 
\end{proof}

Continuing with the hypotheses of Theorem~\ref{thm:palm}, our
assumptions guarantee that the \emph{second moment measure}
$\mu^{(2)}$ of $P$, defined on cylinder sets $A \times A' \subset
\RR^d \times \RR^d$ via $\mu^{(2)}(A \times A') = \int_{\cN^{+}}
\varPhi(A)\, \varPhi(A') \dd P(\varPhi)$, is locally finite. This is a
necessary and sufficient condition for the existence of the first
moment measure of the Palm distribution (as a locally finite
measure). In fact, in the stationary scenario, the autocorrelation of
the process, denoted by $\gamma^{}_{P}$, satisfies $\gamma^{}_{P} =
\mu^{(2)}_\mathrm{red}$, where $\mu^{(2)}_\mathrm{red}$ is the
so-called \emph{reduced second moment measure} of $P$, and this, in
turn, is the same as the intensity of the Palm measure.  We offer a
brief explanation of this (for more details, see
\cite[Prop.~13.2.VI]{DVJ2} or \cite[Ch.~4.5]{SKM}).

First, $\mu^{(2)}_\mathrm{red}$ is obtained from $\mu^{(2)}$ by 
disintegration, via factoring out the translation invariance.
More precisely, following \cite{DVJ2}, $\mu^{(2)}_\mathrm{red}$ is the 
unique positive measure on $\mathbb{R}^d$ such that
\begin{equation} \label{reduced-moment-measure}
   \int_{\RR^d\times\RR^d} h(x,y) \dd \mu^{(2)} (x,y) \; = \; 
   \int_{\RR^d} \int_{\RR^d} h(u, u+v) \dd \mu^{(2)}_\mathrm{red} 
   (v) \dd \lambda(u) \, ,
\end{equation}
for all (real) functions $h\in C_{\mathsf c} (\RR^d\times\RR^d)$. 
When $h = f\otimes g$ is a product function (meaning that 
$h(x,y):=f(x) \ts g(y)$), one finds the relation
\begin{equation} \label{moment-connection}
   \mu^{(2)} (f\otimes g) \, = \, 
   \mu^{(2)}_\mathrm{red} (\tilde{f} * {g})
\end{equation}
via Fubini's theorem. Choosing $g=f$, it is clear
that the measure $\mu^{(2)}_\mathrm{red}$ is positive definite. More
generally, when dealing with complex-valued functions, one has to
consider
\[
   \mu^{(2)} (\bar{f}\otimes g) \, = \, 
   \mu^{(2)}_\mathrm{red} (\tilde{f} * {g})\ts ,
\]
which leads to some technical complications later on. Since we consider
real-valued component processes only, we can stick to the simpler
case of real-valued functions.

The connection of the reduced second moment to the intensity 
measure of the Palm measure comes through applying 
\eqref{PalmGeneral} to a function on $\RR^d\times\cM^+$
defined by
\begin{equation} \label{eq:specialFn}
     (x,\phi) \; \mapsto \;  g(x) \int_{\RR^d} T_x h(y) \dd \phi(y) \, ,
\end{equation}
where $g,h$ are arbitrary, but fixed, non-negative measurable 
functions on $\RR^d$. The left hand side of \eqref{PalmGeneral}
then reads
 \[
 \begin{split}
    \EE_{P} \left( \int_{\RR^d} g(x) \right. &
    \left. \int_{\RR^d}  h(y-x) 
    \dd \varPhi(y)  \dd \varPhi(x) \right)
     \; = \; \EE_{P} \left(\int_{\RR^d} \int_{\RR^d} 
     g(x) \ts h(y-x) \dd \varPhi(y) \dd \varPhi(x) \right) \\[1mm]
    & = \int_{\RR^d \times \RR^d} g(x) \ts h(y-x) \dd \mu^{(2)}(x , y)
    \; = \; \lambda (g) \cdot \mu^{(2)}_\mathrm{red} (h) \, ,
\end{split}
\]
where we employed Fubini's theorem and \eqref{reduced-moment-measure},
while the right hand side reads
\begin{eqnarray*}
       \lefteqn{\pd \int_{\RR^d}  \int_{\cM^+} g(x) \int_{\RR^d} 
       (T_x h)(y) \dd (T_x \phi) (y) \dd P_0 (\phi) 
        \dd \lambda (x) } \\[1mm]
       &=& \pd \int_{\RR^d} \int_{\cM^+} g(x) \int_{\RR^d}  
       h(y) \dd \phi(y) \dd P_0  (\phi)  \dd \lambda (x)
       \; = \; \lambda (g) \cdot \pd I_{P_0} (h)  \, .
 \end{eqnarray*}
Here, we used the notation of the intensity of the Palm measure for its
first moment.  Comparing these two calculations gives  
\begin{equation} \label{eq:moments-versus-Palm}
    \mu^{(2)}_\mathrm{red} \, = \, \pd I_{P_0} \, = \, \gamma^{}_{P} \, . 
\end{equation}

\begin{remark} \textsc{Equivalent definitions of $\mu^{(2)}_{\mathrm{red}}$}.
\label{disintegration-comment} There are several
  different ways to define a reduced measure via disintegration. In
  particular, one could employ $h(u,u\pm v)$ as well as $h(u\pm v,u)$ in
  Eq.~\eqref{reduced-moment-measure}. Using translation invariance of
  Lebesgue measure, this boils down to just two different
  possibilities, the one with $h(u,u+v)$ introduced above and the one
  with $h(u,u-v)$, which is used in \cite[Prop.~I.60]{Karr}. Observing
  the relation
\[
    \widetilde{f * \tilde{g}} \, = \, \tilde{f} * g  
\]
together with $\widetilde{\mu^{(2)}}=\mu^{(2)}$,
one can check that both versions define the same measure,
as the process is restricted to positive (and thus real) random
measures, so that no complex conjugation shows up in the
$\,\widetilde{.}\,$-operation. Alternatively, one can use commutativity
of the convolution together with the symmetry of $\mu^{(2)}$,
which implies $\mu^{(2)} (f\otimes g) = \mu^{(2)} (g\otimes f)$.
\exend \end{remark}

\begin{remark} \textsc{Renewal process revisited.} 
  \label{renewal-revisited}
  Consider a stationary renewal process $\varPhi$ on the real line,
  viewed as a random counting measure, with inter-arrival law
  $\varrho$. The latter is assumed to be a probability measure on
  $\RR_{+}$, with expectation $\int_{\RR_{+}} x \dd\varrho (x) =
  1$. It is well known (compare \cite[Thm.~13.3.I and
  Ex.~13.3(a)]{DVJ2}) that the Palm distribution $P^{}_{0}$, with
  respect to the origin, of (the law of) $\varPhi$ equals the law of
\begin{equation} \label{Palm-renewal}
     \delta^{}_{0} \, + \sum_{i\in\ZZ} \delta^{}_{S_i} \ts ,
\end{equation}   
   where $S^{}_{0}=0$,
\[
      S_i = \begin{cases}
             X^{}_{1} + \ldots + X^{}_{i}  \ts , & \text{if $i \ge 1$}, \\
             X^{}_{i+1} + \ldots + X^{}_{0} \ts , & \text{if $i \le -1$},
         \end{cases}
\]   
and $(X_i)^{}_{i\in\ZZ}$ are i.i.d.\ with law $\varrho$. We see
immediately from \eqref{Palm-renewal} that the first moment measure
$I^{}_{P^{}_{0}}$ of $P^{}_{0}$ is given by formula ~\eqref{auto-1},
see also Proposition~\ref{renewal-1}, and thus recover
Theorem~\ref{non-lattice-gives-ac} as well as Remark~\ref{renewal-ext}
by specialising Theorem~\ref{thm:palm} to the case of a renewal
process on the line.

With the interpretation as a random counting measure, we are no longer
restricted to laws $\varrho$ on $\RR_{+}$. Indeed, when $\varrho$ is
any probability measure on $\RR$ with expectation $1$ (which prevents
the process from being recurrent), the random counting measure almost
surely leads to the autocorrelation given in \eqref{auto-1}, and thus
avoids the complications mentioned after Lemma~\ref{convergence}; see
also \cite[Ch.~XI.9]{Feller} for a systematic exposition, and
\cite[Ex.~8.2(b)]{DVJ1} for comparison.  \exend
 \end{remark}  

\begin{remark} \textsc{Bartlett spectrum.} \label{rem:Bartlett} The
  diffraction measure $\widehat{\gamma}$ of a stationary random
  measure $\varPhi$ (with $\EE \bigl( \vert\varPhi(A)\rvert^{2}
  \bigr)<\infty$ for all compact $A\subset\RR^d$, say) is closely
  related to the so-called \emph{Bartlett spectrum} 
  $\varGamma := (c^{(2)}_{\mathrm{red}})^{\widehat{\hphantom{w}}}$ of
  $\varPhi$ as follows; compare \cite[Ch.~8.2]{DVJ1}.

  Recall that the \emph{covariance measure} $c^{(2)}$ is defined on
  cylinder sets via
\[
  \begin{split}
   c^{(2)}(A \times A') \, & = 
     \int \varPhi(A)\, \varPhi(A') \dd P(\varPhi)  \,
   - \int \varPhi(A) \dd P(\varPhi) \int \varPhi(A') 
     \dd P(\varPhi)  \\[1mm]
   & = \, \mu^{(2)}(A \times A') - 
     \rho^2 \bigl(\lambda \otimes \lambda\bigr) 
   (A \times A') \ts ,
   \end{split}
\]
where $\rho$ is the density of $\varPhi$, compare
\cite[Eq.~(9.5.12)]{DVJ2}. Consequently, the relation between the
reduced second moment measure $\mu^{(2)}_{\mathrm{red}}$ and the
reduced covariance measure $c^{(2)}_{\mathrm{red}}$ of $\varPhi$ is
\begin{equation} \label{eq:c2mu2related}
     c^{(2)}_{\mathrm{red}} = 
     \mu^{(2)}_{\mathrm{red}} - \rho^2 \lambda \ts . 
\end{equation}
Since $\widehat{\gamma}$ is the Fourier transform of
$\mu^{(2)}_{\mathrm{red}}$ and $\varGamma$ that of
$c^{(2)}_{\mathrm{red}}$, Eq.~\eqref{eq:c2mu2related} translates into
\begin{equation} \label{eq:Bartlett-diffraction-relation}
   \varGamma = \widehat{\gamma} - \rho^2 \delta^{}_{0} \ts . 
 \end{equation}
 {}From our perspective, the positive definite autocorrelation measure
 $\gamma$ is a slightly more natural and universal object than the
 inverse Fourier transform $\widecheck{\varGamma}$
 of $\varGamma$, since $\widehat{\gamma}$
 corresponds to a physically observable quantity, namely diffraction.
 Eq.~\eqref{eq:Bartlett-diffraction-relation} gives
 $\widecheck{\varGamma}=\gamma - \rho^{2} \lambda$,
 which is no longer positive definite. Rather, it is tailored to
 situations where $0$ supports the only atom of $\widehat{\gamma}$, as
 in the homogeneous Poisson process; compare Example~\ref{ex-poisson}
 below and the discussion in \cite[p.~305]{DVJ1}, and see
 \cite[Sec.~8.2]{DVJ1} for further explicitly computable examples.
 \exend
\end{remark}

To formulate the standard Poisson process in this setting, let us
start with an intuitive picture.  Imagine independently putting single
points on the sites of $\varepsilon \ZZ^d \subset \RR^d$, each with
probability $\pd\ts \varepsilon^d$, and imagine a process that arises
from this construction in the limit $\varepsilon \to 0$.  For a
rigorous construction, one can start from a tiling of $\mathbb{R}^d$
with translates of $[0,1)^d$ and then proceed, independently for each
cell, as follows:\ Put a Poisson-($\pd$) distributed number of points
in the cell, with their locations independently and uniformly
distributed over the given cell, see \cite[Sec.~2.4.1]{SKM} for
details. Such a more elaborate approach is needed when $d>1$,
as there is no analogue of the renewal process we used for $d=1$.

\begin{example} \textsc{Homogeneous Poisson process}.
\label{ex-poisson}
This process on $\mathbb{R}^d$, with (point) density $\pd$ (compare
Remark~\ref{rem-one}), is a random counting measure $\varPhi$ (with
distribution $P$) such that $\varPhi(A)$ is Poisson-($\pd\ts
\lambda(A)$)-distributed for any measurable $A\subset \mathbb{R}^d$
and that the random variables $\varPhi(A_1), \ldots, \varPhi(A_m)$ are
independent for any collection of pairwise disjoint $A_1, \ldots, A_m
\subset \mathbb{R}^d$.  With this setting, the expectation measure of
the process is given by $\EE_P^{} (\varPhi) = \pd\ts \lambda$.

It is well-known that, under the Palm distribution, a Poisson process
looks like the same Poisson process augmented by an additional point 
at $0$, so that
\begin{equation}  \label{eq:PalmPoisson}
   P^{}_0(\cB) \; = \; \int \mathbf{1}_\cB(\varPhi+\delta_0) 
   \dd P(\varPhi), 
   \qquad \mbox{for } \cB \subset \cN
\end{equation}
(alternatively, write $\mathcal{L} (\varPhi+\delta_0)=P^{}_0$, or $P
* \delta_{\delta_0} = P^{}_0$), by a theorem of Slivnyak, compare
\cite[Ex.~4.3]{SKM}.  This is intuitively obvious from the
approximation via independent coin flips on $\varepsilon\ts \mathbb{Z}^d$
and the idea of obtaining the Palm distribution via conditioning on
the presence of a point at $0$.  Here, this gives
in $I_{P^{}_0} = \delta_0 + I_P = \delta^{}_0 + \pd\ts \lambda$.  
Since homogeneous Poisson processes are stationary and 
ergodic for the translation action of $\RR^d$, we can 
now apply Theorem~\ref{thm:palm}.

Consequently, for almost all realisations $\varPhi$ of a homogeneous
Poisson process of density $\pd$, the autocorrelation measure
and the diffraction measure are given by
\begin{equation} \label{poisson-formulas} 
   \gamma^{}_{P} \,  = \, \pd \, \delta_0 + 
   \pd^2 \ts \lambda \qquad \mbox{and} \qquad
   \widehat{\gamma}^{}_{P} \, = \, \pd^2 \, \delta_0 + 
    \pd \, \lambda\ts ,
\end{equation}
by an application of Eq.~\eqref{poisson-lebesgue}.
This also extends Example~$\ref{poisson-on-the-line}$ to arbitrary
finite values of the point density $\pd$; compare also
\cite[Ex.~8.2(a)]{DVJ1}.
\exend \end{example}

\begin{remark} \textsc{Mat\'{e}rn's hard-core process}.  
  One drawback of the Poisson process (of point density $\rho > 0$ in
  $\RR^{d}$, with $d\ge 2$ say) for applications in physics is the
  missing uniform discreteness. The latter was introduced by
  Mat\'{e}rn through a hard-core condition, realised via a local
  thinning process applied to each realisation, see \cite{Sto} and
  references therein.  Informally, for some fixed radius $R>0$, his
  construction works as follows. Each point of a realisation of a
  Poisson process is equipped with an independent mark that is drawn
  uniformly at random from $[0,1]$ (technically, this is a marked
  Poisson process). Then, only those points $x$ are kept for which
  there exists no point with a smaller mark in $B_R(x)$.  The
  autocorrelation of the modified process is still radially
  symmetric. Moreover, if $R$ is the radius of the hard-core condition
  and $B^{}_{R} = B^{}_{R} (0)$ as before, the autocorrelation for
  distances $r\ge 2R$ is that of a homogeneous stationary Poisson
  process with a new, effective point density
\[
   \rho^{}_{\mathrm{eff}} = \frac{1- e^{-\rho\, \vol (B^{}_{R})}}
             {\vol (B^{}_{R})}\ts . 
\]
In fact, the new autocorrelation almost surely has the form $\gamma =
\rho^{}_{\mathrm{eff}} \ts \delta^{}_{0} + \rho^{2}_{\mathrm{eff}}\ts
\lambda - \nu$, where $\nu$ is a radially symmetric measure with
support $\overline{B^{}_{2R}}$, as follows from \cite[Thm.~1]{Sto}. In
fact, $\nu$ is absolutely continuous with density
$\rho^{2}_{\mathrm{eff}}1^{}_{B^{}_{R} } - \, g$, where $g$ is a
smooth function on $B^{}_{2R} \setminus \overline{B^{}_{R}} $. This
density has a jump (with sign change) at $\lvert x \rvert = R$, but
tends smoothly to $0$ as $\lvert x \rvert \nearrow 2R$.

The diffraction of (this version of) the Mat\'{e}rn model is thus given by
\[
    \widehat{\gamma} = \rho^{2}_{\mathrm{eff}}\ts \delta^{}_{0}
   + \bigl( \rho^{}_{\mathrm{eff}} - h \bigr)\ts \lambda 
   \qquad \text{(a.s)} ,
\]
where $h$ is a smooth (even analytic) function with $h(k) =
\mathcal{O} \bigl(\lvert k \rvert^{-(d+1)/2}\bigr)$ as 
$\lvert k \rvert \to \pm \infty$. Indeed, one has 
\[
    \widehat{1^{}_{B^{}_{R} }} (k) = \Bigl( 
    \frac{R}{\lvert k \rvert} \Bigr)^{d/2}
    J^{}_{d/2} (2\pi\ts \lvert k \rvert \ts R )\ts ,
\]
which is responsible for the above estimate via the exact
asymptotic behaviour of the Bessel function $J^{}_{d/2}$
for large arguments. The remaining contribution, using  the
explicit expression of \cite[Eq.~(3.2)]{Sto} and the reduction
of the radially symmetric Fourier transform to a one-dimensional
Hankel transform, gives another term. It can also be computed 
explicitly, though the resulting formula is too lengthy to write it 
down here. Its decay is as for the previous term, 
by an application of the (refined) Paley-Wiener theorem.
\exend \end{remark}

At this point, let us recall from \cite{BM1} one process that is of
particular interest in the study of aperiodic solids. Unlike the
Poisson process and most of its siblings, it shows a substantial
amount of point spectrum. It is related to Remark~\ref{Lambda-gas},
but based on the cut and project
formalism, for which we refer the reader to \cite{Moody}.

\begin{example} \textsc{Model set based particle gas}.
\label{particle-gas}
Let $\gL\subset \RR^d$ be a regular model set, for simplicity
with internal space $\RR^m$. Let $\cL$ be the lattice in
$\RR^d \times \RR^m$ that is needed for the cut and project
scheme, with projection image $L$ in $\RR^d$. We denote the
corresponding star map (from $L$ into $\RR^m$) by $\star$, 
so that (up to a translation)
\[
     \gL = \{ x\in L \mid x^{\star} \in W \} ,
\]
where $W\subset \RR^m$ is the window of $\gL$. The latter is
assumed to be a compact set with non-empty interior and a
boundary of measure $0$. This guarantees $\gL$ to be a Meyer
set. Let  $f$ now be a continuous function  on $W$,
and consider the weighted Dirac comb
\[
     \omega \, = \sum_{x\in\gL} f(x^{\star})\, \delta_x \ts ,
\]
which is pure point diffractive by the model set theorem
\cite{Hof,Martin2,BM1} with diffraction measure
\[
     \widehat{\gamma_{\omega}} \, = \sum_{k\in M}
     \big\lvert \widehat{f} (-k^{\star}) \big\rvert^2 \, \delta^{}_{k} \ts .
\]     
Here, $M$ is the projection of the dual lattice $\cL^*$ into $\RR^d$,
on which the star map is well-defined, too. It is known as the
\emph{Fourier module} of the model set $\gL$.

Assume now that $0 \le f(y) \le 1$ on $W$, and define a family of
independent binary random  variables $\bigl( U_x\bigr)_{x\in\gL}$, each
taking values $0$ or $1$ with $\PP \bigl( \{ U_x = 1 \} \bigr) = f (x^{\star})$.
The stochastic counterpart $\omega^{}_{\mathrm{s}}$ of $\omega$ is
\[
     \omega^{}_{\mathrm{s}} \, = \sum_{x\in\gL} U_x \, \delta_x \ts ,
\]
which can be interpreted as a particle gas on $\gL$. By 
\cite[Thm.~2 and Eq.~(58)]{BM1},  it almost surely has diffraction
\[
     \widehat{\,\gamma_{\omega^{}_{\mathrm{s}}}\!} \, = \, 
     \widehat{\gamma_{\omega}} + 
     \dens (\gL)\, \overline{V}\, \lambda \ts .
\]
Here, $\overline{V} = \frac{1}{\vol (W)}\int_{W} f(y)
\bigl(1-f(y)\bigr) \dd y$ is the mean variance of the random
variables, averaged over $\gL$, which is a consequence of the uniform
distribution result for model sets \cite{Martin1,Bob}. One can also
calculate the entropy of this system \cite{BM1}.  Moreover, it is not
difficult to restrict the process to produce Meyer sets -- one simply
has to choose a function $f$ that is $1$ on a subset of $W$ with
non-empty interior.  \exend
\end{example}

\subsection{Compound processes} 
Let us now go one step further by adding random clusters to the
picture. To this end, let a stationary, ergodic, point process $\varPhi$
be given, with law $P$, density $\pd$, and locally finite expectation
measure $\EE_P (\varPhi)$.  This is called the \emph{centre process}
from now on.  Moreover, let $\varPsi\in\cM^{+}_{\mathsf{bd}}$ be a
positive random measure with law $Q$, subject to the condition that
both its expected total mass, $m := \EE_Q \bigl( \varPsi(\mathbb{R}^d)
\bigr) > 0$, and the second moment of the total mass, $\EE_Q \bigl(
\bigl(\varPsi(\RR^d)\bigr)^2\bigr)$, are finite. This is the
\emph{component process}. We will also consider signed component
processes $\varPsi$ with values in $\cM^{}_{\mathsf{bd}}$, in which
case we assume that the second moment of the total variation
measure is finite; see the appendix for some details on the required
notions and modifications for signed measures.

A combined cluster process, or \emph{cluster process} for short, is a
combination of a centre process and a component process of cluster
type, and is obtained by replacing each point $x \in \supp (\varPhi)$
by an independent copy of $\varPsi$, translated to that point $x$.  We
denote such a process by the pair $(\varPhi_P,\varPsi_Q)$. As
before, we restrict ourselves to finite clusters here.  Formally, let
$\varPsi_1, \varPsi_2, \dots$ be independent copies of $\varPsi$
(these are the individual \emph{clusters}). When $\varPhi = \sum_i
\delta^{}_{X_i}$, we put
\[ 
    \varPhi_{\sf cl} \; := \; \sum_i T^{}_{X_i} \varPsi_i
    \; = \; \sum_i \delta^{}_{X_i}\! * \varPsi_i \, ,
\] 
and denote the resulting law by $P_{\sf cl}=R$.  Note that, when 
$\varPsi \equiv \delta_0$ is deterministic and concentrated to one 
point, we simply obtain $\mathcal{L} (\varPhi_P,\varPsi_Q) =
\mathcal{L} (\varPhi)$, and the cluster process coincides with
the centre process.

If $\varPsi$ is a counting measure, the cluster process
$(\varPhi_P,\varPsi_Q)$ is again stationary and ergodic,
and its expected density is given by $m \pd$, by 
\cite[Prop.~12.3.IX]{DVJ2}. This property actually holds in 
larger generality, which we need later on.

\begin{prop} \label{ergodic-compound} 
  Let $\varPhi$ be a stationary and ergodic point process with law
  $P$, finite density $\pd$ and locally finite second moments.  Let
  $\varPsi$ be a $(\nts$signed\/$)$ random measure with law $Q$, finite
  mean and finite second moment. Then, the combined cluster process,
  which is a random measure, is again ergodic.
\end{prop}    
\begin{proof}[Sketch of proof]
  If the component process is a (positive) point process as well, this
  result is stated and proved in \cite[Sec.~12.3]{DVJ2}. The necessary
  modifications for an extension to a (possibly signed) random measure
  as component process, which seem to be well-known but which we could
  not explicitly trace in the literature, are provided in the
  appendix.
\end{proof}

The second moment measures of the three processes are connected 
in a way that permits an explicit calculation of the autocorrelation 
$\gamma^{}_{R}$ in terms of $\gamma^{}_{P}$ and various
expectation measures of the component process with law $Q$. 
To employ this powerful connection, we recall another
disintegration formula, this time for any random variable $\varXi$
of the cluster process,
\begin{equation} \label{conditional}
    \EE_{R} (\varXi\ts ) \, = \, \EE_{R} \bigl( \EE_{R}
    (\varXi \mid \text{given the centres}) \bigr)
    \, = \, \EE_{P} \bigl( \EE_{Q}
    (\varXi \mid \text{given the centres}) \bigr) ,
\end{equation}    
which (with obvious meaning) follows from the standard theorems on
conditional expectation.

We are now in the position to use Eq.~\eqref{reduced-moment-measure}
in conjunction with Theorem~\ref{thm:palm} and Eq.~\eqref{conditional}
to calculate $\mu^{(2)}_{\sf cl} = \mu^{(2)}_{R}$, and thus the
autocorrelation of almost all realisations of the cluster process,
where we first concentrate on positive random measures. The extension
to signed measures follows in Section~\ref{extension-signed}. We first
need some technical results.


\begin{lemma} \label{finite-times-lebesgue}
   Let $\lambda$ be Lebesgue measure on\/ $\RR^d$, as before, and 
   $\mu$ a finite Borel measure. Then, one has
   $\mu * \lambda = c\ts\lambda$ with $c=\mu(\RR^d)$.
\end{lemma}
\begin{proof}
Let $g$ be an arbitrary continuous function on $\RR^{d}$, with
compact support. For all $x\in\RR^d$, we have $\lambda (T_{-x} \ts g) 
= (T_x\ts\lambda) (g) = \lambda (g)$ due to translation invariance of 
$\lambda$. The convolution $\mu*\lambda$ is well-defined as $\mu$
is finite while $\lambda$ is translation bounded \cite[Prop.~1.13]{BF}. 
One thus has
\[
  \begin{split}
   \bigl(\mu*\lambda\bigr) (g) & \, = \,
   \int_{\RR^d\times\ts\RR^d} g(x+y)  \dd\lambda (y) \dd\mu (x)
   \, = \, \int_{\RR^d} \lambda (T_{-x} \ts g) \dd\mu (x) \\[1mm]
   & \, = \, \int_{\RR^d} \lambda (g) \dd\mu (x)
   \, = \, \mu(\RR^d)\,\lambda(g) \/ .
  \end{split}
\]
Since $g$ was arbitrary, the claim follows.
\end{proof}


Given a measure $\mu\in\cM^+$ and a continuous function $g$ on 
$\RR^d$ with compact support (possibly complex-valued), we define 
a new function $g_{\mu}$ on $\RR^d$ via
\begin{equation} \label{shorthand}
     g_{\mu} (x) \; := \; (T_x \mu) (g) \ts ,
\end{equation}
which is certainly measurable. It is easy to check that $g_{\mu}$ satisfies
\begin{equation} \label{double-twiddle}
    \widetilde{g_{\mu}} \; = \; \tilde{g}_{\tilde{\mu}} \ts .
\end{equation}

\begin{lemma}\label{convolution-trick}
  Let $\mu\in\cM_{\mathsf{bd}}^{+}$ and let $\gamma$ be a positive,
  translation bounded measure on $\RR^d$. For arbitrary 
  $($possibly complex-valued\/$)$ $f,g \in
  C_{\mathsf{c}} (\RR^{d})$, one has the identity
\[
     \bigl(\mu * \tilde{\mu} * \gamma \bigr) (f \nts * \tilde{g})
     \; = \; \gamma (f_{\mu}\nts * \widetilde{g_{\mu}}) \, .
\]
   This identity also holds when both $\mu$ and $\gamma$ are
   signed measures.
\end{lemma}
\begin{proof}
  Let $f$ and $g$ be $\mu$-measurable functions such that $f \nts
  *\tilde{g}$ is a continuous function with compact support, which
  includes the situation of the claim. One then finds, with Fubini,
  that
\begin{eqnarray*}
    \bigl( \mu * \tilde{\mu} * \gamma \bigr) ( f*\tilde{g})  & = &
    \int \int \Bigl( \,\int f(x+z+\xi) \dd \mu (x) \Bigr)
       \Bigl( \,\int \tilde{g} (y-\xi) \dd\tilde{\mu} (y)  \Bigr)
     \dd \lambda (\xi) \dd \gamma (z) \\
     & = & \int \int \bigl( T_{z+\xi} \mu \bigr) (f) \,
     \bigl( T_{-\xi} \tilde{\mu}\bigr) (\tilde{g}) \dd 
        \lambda (\xi) \dd \gamma (z) \\     
     & = & \int \int f_{\mu} (z+\xi) \, 
            \tilde{g}_{\tilde{\mu}} (-\xi)
            \dd \lambda (\xi) \dd \gamma (z)  \; = \;
        \gamma (f_{\mu} * \widetilde{g_{\mu}} )\, ,         
\end{eqnarray*}
where all integrals are over $\RR^d$ and \eqref{double-twiddle}
was used in the last step.
\end{proof}           

Specialising Lemma~\ref{convolution-trick} to $\gamma = \delta_{0}$
gives the relation
\begin{equation} \label{derived-trick}
     \bigl( \mu * \tilde{\mu} \bigr) (f \nts *\tilde{g}) \, = \,
     \bigl( f_{\mu} \nts * \widetilde{g_{\mu}} \bigr) (0) \, = \,
     \lambda (f_{\mu} \ts \, \overline{\! g_{\mu}\!} \, ) \ts ,
\end{equation}
which simplifies our further discussion. 


\begin{remark} \textsc{Test functions for measures}. \label{rem:tfm}
  Recall that two measures $\mu,\nu \in \cM (\RR^{d})$ are equal when
  $\mu(h)=\nu(h)$ for all $h\in C_\mathsf{c} (\RR^{d})$.  When the
  measures are positive or signed (but not complex), real-valued
  functions suffice. In the latter case, it will be particularly
  helpful to restrict to functions of the form $h = f \nts *g$ with
  $f,g\in C_\mathsf{c} (\RR^{d})$. Since the space $C_\mathsf{c}
  (\RR^{d})$ contains an approximate unit for convolution, the linear
  combinations of such functions are dense in $C_\mathsf{c}
  (\RR^{d})$, so that they suffice to assess equality of measures.
  \exend \end{remark}

\begin{lemma} \label{help-reduce}
  Under our general assumptions on the component process, one has
\[
  \begin{split}
  \bigl(\EE_{Q}(\varPsi\! * \widetilde{\varPsi} \ts) \bigr)  
  (f \nts * \tilde{g} ) & \; = \;
  \lambda \bigl(\EE_{Q} ( f^{}_{\varPsi} \, 
  \,\overline{\! g^{}_{\ts\varPsi}\!}\,) \bigr) 
  \quad \text{and} \\[1mm]
  \bigl( \EE_{Q} (\varPsi) \! * \widetilde{\EE_{Q} (\varPsi)} \bigr) 
  (f \nts * \tilde{g} ) & \; = \;
  \lambda \bigl( f^{}_{\EE_{Q} (\varPsi)} \, 
    \,\overline{\! g^{}_{\ts \EE_{Q} (\varPsi)}\!}\,\bigr) ,
  \end{split}
\]    
where $f,g \in C_{\mathsf{c}} (\RR^d)$,  possibly complex-valued.
\end{lemma}
\begin{proof}
  Let $f$ and $g$ be chosen as in the previous proof, with
  complex-valued functions permitted. For the first claim, observing
  that each realisation of $\varPsi$ is a finite measure, a direct
  calculation with Eq.~\eqref{derived-trick} gives
\[
   \bigl( \EE_{Q} ( \varPsi \! * \widetilde{\varPsi} \ts ) \bigr) 
   (f \nts *\tilde{g}) \, = \,
   \EE_{Q} \bigl(\bigl(\varPsi \! * \widetilde{\varPsi} \ts \bigr) 
   (f \nts *\tilde{g})\bigr) \\
      \, = \, \EE_{Q} \bigl( \lambda ( f^{}_{\varPsi} \, 
            \, \overline{\! g^{}_{\varPsi}\!}\, ) \bigr)
     \, = \, \lambda \bigl(\EE_{Q} ( f^{}_{\varPsi} \, 
        \,\overline{\! g^{}_{\ts\varPsi}\!}\, ) \bigr) ,  
\]
while the second identity simply is Eq.~\eqref{derived-trick} with
$\mu = \EE_{Q} (\varPsi)$, which is a finite measure by assumption.
\end{proof}


Recall that the \emph{covariance} of two real-valued random variables
$X$ and $Y$ related to the law $Q$ (using our general notation as
explained above) is defined as
\begin{equation} \label{def-covar}
    \mathrm{cov}^{}_{\! Q} \ts (X,Y) \; := \;
    \EE_{Q} (X\, Y) - \EE_{Q} (X) \, \EE_{Q} (Y)\, ,
\end{equation}    
where the index $Q$ highlights the reference to the underlying
law $Q$.

\begin{prop} \label{reduction} 
    Let $(\varXi, R)$ be a combined cluster process with stationary
  centre point process $(\varPhi, P)$ and real component process
  $(\varPsi, Q)$, both with the usual assumptions on means and second
  moments as used above.  Then, $\varXi$ is locally square integrable
  in the sense that $\EE_R\big[ ( \varXi(B))^2 \big] < \infty$ for any
  bounded Borel set $B$, and we have the reduction formula
\[
    \mu^{(2)}_{R} (f\otimes g) \; = \;
    \mu^{(2)}_{P} \bigl( f^{}_{\EE_{Q} (\varPsi)} 
     \otimes g^{}_{\ts\EE_{Q} (\varPsi)} \bigr)
    + \pd\, \lambda \bigl( \mathrm{cov}^{}_{\! Q} 
    (f^{}_{\varPsi}, g^{}_{\ts\varPsi}) \bigr) ,
\]
   where $\pd $ is the density of the centre process,
   $f,g$ are continuous with compact support, and the
   covariance is defined as in $\eqref{def-covar}$.
\end{prop}
\begin{proof}
  In order to check that $\EE_R\big[( \varXi(B))^2 \big] < \infty$ for
  bounded, Borel measurable $B \subset \RR^d$, one can trace through
  the steps below, replacing $f$ and $g$ by $\mathbf{1}^{}_{B}$ (the
  corresponding integrals then involve only positive terms and are
  finite by Fubini's theorem). In general, by assumption and the 
  disintegration formula \eqref{conditional}, one finds
\[
   \begin{split}
       \mu^{(2)}_{R} (f\nts \otimes g) & \, = \,
       \int_{\cM^+} \! \varXi (f) \, \varXi (g) \dd R (\varXi)  \\[1mm]
       & \, = \, \int_{\cN^+}  \EE_{Q} \Bigl( \sum_{x,y \in \supp (\varPhi)}
       \varPsi_x (T_{-x} f) \, \varPsi_y (T_{-y} g) \Bigr) \dd P (\varPhi)\, ,
   \end{split}
\]
where $\varPsi_x$ denotes the random measure at centre $x$. Since
$\varPsi_x$ and $\varPsi_y$ are independent for $x\neq y$, the double
sum over the support is split into a sum over the diagonal ($x=y$) and
a sum over all remaining terms ($x\neq y$). Using the linearity of the
expectation operator, the integrand can now be rewritten as a sum over
two contributions, namely
\[
     \sum_{x,y} \EE_{Q} \bigl( \varPsi (T_{-x} f) \bigr)\,
         \EE_{Q} \bigl( \varPsi (T_{-y} g) \bigr)  
         \qquad \text{and}
\]         
\[                
     \sum_{x} \Bigl( \EE_{Q} \bigl( \varPsi (T_{-x} f) \, 
        \varPsi (T_{-x} g)\bigr)
       - \EE_{Q}     \bigl( \varPsi (T_{-x} f) \bigr)\,
                         \EE_{Q} \bigl( \varPsi (T_{-x} g) \bigr) \Bigr) .  
\]
Inserting the first term into the previous calculation leads to the
contribution
\[
      \mu^{(2)}_{P} \bigl( \EE_{Q} (f^{}_{\varPsi}) \otimes 
      \EE_{Q} (g^{}_{\ts\varPsi}) \bigr)
      \; = \; \mu^{(2)}_{P} \bigl( f^{}_{\EE_{Q} (\varPsi)} 
      \otimes g^{}_{\ts\EE_{Q} (\varPsi)} \bigr),
\]
while the second results in
\[
    \EE_{P} (\varPhi) \bigl( \mathrm{cov}^{}_{\! Q} 
     (f^{}_{\varPsi}, g^{}_{\ts\varPsi}) \bigr)
    \; = \; \pd\,\lambda \bigl( \mathrm{cov}^{}_{\! Q} 
     (f^{}_{\varPsi}, g^{}_{\ts\varPsi} ) \bigr) ,
\]
where the last step follows from the stationarity of $(\varPhi, P)$.
\end{proof}   


\begin{theorem} \label{compound-thm}
  Let $\varPhi$ be a stationary and ergodic point process with law
  $P$, finite density $\pd$ and locally finite second moments.
  Let $\varPsi$ be a random measure with law $Q$, finite expectation
  measure and finite second moments. If $(\varXi, R)$ denotes the
  combined cluster process built from the centre process $(\varPhi,
  P)$ and the component process $(\varPsi, Q)$, it is also stationary
  and ergodic.
    
    Moreover, the autocorrelation of the combined process satisfies
\[
     \gamma^{}_{R} \; = \;
     \bigl( \EE_{Q} (\varPsi) * \widetilde{\EE_{Q} (\varPsi)} \bigr) *
     \gamma^{}_{P} + \pd \, \bigl(
     \EE_{Q} (\varPsi \! * \widetilde{\varPsi}) -
     \EE_{Q} (\varPsi) * \widetilde{\EE_{Q} (\varPsi)} \bigr),
\]
    and this is almost surely the natural autocorrelation of a given 
    realisation of the cluster process.
\end{theorem}
\begin{proof}
  Choose two measurable functions $f$ and $g$ such that $f \nts
  *\tilde{g} \in C_{\mathsf{c}} (\RR^{d})$. Then, in
  line with Remark~\ref{rem:tfm} and Eq.~\eqref{moment-connection},
  one finds via Proposition~\ref{reduction} that
\[
  \begin{split}
   \gamma^{}_{R}  (f \nts *\tilde{g}) \, & = \, 
   \mu^{(2)}_{R} (f\nts \otimes g) \; = \;
   \mu^{(2)}_{P} \bigl( f^{}_{\EE_{Q} (\varPsi)} \nts
   \otimes g^{}_{\ts\EE_{Q} (\varPsi)} \bigr)
     +   \pd\,\lambda\bigl( \mathrm{cov}^{}_{\nts Q} 
   (f^{}_{\varPsi}, g^{}_{\ts\varPsi})\bigr) \\
  & = \, \gamma^{}_{P} \bigl( f^{}_{\EE_{Q} 
   (\varPsi)}\nts  * \widetilde{g^{}_{\ts\EE_{Q} (\varPsi)} } 
            \bigr) + \pd\, \bigl( \EE_{Q} 
    (\varPsi\! * \widetilde{\varPsi}) -
            \EE_{Q} (\varPsi)\nts * \widetilde{\EE_{Q} 
    (\varPsi)} \bigr) (f*\tilde{g}) \, ,
  \end{split}
\]
where $\EE_{P} (\varPhi) = \pd \ts\lambda$ due to stationarity of
$(\varPhi, P)$.  The second step makes use of Lemma~\ref{help-reduce}.

The formula for the autocorrelation now follows from
the observation that
\[
    \gamma^{}_{P} \bigl( f^{}_{\EE_{Q} (\varPsi)}\nts * 
       \widetilde{g^{}_{\ts\EE_{Q} (\varPsi)} } \bigr) \; = \;
       \bigl( \EE_{Q} (\varPsi)\nts *\widetilde{\EE_{Q} (\varPsi)}
       \nts * \gamma^{}_{P} \bigr) (f*\tilde{g})\ts ,
\]
which is an application of Lemma~\ref{convolution-trick}.
The remaining claims are clear due to the assumed ergodicity, via an
application of Proposition~\ref{ergodic-compound}.
\end{proof}        

An application of the convolution theorem gives the following
consequence, where also the identity $\widehat{\EE_{Q} (\varPsi)} =
\EE_{Q} (\widehat{\varPsi})$ was used to highlight the structure of
the result.
\begin{coro} \label{compound-diffraction}
   Under the assumption of Theorem~$\ref{compound-thm}$, the diffraction
   measure of the combined cluster process  is given by
\[
   \widehat{\gamma}^{}_{R} \; = \; 
   \big\lvert \EE_{Q} (\widehat{\varPsi}) \big\rvert^2 \cdot
   \widehat{\gamma}^{}_{P} + \pd\, \bigl(
   \EE_{Q} ( \lvert \widehat{\varPsi} \rvert^2 ) -
   \lvert \EE_{Q} (\widehat{\varPsi}) \rvert^2 \bigl) \, \lambda
\]
    which is then almost surely also the diffraction measure of a given
    realisation.
    \qed   
\end{coro}


The result parallels our previous formulas, as was to be expected.
Nevertheless, it does not follow from Theorem~\ref{pprm} in general,
because realisations of stationary point processes in $\RR^d$
generically fail to be FLC sets.
Before we discuss possible generalisations beyond the case of
positive random measures, let us look at some examples.
 
\begin{example} \textsc{Poisson cluster process}.
\label{poisson-cluster}
An important special case emerges when the centre process is the
homogeneous Poisson process of Example~\ref{ex-poisson}, with point
density $\pd$. Let $\gamma^{}_{P}$ and $\widehat{\gamma}^{}_{P}$ be
the corresponding measures.  If we couple a cluster component process
$\varPsi$ to it, with law $Q$ and $m:=\EE_Q (\varPsi) (\RR^d)$ its
expected number of points, our general formula for the compound
process $(\varPhi_{P},\varPsi_{Q})$ applies.  With
Lemma~\ref{finite-times-lebesgue}, the convolution formula can be
simplified, and the result reads as follows.

{}For almost all realisations of a Poisson cluster process
$(\varPhi_{P},\varPsi_Q)$, the natural autocorrelation measure 
exists and is given by
\[ 
    \gamma^{}_{R} = \, \gamma^{}_{P,Q}  = \, 
    (m \pd)^2 \lambda + \pd\ts\, \EE_Q 
    (\varPsi\!\conv\widetilde{\varPsi}) \, ,
\] 
where $\EE_Q (\varPsi\!\conv\widetilde{\varPsi})$ is a finite positive
measure (of expected total mass $\ge m^2$), due to our general
assumption that $\EE_Q \bigl( (\varPsi (\RR^d))^2 \bigr)$ is finite.
Consequently, the diffraction measure is almost surely given by 
\[
   \widehat{\gamma}^{}_{R} \; = \; (m \pd)^2 \delta_0 + 
   \pd\ts \, \bigl(\EE_Q (\varPsi\!\conv\widetilde{\varPsi})
   \bigr)^{\!\widehat{\hphantom{m}}} \!\!  \cdot\lambda,
\]
where $\bigl(\EE_Q (\varPsi\!\ts * \widetilde{\varPsi})
\bigr)^{\!\widehat{\hphantom{m}}}$ is a uniformly continuous
Radon-Nikodym density for Lebesgue measure.  These formulas include
the case of deterministic clusters; compare Example~\ref{det-cluster}.
\exend \end{example}

\begin{remark} \textsc{Random displacement of Poisson processes}.
  An interesting pair of processes is the combination of the
  homogeneous Poisson process from Example~\ref{ex-poisson} with Hof's
  random displacement model from Example~\ref{random-displacement}. A
  simple calculation shows that 
  \[
       \gamma^{}_{R} = \gamma^{(\nu)}_{P} = \gamma^{}_{P}
       \quad \text{and} \quad
       \widehat{\gamma}^{}_{R} = \widehat{\gamma}^{}_{P}
  \]
   in this  case (and, in fact, $R$ and $P$ have the same law here).
  From a physical point of view, this is in line with the
  behaviour of an ideal gas at high temperatures. When the Poisson
  process is a good model for the gas, and random displacement one for
  the disorder due to high temperature, compare the discussion in
  \cite{Hof-random}, the combination should still be an ideal gas --
  and this is precisely what happens, as reflected by the two
  identities.
\exend \end{remark}

\begin{remark} \textsc{Particle gas cluster process}.
It is clear that the particle gas of Example~\ref{particle-gas} satisfies
all requirements for a centre process, so that we can apply the
cluster process machinery to it, too. This produces physically interesting
and relevant examples with a substantial amount of point spectrum.
This observation remains true for more complicated particle gas
models with interactions, under certain conditions on the potential
of the underlying Gibbs measure, say; compare \cite{BZ} for
further details and examples.
\exend
\end{remark}

\begin{example} \textsc{Neyman-Scott processes}.
\label{Neyman-Scott}
Let $K$ be a non-negative random integer with law $\mathcal{L}
(K)=\mu$, mean $m:=\mathbb{E}_{\ts \mu} (K)$ and finite second moment,
$\mathbb{E}_{\ts \mu} (K^2) < \infty$. Now, let $Y_1, Y_2, \dots$ be a
family of $\mathbb{R}^d$-valued i.i.d.\ random variables with common
distribution $\nu$, and independent of $K$. Define the cluster
distribution via
\[ 
   \varPsi \; := \; \sum_{j=1}^K \delta^{}_{Y_j}, 
\]
i.e., a cluster has a random size $K$, while the positions of its
atoms are independently drawn from the probability distribution $\nu$.
The induced distribution for $\varPsi$ is again called $Q$. With a
calculation similar to the one in Example~\ref{random-displacement},
one finds
\[
   \EE_Q (\varPsi) (A) \, = \, \EE_Q \Bigl( \sum_{i=1}^{K}
   \mathbf{1}_{A} (X_i)\Bigr) \, = \, \EE_{\mu} \Bigl(
    \sum_{i=1}^{K} \int_{\RR^d} \mathbf{1}_{A} (x_i) 
    \dd\nu (x_i)\Bigr) \, = \, \EE_{\mu} \bigl(K\!\cdot\nu(A)\bigr)
    \, = \, m\ts\ts\nu (A)
\]
for $A\subset\RR^d$ Borel, so that $\EE_Q (\varPsi) = m\ts\nu$
and $\EE_Q (\varPsi) * \EE_Q (\widetilde{\varPsi}) = 
m^2 \ts (\nu*\widetilde{\nu}\ts)$. Moreover, one has
\[
   \EE_Q (\varPsi\! *\widetilde{\varPsi}) (A) \, = \,
   \EE_Q \Bigl(\sum_{k,\ell=1}^{K} \mathbf{1}_{A}
   (X_{k} - X_{\ell}) \Bigr) \, = \,
   m\ts\ts\delta_0 (A) + \EE_{\mu} \bigl(K(K-1)\bigr)
   (\nu * \widetilde{\nu}\ts) (A) \/ ,
\]
which gives $\EE_Q (\varPsi\! *\widetilde{\varPsi}) =
m\ts\ts\delta_0 + \EE_{\mu} \bigl(K(K-1)\bigr) (\nu * \widetilde{\nu})$,
so that the general formulas from Theorem~\ref{pprm}
can now be applied again. Note that
$\EE_{\mu} \bigl(K(K-1)\bigr) = \EE_{\mu} (K^2) - m$.

If the centre process is once more the homogeneous Poisson process
with mean (point) density $\pd$, Lemma~\ref{finite-times-lebesgue}
gives similar simplifications as in Example~\ref{poisson-cluster}.
Consequently, for the resulting law $R$, the autocorrelation is almost
surely given by
\[ 
   \gamma^{}_{R} \; = \; 
      (m \pd)^2 \ts\ts \lambda + m\pd\ts\ts \delta_0 +
         \pd\ts \bigl(\EE_{\ts \mu} (K^2) - m\bigr)\ts 
       (\nu * \widetilde{\nu}\ts) \, , 
\]
whence the corresponding diffraction measure is given by 
\[ 
    \widehat{\gamma}^{}_{R} \; = \; 
    (m \pd)^2 \ts\ts \delta_0 + \pd \ts \bigl( m +
     ( \mathbb{E}_{\ts \mu} (K^2) - m )
     \lvert \widehat{\nu} \rvert^2 \bigr) \, \lambda \, ,
\]
which is an interesting extension of the Poisson process;
compare \cite[Ex.~8.2(f)]{DVJ1} for a circularly symmetric
case in $\RR^{2}$.
\exend \end{example}

\subsection{Autocorrelation for signed (ergodic) processes}
\label{extension-signed}

It is intuitively clear that the results of this section are not
really restricted to point processes or positive measures for the
clusters.  Here, we sketch how they can be adapted to the situation of
signed random measures.  Consider a stationary, possibly signed,
random measure $\varPsi$ (with law $Q$ and `finite second moments',
meaning that $\EE_Q \bigl( ( \lvert \varPsi \rvert (A))^{2} \bigr) <
\infty$ holds for any bounded $A \subset \RR^d$), with second moment
measure $\mu^{(2)}$, defined as before via bounded $f$ of compact
support as
\[ 
    \int_{\RR^d \times \RR^d} f(x,y) \dd \mu^{(2)}(x,y) 
    \; = \;  \EE_Q \Bigl( \int_{\RR^d \times\ts \RR^{d}} f(x,y) 
    \dd\varPsi(x) \dd\varPsi(y) \Bigr).
\] 
The reduced second moment measure $\mu^{(2)}_{\mathrm{red}}$ 
on $\RR^d$ with the property
\begin{equation}  \label{eq:mu2red}
   \mu^{(2)}_{\mathrm{red}} (f * \tilde{g}) 
    \; = \; \mu^{(2)}(f \otimes g)
\end{equation} 
is defined in complete analogy to the positive case. 
The analogue of Theorem~\ref{thm:palm}  is: 

\begin{theorem} \label{thm:palm.variant} 
  Let $\ts\varPhi$ be a stationary and ergodic, random, signed measure
  with distribution $P$.  Assume that $\ts\varPhi$ has finite second
  moments in the sense that\/ $\EE_P \bigl(( \lvert \varPhi \rvert (A)
  )^{2} \bigr) < \infty$ for any bounded measurable set $A \subset
  \mathbb{R}^d$ $($which follows, for example, from $\EE_P \bigl( \bigl(
  \lvert \varPhi \rvert (B_r (x))\bigr)^{\nts 2} \bigr) < \infty$ for all
  $x\in\RR^d$ and some open ball $B_r)$.  Let $\varPhi_n :=
  \varPhi|_{B_n}$ denote the restriction of\/ $\varPhi$ to the ball of
  radius $\ts n$ around\/ $0$.  Then, the natural autocorrelation of
  $\ts\varPhi$, which is defined with an averaging sequence of nested,
  centred balls, almost surely exists and satisfies
\[ 
   \gamma^{(\varPhi)}_{P} \, := \,
   \lim_{n\to\infty} \, \frac{\varPhi_n \! * 
   \widetilde{\varPhi_n}} {\lambda(B_n)} \; = \;
   \lim_{n\to\infty}\, \frac{\varPhi_n \! * 
   \widetilde{\varPhi}} {\lambda(B_n)}
   \; = \;  \mu^{(2)}_\mathrm{red}
   \; = \; \gamma^{}_{P}\, ,
\] 
where the limit refers to the vague topology on $\cN$.  Here,
$\mu^{(2)}_\mathrm{red}$ is the reduced second moment measure 
of $P$ according to $\eqref{eq:mu2red}$. 
\end{theorem} 

\begin{proof}
The proof is a variation of that of Theorem~\ref{thm:palm}. 
Fix a continuous function $h :\, \RR^d \to \RR$ with 
compact support. We have to check that 
\begin{equation} 
\frac{1}{\lambda(B_n)} \bigl( \varPhi_n \! * 
   \widetilde{\varPhi_n}\,\bigr)(h)\; \xrightarrow{\, n\to\infty\,} 
   \; \mu^{(2)}_\mathrm{red} (h)
   \qquad \mbox{(a.s.)}. 
\end{equation}

Let $\varPhi$ be an ergodic, random, signed measure 
as above and $F$ an ergodic random function on $\RR^d$, 
the latter with the property that 
\begin{equation} 
   \EE_{P} \bigl( \int_{A} \lvert F(x) \rvert \dd 
   \lvert \varPhi \rvert (x) \bigr)
   \, < \, \infty 
\end{equation}
for any bounded measurable $A \subset \RR^d$. We can then define 
an additive covariant spatial process $X_A$ in the sense of 
\cite{NZ79}, indexed by bounded measurable subsets $A$, via 
\[ 
    X_A \, := \, \int_A F(x) \dd\varPhi(x). 
\] 
Note that ergodicity of $\varPhi$ and $F$ implies 
that $(X_A , \RR^d)$ is again ergodic, meaning that the 
shift-invariant $\sigma$-field is trivial. 
Now, \cite[Cor.~4.9]{NZ79} yields 
\[ 
   \lim_{n\to\infty} \frac{1}{\lambda(B_n)} X_{B_n} 
   \, = \, \EE_{P} \Bigl( \frac{1}{\lambda(B_1)} X_{B_1} \Bigr)
   \qquad \mbox{(a.s.)}.
\]
Applying this to $\varPhi$ as in the theorem, together
with $F(x):= \int_{\RR^d} h(x-y) \dd\varPhi(y)$, yields 
\begin{align*} 
     \lim_{n\to\infty} & \frac{1}{\lambda(B_n)} 
     \int_{B_n} F(x) \dd\varPhi(x) 
     \, =  \, \lim_{n\to\infty} \frac{1}{\lambda(B_n)} 
     \bigl( \varPhi_n * \widetilde{\varPhi} \bigr)  (h) \\
 & =  \, \EE_{P} \Bigl( \frac{1}{\lambda(B_1)} 
     \int_{B_1} \int_{\RR^d} h(x-y) \dd\varPhi(y) \dd\varPhi(x) \Bigr) \\
 & = \, \frac{1}{\lambda(B_1)} \int_{\RR^d \times\ts \RR^d} 
       \mathbf{1}_{B_1}(x) \, h(x-y) \dd\mu^{(2)}(x,y) \\
 & =  \, \frac{1}{\lambda(B_1)} 
        \int_{\RR^d} \int_{\RR^d}\mathbf{1}_{B_1}(x) \, h(z) 
        \dd\mu^{(2)}_\mathrm{red}(z) \dd x 
       \,  =   \int_{\RR^d} h \dd\mu^{(2)}_\mathrm{red}
\end{align*}
almost surely, which is almost the claim. 
The difference between $\varPhi_n * \widetilde{\varPhi}$ 
and $\varPhi_n * \widetilde{\varPhi_n}$ 
can be treated as in the proof of Theorem~\ref{thm:palm}. 
\end{proof}

Combining Proposition~\ref{prop:signedergodic} and
Theorem~\ref{thm:palm.variant}, and observing that the calculations in
the proof of Proposition~\ref{reduction} carry over literally to the
signed case, we obtain
\begin{coro}  \label{coro:signed}
   The statements of Theorem~$\ref{compound-thm}$ and 
   Corollary~$\ref{compound-diffraction}$ remain true 
   for cluster processes with signed clusters.  \qed 
\end{coro}

\begin{example} \textsc{Signed Poisson process.} \label{ex-signed-poisson}
If we combine the homogeneous Poisson process of Example~\ref{ex-poisson}
with the random weight model of Example~\ref{random-weight}, and
chose weights $1$ and $-1$ with equal probability, Corollary~\ref{coro:signed}
implies the almost sure diffraction
\[
    \widehat{\gamma} = \rho \ts \lambda\ts .
\]
In particular, one has $\widehat{\gamma}=\lambda$ for density $\rho =
1$, which makes this signed Poisson point set, on the level of the
$2$-point correlations, indistinguishable from the signed Bernoulli
sequence (or process) on $\ZZ^d$. 
This is remarkable in view of the rather different
geometric structure and demonstrates the intrinsic difficulty of the
corresponding inverse problem.  
\exend \end{example}

\subsection{Equilibria of critical branching Brownian motions in 
$d \ge 3$} 
Consider a system of particles performing independent Brownian motions
in $\mathbb{R}^d$, $d \ge 3$ (for ease of comparison with the cited
literature, we assume that the variance parameter is $\sigma^2=2$).

Additionally, each particle, after an exponentially distributed
lifetime with parameter $V$, either doubles or dies, where each
possibility occurs with probability $1/2$. In the situation of a birth
event, the daughter particles appear at the position of the mother.
Note that if we start with a finite number of particles, the expected
number of particles is preserved for all time, as the expected number
of offspring equals $1$.  This is what `critical' in the name refers
to.  Imagine we start such a system from a homogeneous Poisson process
of density $\pd$, denote by $\varPhi_t$ the random
configuration observed at time $t \ge 0$, and its distribution by
$P_t$.  Here, $P_t$ is stationary with density $\pd$, see \cite{GW}
and the references given there for background.

It follows from \cite[Thm.~2.3]{GW} that the first moment measure of
the Palm distribution of $P_t$ is given by
\begin{equation} \label{eq:BB1} 
    I_{(P_t)_0} \; = \; \delta_0 + (\pd + f_t) \lambda \, , 
\end{equation}
where 
\[ 
   f_t(x) \; = \;
   V\int_0^t \int_{\mathbb{R}^d} p_s(0,y)\, p_s(y,x) \dd y \dd s 
       = \frac{V}{2} \int_0^{2t} p_{u}(0,x) \dd u \, ,
\]
with $p_t(x,y) = (4\pi t)^{-d/2} \exp\big(-|x-y|^2/(4t)\big)$ the
$d$-dimensional Brownian transition density (with variance parameter
$2$).  As explained in \cite{GW}, there is a genealogical
interpretation behind (\ref{eq:BB1}): In view of the interpretation of
the Palm distribution as the configuration around a typical
individual, $\delta_0$ is the contribution of this individual, $f_t \,
\lambda$ that from its relatives in the family decomposition of the
branching process, and $\pd \ts\ts \lambda$ is the contribution from
unrelated individuals.

Furthermore, by \cite[Thm.~2.2]{GW}, $P_t$ converges (vaguely) towards
$P_\infty$, which is the unique, ergodic, equilibrium distribution of
density $\pd$ (cf.~\cite{BCG97} for uniqueness), and the limit
$t\to\infty$ can be taken in (\ref{eq:BB1}) to obtain
\[ 
   I_{(P_\infty)_0} \; = \; \delta_0 + (\pd + f_\infty) \lambda \, , 
\]
where 
\[ 
     f_\infty(x) \, = \, 
     \frac{V}{2} \int_0^{\infty} p_{u}(0,x) \dd u 
     \, = \, \frac{V}{2}\, 
     \frac{\Gamma\big(\frac{d-2}{2}\big)}{4\pi^{d/2}}
     \,  \frac{1}{|x|^{d-2}}
\]
is (up to the prefactor $V/2$) the Green function of Brownian motion. 
Thus, using Lemma~\ref{lemma-1}, we have 
\begin{theorem} \label{cbbm}
    Let $\varPhi_\infty$ be a realisation of the critical branching
    Brownian motion, from the equilibrium distribution $P_\infty$.  The
    autocorrelation is then almost surely given by
\[ 
   \gamma \; = \;  \pd\ts \delta_0 + \pd\ts(\pd + f_\infty) \lambda \, , 
\]
   while 
\[ 
   \widehat{\gamma} \; = \;
   \pd^2 \delta_0 +  \pd\Big( 1 + \frac{V}{2} \,
       \frac{1}{4\pi^2|k|^{2}}\Big) \,  \lambda \, . 
\]
   is the corresponding diffraction measure.   \qed
\end{theorem}

\begin{remark} \textsc{Extension of Theorem~\ref{cbbm}}.
  One can also consider the scenario where, instead of Brownian
  motion, particles move during their lifetime according to a
  symmetric, stable process of index $\alpha \in (0,2]$ in
  $\mathbb{R}^d$ ($\alpha=2$ corresponds to Brownian motion).  Such
  processes have discontinuous paths, and their transition density
  $p^{(\alpha)}_t(x,y) = p^{(\alpha)}_t(0,y-x)$ satisfies
\[ 
    \int_{\mathbb{R}^d} e^{i k \cdot x} p^{(\alpha)}_t(0,x) 
    \dd x   \; = \;  \exp(-t |k|^\alpha) 
\]
(in general, no explicit form of $p^{(\alpha)}_t$ is known). By
\cite[Thm.~2.2]{GW}, non-trivial equilibria exist if the spatial
dimension $d$ satisfies $d > \alpha$. In this case, a reasoning
analogous to that above yields the following: The autocorrelation of a
realisation $\varPhi^{(\alpha)}_\infty$ of the equilibrium of a system
of critical, branching, symmetric $\alpha$-stable processes (with
density $\pd$) is almost surely given by
\[ 
   \gamma \; = \; \pd\ts \delta_0 + \pd\ts 
      (\pd + f_\infty^{(\alpha)}) \lambda\, ,  
\]
where 
\[ 
    f_\infty^{(\alpha)}(x) \, = \, 
    \frac{V}{2} \int_0^\infty p^{(\alpha)}_u(0,x) \dd u 
    \, = \, \frac{V}{2}\, \frac{\Gamma((d-\alpha)/2)}{2^\alpha \pi^{d/2}
    \Gamma(\alpha/2)} \, \frac{1}{|x|^{d-\alpha}}
\]
(for the form of the Green function of the symmetric $\alpha$-stable
process, see \cite[Ex.~1.7]{BG}).  Hence, the diffraction measure 
is almost surely given by
\[ 
    \widehat{\gamma} \; = \; \pd^2\delta_0 + 
    \pd \Big( 1 + \frac{V}{2} \frac{1}{(2\pi)^\alpha |k|^\alpha}
                    \Big) \lambda \, ,
\]
by another application of Lemma~\ref{lemma-1}.
\exend \end{remark}

Note that, due to the independence properties of the branching
mechanism, these equilibria can also be considered as Poisson cluster
processes.  In contrast to the scenario considered above, clusters in
$\varPhi_\infty$ are infinite, and the spatial correlation decays only
algebraically (without being integrable).

\section{Outlook}

This article demonstrates that various aspects of mathematical
diffraction theory for random point sets and measures can be
approached systematically with methods from point process theory, as
was originally suggested in \cite{Gou1}.  At the same time, the
approach is sufficiently concrete to allow for many explicitly
computable examples, several of which were presented above.  They
comprise many formulas from the somewhat scattered literature on this
subject in a unified setting. There are, of course, many more
examples, but we hope that the probabilistic platform advertised here
will prove useful for them as well.

The next step in this development needs to consider point processes
and random measures with interactions, such as those governed by Gibbs
measures. First steps are contained in
\cite{Hof-random,BH,Gou1,K1,K2,BS,DM,BZ} and indicate that both
qualitative and quantitative results are possible, though some further
development of the theory is needed.

A continuation along this path would also make the results more
suitable for real applications in physics and crystallography, though
it is largely unclear at the moment what surprises the corresponding
inverse problem might have to offer here.

\section*{Appendix:\ Ergodicity for cluster processes with 
signed random measures}

Let $\cM=\cM (\RR^d)$ be the  space of (locally finite) real or signed
measures on $\RR^d$, equipped with the topology of vague convergence,
with $\cM^{+}=\cM^{+} (\RR^d)$ denoting the subspace of positive measures.
Let $\varSigma_{\cM}$ denote the Borel $\sigma$-algebra of $\RR^d$. 
Note that the latter is also generated by the
mappings $\cM \ni \mu \mapsto \mu(A)$, for bounded
and measurable sets $A \subset \RR^d$.  Recall that any $\mu \in
\cM$ admits a unique Hahn-Jordan decomposition 
\[ 
   \mu \, = \,  \mu_+ - \mu_-  \, , \quad 
   \mbox{with $\mu_+$, $\mu_- \in \cM^{+}$
   mutually singular}.
\] 
The mappings $\mu \mapsto \mu_+$ and $\mu \mapsto \mu_-$ are
$\varSigma_{\cM}$-measurable.  We write $\lvert \mu \rvert := \mu_{+}
+ \mu_{-} \in \cM^{+}$ for the total variation measure of $\mu$.  A
\emph{random signed measure} $\varPhi$ is a random variable with
values in $(\cM, \varSigma_{\cM})$.  In the context of signed random
measures, it is convenient to work with the \emph{characteristic
  functional}
\begin{equation} 
  \varphi^{}_{\varPhi}(h)  \, := \,
  \EE\Big[ \exp\bigl(i {\textstyle \int h \dd\varPhi}\bigr) \Big],
\end{equation}
which is defined for any $h \! : \, \RR^d \to \RR$ that is bounded and
measurable with compact support. Here and below, we suppress $\RR^d$
as the integration region. In analogy to the Laplace functional for
positive random measures, the distribution of $\varPhi$ is determined
by $\varphi^{}_\varPhi$.

Here, we are interested in signed cluster processes: Let $\varPhi$ be a
stationary counting process with finite intensity $\rho$, and
$\varPsi_j$ (with $j\in\NN$) independent (and independent from $\varPhi$),
identically distributed, random, signed measures such that
$\EE\big[|\varPsi_1|\big]$ is a {\em finite}\/ measure.  Then, given a
realisation $\varPhi = \sum_j \delta_{X_j}$, where $X_j$ are the
positions of the atoms of $\varPhi$ (in some enumeration), the cluster
process is defined as
\begin{equation}\label{eq:Xidef1} 
     \varXi \, := \, \sum_j T^{}_{X_j} \varPsi_j \, . 
\end{equation}
Note that for any bounded $B \subset \RR^d$,
\[
     \EE\big[ |\varXi(B)| \big] 
     \, \le \, \EE\Big[ \sum_j |\varPsi_j|(B-X_j) \Big] 
     \, = \, \rho \int_{\RR^d} \int_{B-x} \dd\EE\big[|\varPsi_1|\big] \dd x 
     \, = \, \rho \big(\EE\big[|\varPsi_1|\big] \conv {\lambda} \big)(B) 
     \, < \, \infty \, ,
\]
so that \eqref{eq:Xidef1} is indeed well-defined. 

\begin{lemma} 
\label{lemma:signedergodic}
Let $\varPsi$ be a signed random measure on $\RR^d$. The following 
are equivalent:
\begin{enumerate}
\item $\varPsi$ is ergodic.
\item For any $U, V \in \varSigma_{\cM},$, 
\[
   \lim_{n\to\infty} \frac{1}{\lambda(B_n)} 
   \int_{B_n} \bigl( \PP\big( \varPsi \in U \cap T_x V \big) 
            -\PP(\varPsi \in U) \, \PP(\varPsi \in V)\bigr) \dd x = 0\, .
\]
\item For any $g, h \! : \, \RR^d \to \RR$ measurable with compact support, 
\[
  \lim_{n\to\infty} \frac{1}{\lambda(B_n)} 
  \int_{B_n}  \bigl(\varphi_\varPsi(g + T_xh) - 
        \varphi^{}_\varPsi(g)\, \varphi^{}_\varPsi(h) \bigr) \dd x = 0\, .
\]
\end{enumerate} 
Furthermore, it suffices to restrict to $U, V$ to a 
semiring which generates $\varSigma_{\cM}$ in $(2)$, and  
it suffices to restrict to continuous $g, h$ with compact support
in $(3)$. 
\end{lemma}
\begin{proof} This is a straightforward adaptation of the proofs of
  Propositions~12.3.III and 12.3.VI and Lemma~12.3.II of
  \cite{DVJ2} to the signed case.
\end{proof}

The following result is an analogue \cite[Prop.~12.3.IX]{DVJ2} 
for the signed measure case. Since we have not been able to find a
proof in the literature, we provide a sketch.
\begin{prop} \label{prop:signedergodic}
   Let $\varPhi$, $\varPsi_j$, and $\varXi := \sum_j T^{}_{X_j} \varPsi_j$ 
   be as above. If $\varPhi$ is ergodic, then $\varXi$ is ergodic as well.
\end{prop}
\begin{proof}[Sketch of proof] 
We verify condition~(3) from Lemma~\ref{lemma:signedergodic}. 
Observe that for any $f \! : \, \RR^d \to \RR$ with compact support and 
any $\varepsilon > 0$, we can find $R < \infty$ 
such that 
\begin{equation} \label{eq:contribfaraway}
   \PP\Bigg( \sum_{j\, :\, \lvert X_j \rvert \ge R} 
         \bigg| \int f \dd (T_{X_j} \varPsi_j) \bigg| 
         \, \ge \, \varepsilon \Bigg) \le \varepsilon\,. 
\end{equation} 
To check (\ref{eq:contribfaraway}), let
$R'$ be large enough so that $\supp(f) \subset [-R',R']^d$, and note
that for $R > R'$, the left-hand side of (\ref{eq:contribfaraway}) 
is bounded by 
\[
  \PP\Bigg( \sum_{j\, :\, \lvert X_j  \rvert_\infty \ge R} \hspace{-1em}
         |\varPsi_j|\big([-R',R']^d+X_j\big) \ge
         \frac{\varepsilon}{||f||_\infty} \Bigg) 
         \, \le \,   \frac{||f||_\infty}{\varepsilon} \,
         \EE \Bigg[ \sum_{j\, :\, \lvert X_j \rvert_\infty \ge R} \hspace{-1em}
               |T_{X_j}\varPsi_j|\big([-R',R']^d\big) \Bigg].
\]
The expectation on the right-hand side above equals
\[
\begin{aligned}
 \rho \hspace{-1em} 
      \int\limits_{\RR^d \setminus [-R,R]^d} \int\limits_{\RR^d} &
      \mathbf{1}_{[-R',R']^d}(x-y) 
      \dd \EE\big[ |\varPsi_1| \big](y) \dd x \\
     & \le  \rho (2R')^d \, \EE\big[ |\varPsi_1| \big]
           \big( \RR^d \setminus [-(R-R'),(R-R')]^d\big), 
\end{aligned}
\]
which converges to $0$ as $R\to\infty$ because $\EE\big[ |\varPsi_1|
\big]$ is a finite measure.

Let $g, h \! : \, \RR^d \to \RR$ continuous with compact support and define
\[ 
   G(\varPhi) := \, \EE\Bigl[ \exp\bigl( i {\textstyle 
      \int g \dd \varXi}\bigr) \,
              \Big| \, \varPhi \Bigr],     \quad 
   H(\varPhi) := \, \EE\Bigl[ \exp\bigl( i {\textstyle 
       \int h \dd \varXi}\bigr) \,
              \Big| \, \varPhi \Bigr]. 
\]

Decompose 
\begin{eqnarray*} 
   \int (g + T_x h) \dd \varXi  & = & 
   \sum_{j \, : \, X_j \in [-R,R]^d} \int T_{X_j} g \dd \varPsi_j  \;
   + \!\!  \sum_{j \, : \, X_j \not\in [-R,R]^d} 
   \int T_{X_j} g \dd \varPsi_j \\
   && {} + \!\!
   \sum_{j \, : \, X_j \in [-R,R]^d-x} \int T_{X_j+x} h 
        \dd \varPsi_j \;
   + \!\! \sum_{j \, : \, X_j \not\in [-R,R]^d-x} 
    \int T_{X_j+x} h \dd \varPsi_j \ts ,
\end{eqnarray*} 
and choose $R$ so large that (\ref{eq:contribfaraway}) is fulfilled
for $f=g$ and $f=h$. Recall that, for any real-valued random variables
$X$, $Y$ with $\PP(|Y|\ge\varepsilon)\le\varepsilon$, we have
\[
\Big| \EE\,e^{i (X+Y)} - \EE\,e^{i X} \Big| 
\le \EE\, \Big| e^{i (X+Y)} - e^{i X} \Big|
\le \EE\Big[ \big|e^{i X}\big| \, \big|e^{i Y}-1\big|\Big] 
\le \varepsilon+\PP(|Y|\ge\varepsilon)
\le 2\varepsilon. 
\]
For $A \subset \RR^d$, write $\varXi_A := \sum_{j \, : \, X_j \in A}
T^{}_{X_j} \varPsi_j $ for the random measure which consists of
clusters with centres in $A$.  For $x \in \RR^d \setminus [-2R,2R]^d$,
we then have
\begin{eqnarray*}  
   \lefteqn{ \Big| \EE\left[  \exp\big({\textstyle i 
   \int (g + T_x h) \dd \varXi}\big)
                   \right] - \EE\left[ G(\varPhi) 
       H(T_x \varPhi) \right] \Big| } \\
      & \le & \Big| \EE\left[  \exp\big({\textstyle i 
    \int (g + T_x h) \dd \varXi}\big)
                          \right]
              -\EE\left[ \exp\left(  i  {\textstyle 
	    \int  g \dd \varXi_{[-R,R]^d} }  + i {\textstyle 
             \int T_{x} h \dd \varXi_{[-R,R]^d-x}} \right) \right] \Big| \\
     && {} +  \Big| \EE\left[ \EE\left[ \left. \exp\left( 
                i  {\textstyle  \int  g \dd \varXi_{[-R,R]^d} } 
	    + i {\textstyle  \int T_{x} h \dd \varXi_{[-R,R]^d-x}} \right)  
	  \right| \varPhi \right] \right] 
           - \EE\left[ G(\varPhi) H(T_x \varPhi) \right] \Big|. 
\end{eqnarray*} 
The first term on the right-hand side is bounded by $2\varepsilon$.
Observing that the conditional expectation in the second term 
is in fact a product because clusters with centres in disjoint regions 
are (conditionally) independent, we can bound the second term from 
above by
\[ \begin{aligned} 
&\Big| 
 \EE\left[ \EE\left[ \left. \exp\left( 
                           i  {\textstyle 
			     \int  g \dd \varXi_{[-R,R]^d}} \right) 
                       \right| \varPhi \right] 
    \left( \EE\left[ \left. \exp\left( 
                           i  {\textstyle 
			     \int  T_x h \dd \varXi_{[-R,R]^d-x}} \right) 
                       \right| \varPhi \right]
           - H(T_x \varPhi)
    \right) \right] \Big| \\
& \hspace{2em} {} + \Big| 
 \EE\left[ 
    \left( \EE\left[ \left. \exp\left( 
                           i  {\textstyle 
			     \int  g \dd \varXi_{[-R,R]^d}} \right) 
                       \right| \varPhi \right]
           - G(\varPhi)
    \right) H(T_x \varPhi) \right] \Big| \\
& \hspace{0.5em} 
\le \; \EE \left| \exp\left( 
                           i  {\textstyle 
			     \int T_x h \dd \varXi_{[-R,R]^d-x}} \right) 
                         - \exp\left( 
                           i  {\textstyle 
			     \int T_x h \dd \varXi} \right)
                       \right|   \\
& \hspace{2em}  + \EE \left| \exp\left( 
                           i  {\textstyle 
			     \int g \dd \varXi_{[-R,R]^d}} \right) 
                         - \exp\left( 
                           i  {\textstyle 
			     \int g \dd \varXi} \right) \right|,
\end{aligned} \] 
which is not more than $2\varepsilon$.

Thus, using the relation $\EE\left[ \EE\big[ \exp\big({\textstyle i
    \int (g + T_x h) \dd \varXi}\big) \, \big| \, \varPhi \big]\right]
= \varphi^{}_{\varXi}(g+T_x h)$ together with $\EE \left[G(\varPhi)\right] =
\varphi^{}_{\varXi}(g)$ and $\EE \left[H(\varPhi)\right] = 
\EE\left[ H(T_x \varPhi)\right] = \varphi^{}_{\varXi}(h)$, we obtain
\begin{equation}
\begin{aligned}
\limsup_{n\to\infty} \,& \frac{1}{\lambda(B_n)} 
   \left|  \int_{B_n}  \bigl(\varphi^{}_\varPsi(g + T_xh) - 
   \varphi^{}_\varPsi(g)\, \varphi^{}_\varPsi(h) \bigr) \dd x \,  
   \right| \\  & \le \,
\limsup_{n\to\infty} \frac{1}{\lambda(B_n)} 
  \left| \int_{B_n} \bigl( \EE\big[G(\varPhi) H(T_x \varPhi)\big] 
             -\EE\big[G(\varPhi)\big] \EE\big[H(\varPhi)\big] \bigr)
                     \dd x \,\right| + 4\varepsilon 
\, = \, 4\varepsilon 
\end{aligned} 
\end{equation}
by ergodicity of $\varPhi$ (in order to deduce this literally from
statement (2) in Lemma~\ref{lemma:signedergodic}, one can for instance
discretise the support of $g$ and $h$ and approximate $G(\varPhi)$,
$H(\varPhi)$ with functions depending only on the random vector
$(\varPhi(c_i))_{1\le i \le N}$, where $\{c_i \mid 1\le i \le N\}$ is a
collection of disjoint (small) cubes). Finally, take $\varepsilon\to 0$
to conclude.
\end{proof}

\bigskip

\section*{Acknowledgements}
This work was supported by the German Research Council (DFG),
within the CRC 701, by the Natural Sciences and Engineering Research
Council of Canada (NSERC), and by the RiP program at Oberwolfach.
We thank the referees for their thorough analysis of the paper and
for making useful suggestions that have helped to improve it.

\end{document}